\documentclass[twocolumn,twocolappendix]{aastex63}

\usepackage{amsmath,amsopn,amsxtra,txfonts}
\usepackage{comment} 
\usepackage{natbib} 
\usepackage{hyperref}

\usepackage{threeparttablex}

\newcommand{\nsightlines}{8}

\newcommand{\ha}{H$\alpha$}

\newcommand{\hi}{H\textsc{~i}}
\newcommand{\hii}{H\textsc{~ii}}
\newcommand{\kms}{ \ifmmode{\rm km\thinspace s^{-1}}\else km\thinspace s$^{-1}$\fi}
\newcommand{\K}{\ensuremath{ \, \mathrm{K}}}
\newcommand{\pc}{\ensuremath{ \, \mathrm{pc}}}
\newcommand{\kpc}{\ensuremath{\, \mathrm{kpc}}}
\newcommand{\cm}{\ensuremath{ \, \mathrm{cm}}}

\newcommand{\s}{\ensuremath{ \, \mathrm{s}}}

\newcommand{\R}{\ensuremath{ \, \mathrm{R}}}

\newcommand{\vlsr}{\ifmmode{{v}_{\rm{LSR}}}\else ${v}_{\rm{LSR}}$\fi}

\newcommand{\dg}{\ifmmode{^{\circ}}\else $^{\circ}$\fi}

\begin{document}

\author[0000-0003-0536-3081]{Suraj Poudel}
\affiliation{Department of Physics \& Astronomy, Texas Christian University, Fort Worth, TX 76129, USA}

\author{April Horton}
\affiliation{Department of Physics \& Astronomy, Texas Christian University, Fort Worth, TX 76129, USA}

\author[0000-0003-2308-8351]{Jo Vazquez}
\affiliation{Department of Physics \& Astronomy, Texas Christian University, Fort Worth, TX 76129, USA}

\author[0000-0001-5817-0932]{Kathleen A. Barger}
\affiliation{Department of Physics \& Astronomy, Texas Christian University, Fort Worth, TX 76129, USA}

\author[0000-0003-4237-3553]{Frances H. Cashman}
\affiliation{Space Telescope Science Institute, 
3700 San Martin Drive,
Baltimore, MD 21218, USA}
\affil{Department of Physics, Presbyterian College, Clinton, SC 29325, USA}

\author[0000-0003-0724-4115]{Andrew J. Fox}
\affil{AURA for ESA, Space Telescope Science Institute, 3700 San Martin Drive, Baltimore, MD, 21218, USA}
\affil{Department of Physics \& Astronomy, Johns Hopkins University, 3400 N. Charles St., Baltimore, MD 21218, USA}

\author[0000-0001-9158-0829]{Nicolas Lehner}
\affiliation{Department of Physics and Astronomy, University of Notre Dame, Notre Dame, IN 46556, USA}

\author[0000-0001-9982-0241]{Scott Lucchini}
\affiliation{Center for Astrophysics $|$ Harvard \& Smithsonian, 60 Garden Street, Cambridge, MA 02138, USA}

\author[0000-0002-7955-7359]{Dhanesh Krishnarao}
\affiliation{Department of Physics, Colorado College, Colorado Springs, CO 80903, USA}

\author[0000-0003-2730-957X]{N.\ M.\ McClure-Griffiths}
\affiliation{Research School of Astronomy \& Astrophysics, Australian National University, Canberra, Australia}

\author{Elena D'Onghia}
\affiliation{Department of Astronomy, University of Wisconsin, Madison, WI 53706, USA}

\author[0000-0002-7982-412X]{Jason Tumlinson}\affiliation{Space Telescope Science Institute, 
3700 San Martin Drive,
Baltimore, MD 21218, USA}

\author[0000-0001-6398-5466]{Ananya Goon Tuli}
\affiliation{Department of Physics and Astronomy, University of Notre Dame, Notre Dame, IN 46556, USA}

\author{Lauren Sdun}\affiliation{Department of Physics \& Astronomy, Texas Christian University, Fort Worth, TX 76129, USA}

\author{Stone Gebhart}\affiliation{Department of Physics \& Astronomy, Texas Christian University, Fort Worth, TX 76129, USA}

\author{Katherine Anthony}
\affiliation{Department of Computing, Mathematics and Physics,
Messiah University, 
Mechanicsburg, PA, 17055, USA}
\affiliation{Department of Physics \& Astronomy, Texas Christian University, Fort Worth, TX 76129, USA}

\author{Bryce Cole}\affiliation{Department of Physics \& Astronomy, Texas Christian University, Fort Worth, TX 76129, USA}

\author[0000-0002-1272-3017]{Jacco Th. van Loon}
\affiliation{Lennard-Jones Laboratories, Keele University, ST5 5BG, UK}

\author[0000-0001-6326-7069]{Julia Roman-Duval}
\affiliation{Space Telescope Science Institute, 3700 San Martin Drive, Baltimore, MD 21218, USA}

\author[0000-0003-0742-2006]{Yik Ki Ma}
\affiliation{Research School of Astronomy \& Astrophysics, Australian National University, Canberra, Australia}

\author[0000-0001-6846-5347]{Callum Lynn}
\affiliation{Research School of Astronomy \& Astrophysics, Australian National University, Canberra, Australia}

\author{Min-Young Lee}
\affiliation{Korea Astronomy and Space Science Institute
776 Daedeok-daero, Yuseong-gu, Daejeon 34055, Republic of Korea}

\author[0000-0002-4814-958X]{Denis Leahy}
\affiliation{Department of Physics and Astronomy,
University of Calgary,
Calgary, Canada, T2N 1N4}

\title{The Gaseous Blowout of the 30 Doradus Starburst Region in the LMC\footnote{Based on observations made with the NASA/ESA \textit{Hubble Space Telescope}, obtained at the Space Telescope Science Institute, which is operated by the Association of Universities for Research in Astronomy, Inc. under NASA contract No. NAS5-26555.}}

\begin{abstract}
Widespread galactic winds emanate from the Large Magellanic Cloud (LMC), with the 30~Doradus starburst region generating the fastest and most concentrated gas flows. We report on the gas distribution, kinematics, and ionization conditions of the near-side outflow along 8~down-the-barrel sightlines using UV absorption-line observations from the HST's ULLYSES program for this region along with H\textsc{~i} 21-cm observations from the GASS and GASKAP surveys. We find that within $\Delta\theta\lesssim1.7\arcdeg$ from the center of 30~Doradus, the wind reaches maximum speeds of $100-150\,\text{km}\,\text{s}^{-1}$ from the LMC's disk. The total integrated column densities of low-ions (O\textsc{~i}, Si\textsc{~ii}, and Fe\textsc{~ii}) in the blueshifted wind, up to $v_{\rm LSR}=150\,\text{km}\,\text{s}^{-1}$, are highest near the center and decline radially outward. We estimate an outflow mass of $M_{\rm outflow,\,Si\textsc{~ii}}\approx(5.7-8.6)\,\times 10^{5} M_{\odot}$, outflow rate of $\dot{M}_{\rm outflow}\gtrsim0.02 M_{\odot}\,\text{yr}^{-1}$, and mass loading factor of $\eta\gtrsim0.10$ within $\Delta\theta\lesssim0.52\arcdeg$ from the center of 30~Doradus. The observed ion ratios---together with photoionization modeling---reveal that this wind is roughly $40-97\%$ photoionized. The metallicities and dust depletion patterns of the high-velocity absorbers at $v_{\rm LSR}\approx+120\,\text{km}\,\text{s}^{-1}$ can be explained by either a foreground Milky Way (MW) halo cloud or an outflow from the LMC. For the high-ions, Si\textsc{~iv} and C\textsc{~iv} are broader and kinematically offset from the low-ions, suggesting turbulent mixing layers (TMLs) existing in the wind. Finally, our hydrodynamical simulations of the Magellanic Clouds (MCs) and MW system suggest that the Magellanic Corona can protect the LMC winds from the ram-pressure forces exerted by the MW's halo.
\end{abstract}

\section{Introduction}

Galactic winds are high-speed outflows of gas and dust that emanate from galaxies at large scales. They are driven by a variety of processes, including the energy and momentum generated by supernovae, active galactic nuclei (AGN), stellar winds, and other sources of galactic feedback. 
The interplay between outflows associated with galactic winds and inflows represents a delicate balance that is a major part of the baryon cycle.
Star-formation feedback can induce large-scale gas circulation in the circumgalactic and interstellar medium (CGM \& ISM) that help to regulate the galaxy's star-formation rate and chemical composition (\citealt{2005ARAA..43..769V}, \citealt{2008MNRAS.387..577O}).

Despite the importance of galactic winds, our understanding of their behavior and effects on both small (10s of \pc) and large (10s of \kpc) scales is  limited. Galactic winds are diffuse and faint and are therefore difficult to detect. Absorption-line studies are far more sensitive than imaging and emission-line studies to low-density material as they scale linearly with density as opposed to the density squared, but are unable to map the wind's properties. Further, degeneracies in the measurable properties of CGM material typically limit our ability to determine its origin, whether from an AGN, stellar feedback, a neighboring galaxy, halo-gas condensations, or the intergalactic medium. Accurately establishing the presence of outflows is challenging except at the disk and halo interface. 

The LMC is an ideal galaxy for studying the impact of stellar feedback due to its proximity at  $d_\sun=50\,\kpc$ away (\citealt{2014AJ....147..122D} and references therein) and its face-on orientation at an inclination angle of $i \approx23.4\arcdeg$ \citep{2022ApJ...927..153C}. 
This galaxy shows no sign of AGN activity in X-ray emission (e.g., \citealt{2012ApJ...746...27K}), which indicates that its galactic winds are mostly powered by its stars. 
Elevated stellar activity is observed extensively throughout the LMC's disk, which encompasses 400 H\textsc{~ii} regions (e.g., \citealt{1976A&AS...23..181D}, \citealt{2012ApJ...755...40P}) and more than 50~supernova remnants (SNRs: \citealt{2016A&A...585A.162M,2017ApJ...837...36L}).

Several UV absorption-line studies have detected a large-scale outflow from the LMC, including \citet{2007MNRAS.377..687L}, \citet{2016ApJ...817...91B}, \citet{2024arXiv240204313Z}, and references therein. Additionally, the \citet{2003MNRAS.339...87S} and \citet{Ciampa2020} studies have mapped this wind in 21-cm H\textsc{~i} and \ha\ emission, respectively. \citet{2016ApJ...817...91B} confirmed that these outflows are associated with a galactic wind due to their symmetry on both sides of the galaxy and estimated that the winds have a mass of M$_{\rm wind}\geq 10^{7}M_{\odot}$. \citet{Ciampa2020} established that the 30~Doradus region of the LMC---the region with the most extreme stellar activity---is producing the fastest and most concentrated winds. 

In this study, we investigate the properties of the winds produced by the 30~Doradus starburst using UV absorption- and \hi\ 21-cm emission line observations toward LMC disk stars. In Section~\ref{section:observations}, we describe these observations and their reduction. 
We outline our process for identifying the LMC's wind in Section~\ref{section:WindIdentification}. 
In Section~\ref{section:Analysis}, we summarize our technique for flux normalization, fitting Voigt profiles to the absorption features, and handling overly saturated and blended lines. 
We analyse the gas distribution in Section~\ref{sec:gas_distribution} followed by gas kinematics in Section~\ref{sec:gas_kinematics}. We explore the ionization state of the wind using ion ratios and \textsc{Cloudy} radiative transfer simulations in Section~\ref{sec:ionization}. In Section~\ref{section:simulation}, we discuss the results from the high-resolution simulation of the LMC. We discuss our results and put them in context of previous studies in Section~\ref{section:Discussion} and summarize our main conclusions in Section~\ref{section:Summary}.

\section{Observation and Data Reduction} \label{section:observations}

\begin{figure*}
  \centering
  \includegraphics[width=1\textwidth]{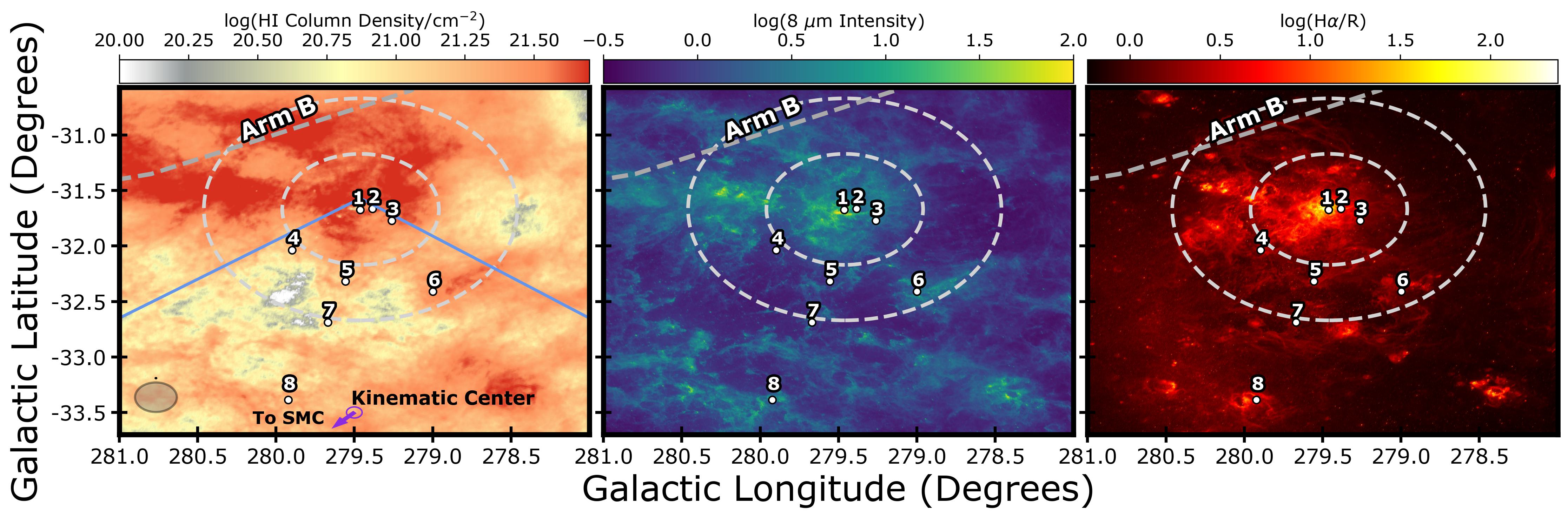}
  \caption{Left: Integrated H\textsc{~i} emission map of the LMC's disk over the range of $+185\le\  v_{\rm LSR}\le+335\,\kms$ using GASKAP observations. Middle: $8~\mu m$ emission using Spitzer observations. Right: \ha\ emission of this  from the Magellanic Cloud Emission-line Survey (MCELS; \citealt{1999IAUS..190...28S}). The positions of our \nsightlines\ stellar targets are marked with white points and have labels that match the indices listed in Table~\ref{table:targets}. The two dashed circles are centered at 30~Doradus and have radii of 0.5~and 1~degree. The small purple circle at the bottom of the \hi\ map represents the kinematic \hi\ center of the LMC and the arrow indicates the direction of the SMC. The dashed-gray diagonal line at the top left of the map tracks the position of a gaseous \hi\, overdensity filament known as Arm~B \citep{1998ApJ...503..674K,2003MNRAS.339...87S}. We choose background stellar targets that are confined within a downward angular cone centered at the 30~Doradus (marked by the two solid lines) to avoid the \hi\ overdensity filament. The gray circle positioned at the bottom left of the left-hand panel depicts the angular resolution of GASS at $\Delta\theta_{\rm GASS,\,res}=16\arcmin$ and the tiny black dot immediately above it represents the GASKAP angular resolution at $\Delta\theta_{\rm GASKAP,\,res}=30\arcsec$.}
  \label{fig:hi_target_map}
\end{figure*}

We explore the galactic wind along \nsightlines~sightlines within 1.7~degrees of 30~Doradus (see Table~\ref{table:targets} and Figure~\ref{fig:hi_target_map}). We choose these sightlines to lie preferentially to one side of 30~Doradus to avoid possible Magellanic tidal structures, such as the Magellanic Stream \citep{2016ARA&A..54..363D}. Below, we discuss the observational details and data reduction steps for the UV absorption lines and radio emission line data utilized in this work.

\subsection{UV Absorption Lines Data}\label{section:uv_absorption}

\begin{deluxetable*}{c c c c c c c c c c}
\tablecaption{Target Properties}
\label{table:targets}
\tablecolumns{10}
\tablewidth{0pt}
\tablehead{
\colhead{ID} \vspace{-0.2cm} & \colhead{Background} & \colhead{Stellar} & $l$ & $b$ & \colhead{$\Delta\theta_{\rm 30\,Dor}$\tablenotemark{a}} & \colhead{$\Delta D_{\rm 30\,Dor}$\tablenotemark{a}} & \colhead{UV Observing\tablenotemark{m}}  \\
\colhead{} & \colhead{Target} & \colhead{Type} & \colhead{$(^\circ)$} & \colhead{$(^\circ)$} & \colhead{$(^\circ)$} & \colhead{(kpc)} & \colhead{Modes} 
}
\startdata
(1) & BAT99\,105  &  O2 If* & $279\fdg46$ & $-31\fdg67$ & $3.4\times10^{-3}$ & $2.9\times10^{-3}$ & HST/STIS\,E140M+FUSE  \\           
(2) & Sk$-$69$^\circ$246 &  WN5/6h+WN6/7h & $279\fdg38$ & $-31\fdg66$ & $6.5\times10^{-2}$ & $5.6\times10^{-2}$ & HST/STIS\,E140M\,E230M+FUSE  \\ 
(3) & Sk$-$68$^\circ$135 & ON9.7Ia+ & $279\fdg26$ & $-31\fdg77$ &$1.9\times10^{-1}$ & $1.7\times10^{-1}$& HST/STIS\,E140M\,E230M+FUSE\\
(4) & BI\,214 & O6.5(n)(f)p & $279\fdg89$ & $-32\fdg04$ &$5.2\times10^{-1}$ & $4.5\times10^{-1}$ & HST/STIS\,E140M \\
(5) & Sk$-$69$^\circ$175 & WN11h & $279\fdg55$ & $-32\fdg32$ &$6.5\times10^{-1}$ & $5.7\times10^{-1}$& HST/STIS\,E140M\,E230M+FUSE \\
(6) & Sk$-$68$^\circ$112 & O7.5(n)(f)p & $278\fdg99$ & $-32\fdg41$ &$8.4\times10^{-1}$ & $7.3\times10^{-1}$& HST/STIS\,E140M\\
(7) & BI\,173 & O8II &  $279\fdg66$ & $-32\fdg68$ & $1.0$ & $9.0\times10^{-1}$ & HST/STIS\,E140M\,E230M+FUSE\\
(8) & Sk$-$69$^\circ$104 & O6Ib(f) &  $279\fdg92$ & $-33\fdg38$ & $1.7$ & $1.5$ & HST/STIS\,E140M\,E230M+FUSE 
\enddata
\tablenotetext{a}{These angular and projected distance offsets are from the center of 30~Doradus at $ (l,\,b) = (279\fdg46,\,-31\fdg67)$ at an assumed distance of $d_\odot=50\,\kpc$ to our LMC target.}
\tablenotetext{m}{We assumed the following widths for Gaussian shaped LSF for the observing modes: $\rm FWHM_{\rm res}\approx6.5\,\kms$ (R$\approx$45,800) for STIS/E140M, $\rm FWHM_{\rm res}\approx10\,\kms$ (R$\approx$30,000) for STIS/E230M, and $\rm FWHM_{\rm res}\approx20\,\kms$ (R$\approx$15,000) for FUSE.}
\end{deluxetable*}

All of our UV absorption-line observations are from \textit{HST} Ultraviolet Legacy Library of Young Stars as Essential Standards (ULLYSES: \citealt{2020RNAAS...4..205R}) director's discretionary program. This science-ready spectroscopic library includes observations of high-mass stars in the LMC with typical signal-to-noise ratios per resolution element for the stellar continuum of $20\lesssim S/N\lesssim30$. For our sample, we selected the background O-stars from the data release (DR)~5 catalogue as this was the release that was available at the time of our selection. Although we initially used that DR, we reran our analysis scripts on the observations from DR6 once they became available and present that analysis here. For this study, we only included the \textit{HST} Space Telescope Imaging Spectrograph (STIS) observations for their higher spectroscopic resolution at a full width at half maximum of $\rm FWHM_{STIS}\approx 6.5-10\,\kms$ compared to $\rm FWHM_{COS}\approx 17\,\kms$ for the Cosmic Origin Spectrograph. These sightlines were confined opposite to an \hi\, overdensity filament known as Arm B (See Figure \ref{fig:hi_target_map}) to minimize potential contamination. \citet{2008ApJ...679..432N} performed a Gaussian decomposition of the LAB \hi\ survey ($\Delta\theta_{\rm LAB,\,res}=0.6\arcdeg$; \citealt{2005A&A...440..775K}) and kinematically traced Arm~B back to the region near 30~Doradus and then to the Magellanic Stream. With these criteria, there were a total of 10 O-stars; however, two sightlines were removed from the final analysis due to having a noisy continuum, resulting in a total sample size of 8. Out of our 8~background O-stars, there are~2 Wolf–Rayet (WR) stars (see Table~\ref{table:targets}).

The STIS observations have the additional benefit of being less affected by atmospheric airglow that contaminates the O\textsc{~i}\,$\lambda1302$ and Si\textsc{~ii}\,$\lambda1304$ lines due to its much smaller aperture. 
All of our targets were observed with the E140M diffraction grating centered at $\lambda_{\rm center}=1{,}425\,\AA$, which covers a wavelength range of $1{,}140\lesssim\lambda\lesssim1{,}735\,\AA$. For 5 out of the 8 targets, the ULLYSES survey additionally observed them with the E230M grating, which covers a wavelength range of $1{,}700\lesssim\lambda\lesssim3{,}100\,\AA$.
All UV observing modes are listed in Table~\ref{table:targets}.

The data reduction procedure for the ULLYSES data products is described in \citet{2020RNAAS...4..205R} and at the survey's website\footnote{\url{https://ullyses.stsci.edu}}. 
Here we provide a brief summary of their procedure. The STIS observations were run through the calibration pipeline (CalSTIS), which calibrated the flux and the wavelength and applied any wavelength shifts to align the spectra. 
Observations that were taken with the same instrument and diffraction grating were co-added. Observations taken with different instruments or diffraction gratings were abutted when the spectra did not overlap such that the wavelength arrays and dispersion are discontinuous at the transition point; overlapping spectra were truncated then abutted at a transition wavelength and have a spectral sampling that is discontinuous at that transition. 

The ULLYSES archive additionally provides science-ready FUSE spectral products that we incorporate when available, as it also covers the O\textsc{~i} $\lambda1039$ absorption features, which can be used when O\textsc{~i} $\lambda1302$ is saturated.
Of our \nsightlines~targets, 5~of them were also observed with FUSE in the far UV which has a kinematic resolution of $\rm FWHM_{FUSE}\approx 20\,\kms$ and with typical signal-to-noise ratios per resolution element for the stellar continuum of $S/N\approx10$ (see Table~\ref{table:targets}).


\subsection{Radio Emission Line}\label{subsection:radio_data}

To investigate the neutral hydrogen gas phase in the LMC and its wind, we used the H\textsc{~i} 21-cm emission-line data from two  surveys that offer different advantages in terms of sensitivity and resolution. We utilized the sensitive Galactic Australian Square Kilometre Array Pathfinder (GASKAP) survey \citep{2013PASA...30....3D,2022PASA...39....5P} to explore the relatively bright gas within the disk of the LMC. The GASKAP-HI Pilot II data of the LMC that we used here has a sensitivity of $\log{\left(N_{\rm H\textsc{~i},\,3\sigma}/\cm^{-2}\right)}=20.0$ for typical clouds with $\rm FWHM=30\,\kms$, and a spatial resolution of $\Delta\theta_{\rm res}=30\arcsec$.  The raw kinematic resolution of these data is $\Delta v_{\rm bin}=0.2\,\kms$, we binned the datacubes to $\Delta v_{\rm bin}=1\,\kms$ to smooth stochastic variations associated with noise. Further, we constrain the kinematic extent of the LMC's H\textsc{~i} disk along our sightlines using this survey as it has a higher spatial resolution; the smaller beam of GASKAP translates to less blending of emission from gas surrounding our UV sightlines (i.e., beam smearing), which broadens the emission due to variations in gas motion.

Because GASKAP lacks the surface brightness sensitivity to detect the LMC's wind, we also employ data from the Parkes Galactic All-Sky Survey (GASS; \citealt{2010AA...521A..17K}) to measure H\textsc{~i} column densities for the wind. The GASS survey is 63 times more sensitive than GASKAP at $\log{\left(N_{\rm H\textsc{~i},\,3\sigma}/\cm^{-2}\right)}=18.2$ for lines of $\rm FWHM=30\,\kms$, but has a much larger spatial resolution of $\Delta\theta_{\rm res}=16\arcmin$ (32 times larger than GASKAP, see left-hand panel of Figure~\ref{fig:hi_target_map}).

\section{Wind identification}\label{section:WindIdentification}

Because of the physical placement of the targets within the disk of the LMC, we are only sensitive to the nearside portion of the LMC's galactic wind, which is kinematically blueshifted. 
However, besides the LMC wind, numerous absorbers such as MW's ISM, intermediate-velocity clouds (IVCs; $+30 \lesssim \vlsr \lesssim +90 \,\kms$), high-velocity clouds (HVCs; $+90 \lesssim \vlsr \lesssim +175 \,\kms$), and LMC's ISM are found between the systemic velocities of the MW and LMC (\citealt{2019ApJ...871..151R, 2007MNRAS.377..687L} and references therein). Additionally, there can also be circumstellar absorbers associated with the background stellar targets that are embedded in the LMC's disk and also absorption features associated with coronal material in the MW's and LMC's halo \citep[see][for a recent discovery of the Magellanic Corona]{2022Natur.609..915K}. 
In the sections below, we outline our method for identifying the LMC's wind.

\subsection{Checking the stellar environment}
\label{subsection:stellar_enviornment}
Our UV background targets for this study include O~and Wolf-Rayet (WR) stars that are luminous, hot, and massive. Their high luminosities translate into higher radiation pressures at the stellar surfaces, which can drive local stellar winds. 
The radiation and cosmic rays that emanate from their stellar surfaces ionize the surrounding circumstellar material, creating \hii\ regions. 
The approaching side of these \hii\ regions will be blueshifted relative to the star and can be kinematically confused with the LMC wind. 
For example, \citet{2016ApJ...817...91B} found that the expanding \hii\ region surrounding WR~star HD33133---a star that lies in a relatively quiescent quadrant of the LMC---reaches out to about $100\,\kms$ from the LMC's systemic velocity in that direction. 
In active regions of star formation, such as the core of 30~Doradus, stellar contamination can be notably pronounced as the \hii\ shells from either stellar winds or supernovae explosions can overlap \citep{2013A&A...558A.134D, 2013A&A...550A.108V}. 
As a consequence, certain absorption features detected in the spectra could potentially arise from the wind nebula enveloping the star rather than from the LMC's galactic winds.

To distinguish the absorption components originating from the stellar nebula, we employed stellar catalogs, such as the one compiled by \citet{2002ApJS..139...81D}. 
This resource allowed us to investigate the circumstellar medium encompassing these background stars. 
Specifically, the work by \citet{2002ApJS..139...81D} presented \ha\ images of the regions surrounding four sightlines studied in this paper: Sk$-$68$^\circ$135, Sk$-$69$^\circ$175, Sk$-$69$^\circ$246, and BI\,173. 
These images illuminated the morphology of warm ionized gas in the proximity of these stars. 
However, we found no morphological indications of wind nebulae encircling these stars. Additionally, we meticulously examine the locations of our sightlines using the Magellanic Cloud Emission Line Survey (MCELS) in the bright emissions of \ha\,[6563], [S\textsc{~ii}] 6724, and [O\textsc{~iii}] 5007 from the interstellar gas (see right panel in Figure \ref{fig:hi_target_map} for the \ha\, image). We do not observe circumstellar structures in the gas surrounding most of our targets, with the exception of the sightlines toward BAT99\, 105 and Sk$-$69$^\circ$104. In the direction of these stellar targets, the \ha\ emission is elevated and more structured (see the last panel in Figure~\ref{fig:hi_target_map}). However, the spectra for these two stars do not contain features that are more distinct than those of the other stars, possibly suggesting that the LMC wind largely dominates the absorption associated with the circumstellar structures.

\subsection{Conversion of reference frames}
\label{subsection:reference_frames}
To assist in the identification and separation of the MW's HVCs, LMC's winds, and LMC's ISM, we explore the UV spectra in both the kinematic Local Standard of Rest (LSR) and the LMC Standard of Rest (LMCSR) frame. We utilize the LMCSR frame to identify wind components with respect to the center of the LMC. We converted the ULLYSES observations from the heliocentric reference frame to the kinematic LSR frame by taking into account the velocity offset of the Sun with respect to the LSR, which is approximately $\Delta v_{\rm LSR,\,offset}\approx 20\,\kms$ toward the direction of $(\text{RA}, \text{Dec})=(18h, +30^\circ)$. 

We convert to the LMCSR reference frame by first estimating a central velocity offset. This offset is a function of the line of sight Galactic longitude ($l$) and Galactic latitude ($b$) and is calculated as follows from \citet{Ciampa2020}:
\begin{equation}
\frac{\Delta  v_{\text{LMCSR}}}{\kms} = 262.55 - 3.25(l - 280) + 3.66(b + 33)
\end{equation}
We subtract this offset from the LSR velocity to convert into the LMCSR reference frame.  

\subsection{Separating the LMC's wind from the LMC's disk}
\label{subsection:wind_disc}

\begin{figure*}
  \centering
  \includegraphics[width=0.48\textwidth]{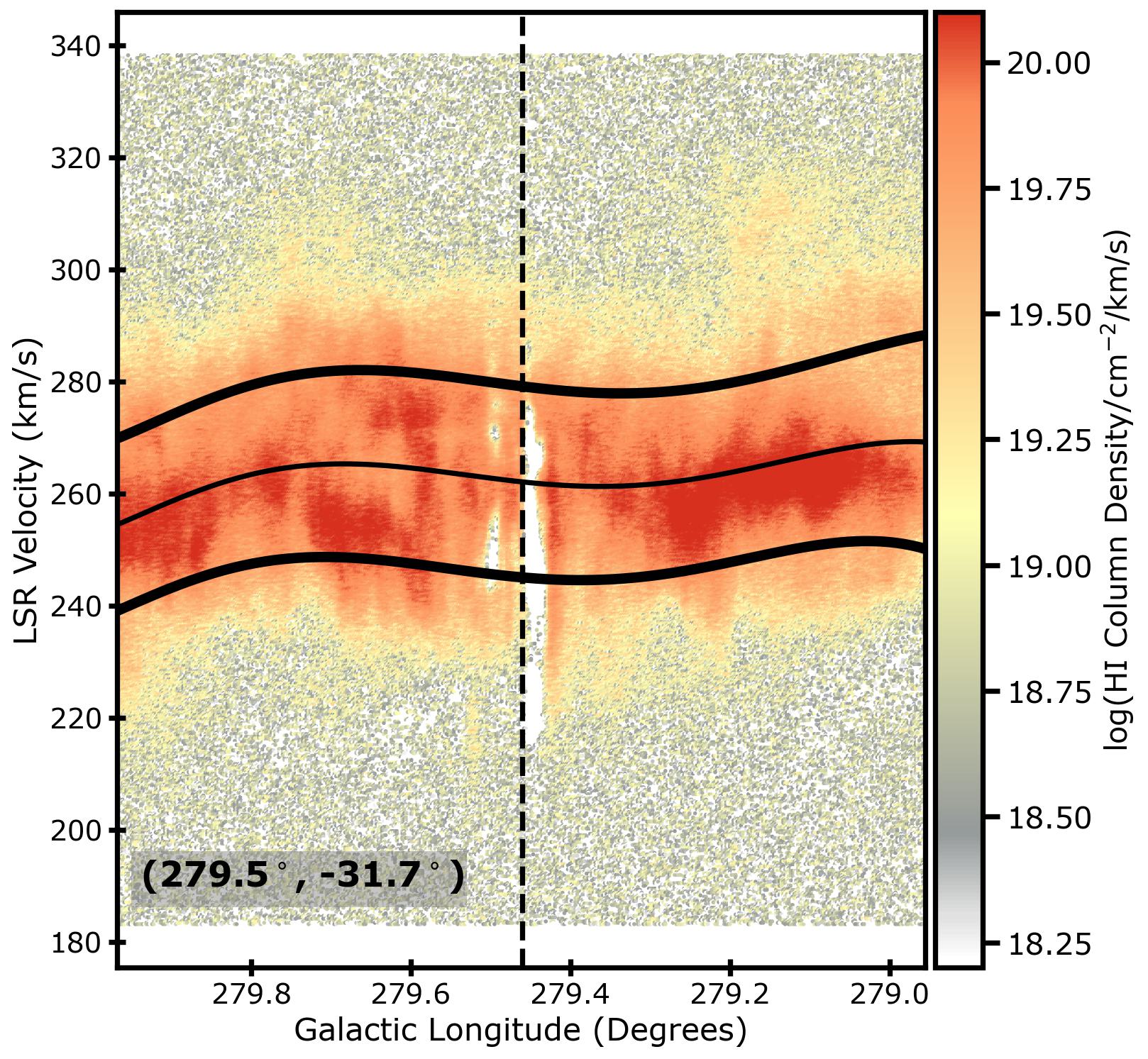}
  \includegraphics[width=0.48\textwidth]{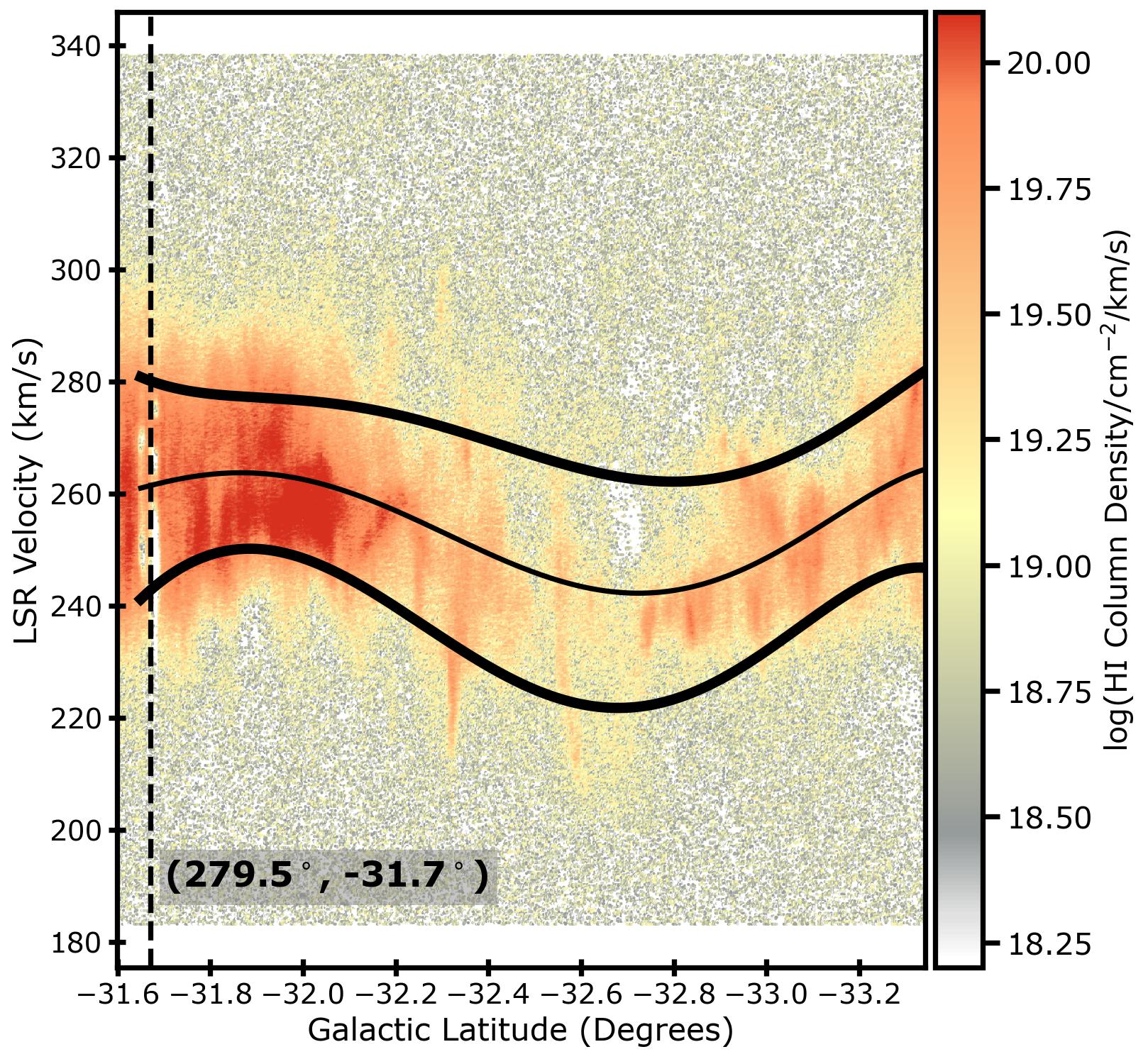}
  \caption{Position-velocity maps of the H\textsc{~i} column density in the LMC's disk using GASKAP observations for a positional slice through the total longitude and latitude space comprised by the 8 sightlines. The left panel includes a horizontal slice at Galactic latitude ($-31.7^\circ$) and the right panel, a vertical slice at Galactic longitude ($279.5^\circ$) of the sightlines. The vertical black dashed line represents the position of 30~Doradus. The thin solid black line is the kinematic center across the sliced region and the thick solid black lines at the top and bottom mark the boundary of the LMC's disk.}
  \label{fig:width_method_0}
\end{figure*}

The structure of the \hi\ in the LMC's disk varies a lot over kiloparsec scales (roughly 1-degree at the distance of this galaxy). We explored this using position-velocity plots that we generated with the GASKAP dataset in which we sliced through the disk in both Galactic longitude and latitude (see Figure~\ref{fig:width_method_0}). Although we can characterize the global behavior of the disk by finding trends in the emission weighted kinematic means and widths (i.e., first and second moments)  in these slices, there is too much stochastic variation in the \hi\ emission for these widths to be a good representation of the gas along the probed sightlines. 

\begin{figure}
  \centering
  \includegraphics[width=0.425\textwidth, trim={6 30 0 0}, clip]{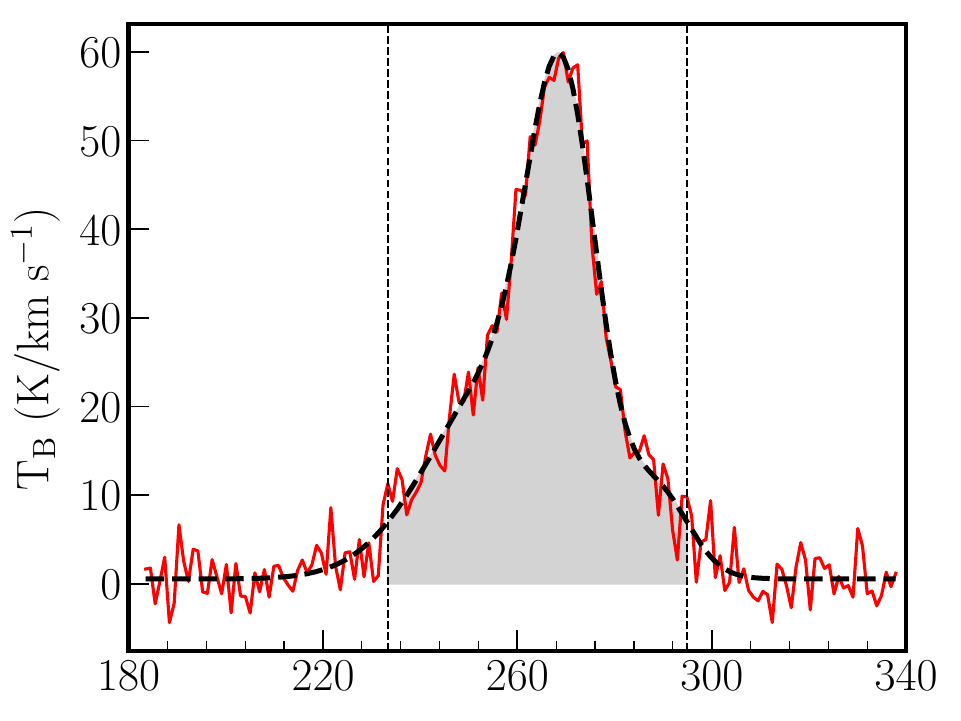}\vspace{0pt}
  \includegraphics[width=0.40\textwidth, trim={-3 0 12 10}, clip]{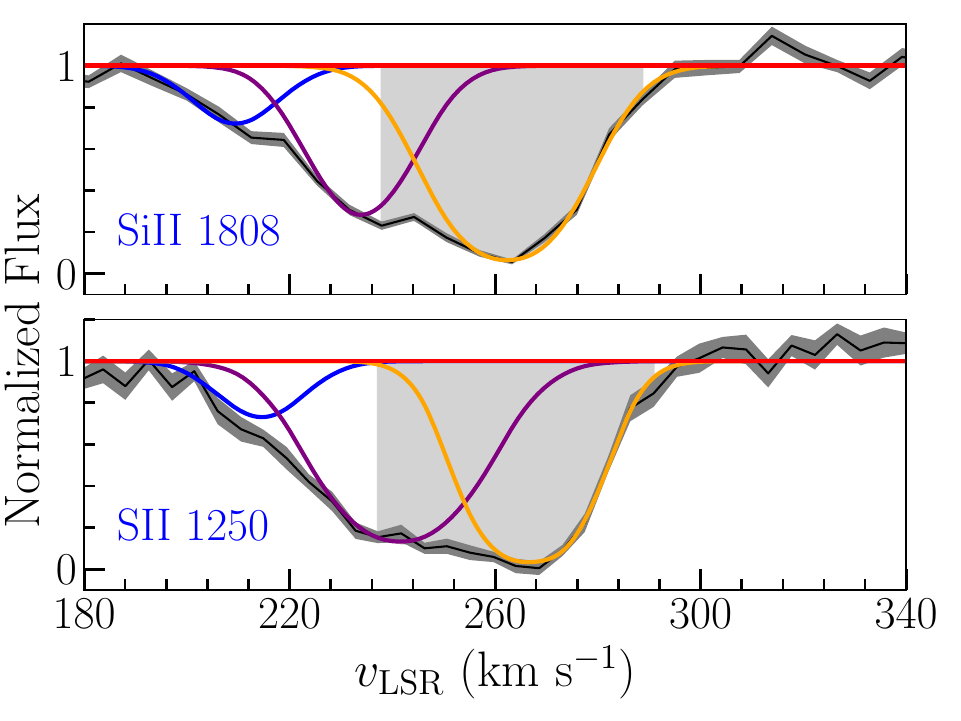}
  \caption{Determination of the kinematic width of the LMC's disk by comparing the H\textsc{~i} emission line with the UV absorption lines for the Sk$-$69$^\circ$246 sightline, which we have shaded in gray. In the upper panel, the red curve represents the H\textsc{~i} emission data from GASKAP, and the black-dashed curve is the corresponding multi-Gaussian fit to the data. The shaded region between the two vertical black dashed lines represents the estimated LMC disk width, as described in Section~\ref{subsection:wind_disc}. The lower panels display examples of weak UV absorption lines associated with this sightline. The horizontal line in red is drawn to show the position of the continuum and the profiles in different colors represent the individual components from the Voigt profile fitting (see Section \ref{subsec:Voigt}).} 
  \label{fig:width_method}
\end{figure}

We therefore estimate the kinematic width of the \hi\ disk using the closest GASKAP spectrum to each of our background stellar targets. We defined the kinematic boundaries of the disk to be where the \hi\ emission is at least $3\sigma$ above the RMS noise in the continuum, as illustrated in Figure~\ref{fig:width_method}. The widths obtained through this method generally match well with those of low-ions. In a few cases, there were slight inconsistencies of about 5~to 10~\kms\, when comparing the \hi\ emission with the weak UV absorption metal lines, which may be attributed to the larger H\textsc{~i} beam. Additionally, the H\textsc{~i} emission probes gas through the entire thickness of the disk, while low-ions probe only the gas foreground to the star. While we use this definition for consistency across all sightlines, we will also test different velocity cutoffs, particularly on the blue side of the disk, to determine if this causes any significant changes in the results we obtained about the wind. For example, see Section~\ref{section:Discussion}, where we estimate the systematic error that might affect the outflow mass and outflow rate of the wind. For most other results, we do not find any significant changes. Moreover, the kinematic widths determined using neutral hydrogen may not accurately represent the characteristics of the disk for the high-ions. The $v_{\mathrm{LSR}}$ velocities for the left (blueshifted) and right (redshifted) disk boundaries that we define for all 8~sightlines that we probe with UV~observations are listed in Table~\ref{tab:boundaries}. We shade the region that we associate with the LMC's disk as gray in all of the plotstacks (see Figure~\ref{fig:SK-69d246_lines} for example).

\begin{table}[htbp]
\centering
\caption{LMC's \hi\ disk boundaries}
\label{tab:boundaries}
\begin{tabular}{cccc}
\hline
\hline
ID & Sightline & Left Boundary  & Right Boundary \\ 
   &       & ($v_{\mathrm{LSR}}$/km s$^{-1}$) & $(v_{\mathrm{LSR}}$/km s$^{-1}$) \\
\hline
(1) & BAT99\,105 & +243.1 & +299.8 \\
(2) & Sk$-$69$^\circ$246 & +233.4 & +294.9 \\
(3) & Sk$-$68$^\circ$135 & +233.4 & +292.9 \\
(4) & BI\,214 & 238.2 & +276.3 \\
(5) & Sk$-$69$^\circ$175 & +232.4 & +285.0 \\
(6) & Sk$-$68$^\circ$112 & +238.2 & +281.2 \\
(7) & BI\,173 & +203.1 & +249.0 \\
(8) & Sk$-$69$^\circ$104 & +225.5 & +276.3 \\ \hline
\end{tabular}
\end{table}

\subsection{Separating the LMC's wind from Milky Way's HVCs}
\label{subsection:HVC_wind}
Numerous absorption features lie between the MW's and LMC's systemic velocities, corresponding to redshifted material with respect to the MW and blueshifted with respect to the LMC. In general, the absorption at $\vlsr \lesssim +90 \,\kms$ corresponds the MW ISM and MW IVCs. However, the origin of the HVCs at $+90 \lesssim \vlsr \lesssim +175 \,\kms$ remains unclear. Some studies conclude that high-velocity gas toward the LMC is associated with the MW halo \citep{1981ApJ...243..460S, 1990A&A...233..523D, 1999Natur.402..386R, 2015A&A...584L...6R}. Some of this high-velocity absorption is detected many degrees away from the  Magellanic Clouds. Most of the MW's HVCs are found to be within a distance of 10--20 $\kpc$ \citep{2011ApJ...727...46L, 2022MNRAS.513.3228L}. On the other hand, \citet{2009ApJ...702..940L} conducted an investigation of the gas in front of the LMC using FUSE spectra of 139 early-type stars in the LMC as background targets and H\textsc{~i} 21-cm emission data. Their examination of the HVCs spanning over the velocity range of $+90 \lesssim v_{\rm LSR} \lesssim 175\, \kms$ suggested an LMC origin based on their distribution, metallicity, kinematics, and dust depletion patterns. In the \citet{Ciampa2020} \ha\ emission-line study of the LMC's galactic winds, they found that the 30~Doradus region produces winds with the greatest speeds $v_{\rm LMCSR} > -175\,\kms$ compared to the bulk material in the galaxy's global winds at $v_{\rm LMCSR} < -110\,\kms$. Winds at these extreme speeds exceed the escape velocity of the galaxy of $v_{\rm LMCSR,\,esc}\approx 90\,\kms$ \citep{2016ApJ...817...91B}. \citet{Ciampa2020} further found that \ha\ morphology of the global wind is asymmetric such that the wind is more concentrated in the 30~Doradus quadrant of the galaxy. Given that absorption in the range $+90\, \kms \lesssim v_{\rm LSR} \lesssim +175\, \kms$ has been the subject of debate regarding their origins, whether from the MW's HVC or the LMC's wind, we thoroughly analyze them to investigate their potential origins and properties in more detail in Sections~\ref{sec:gas_distribution}, \ref{sec:gas_kinematics}, and \ref{sec:ionization}.\\

\section{UV Absorption-lines Analysis}\label{section:Analysis}

We explore LMC outflow and coincidental inflow absorption features associated with the following spectral transitions that are covered by the \textit{HST}/STIS observations: O\textsc{~i}\,$\lambda1302$, S\textsc{~ii}\,$\lambda$$\lambda1250$, 1253, Fe\textsc{~ii}\,$\lambda1608$, 2249, 2260, 2344, Al\textsc{~ii}\,$\lambda1670$, Al\textsc{~iii}\,$\lambda$$\lambda1854$, 1862, Si\textsc{~ii}\,$\lambda$$\lambda1526$, 1808, Si\textsc{~iv}\,$\lambda1393$, 1402, and C\textsc{~iv} $\lambda$$\lambda1548$, 1550. 
Additionally, in some cases, we also explore the weaker O\textsc{~i}\,$\lambda1039$ absorption features covered by FUSE because the O\textsc{~i}\,$\lambda1302$ is often saturated. We do not examine absorption for the C\textsc{~ii}\,$\lambda1334$ transition because the MW's C\textsc{~ii}\,$\lambda1335$ absorption overwhelms any signatures of the LMC's wind. 
We also exclude the Si\textsc{~iii}\,$\lambda1206$ line as it is in the wing of the MW's Ly$\alpha$ signature where the S/N is too low to reliably use that transition.\\

\subsection{Continuum Normalization}
\label{subsec:cont}

\begin{figure*}
  \centering
  \includegraphics[width=0.99\textwidth]{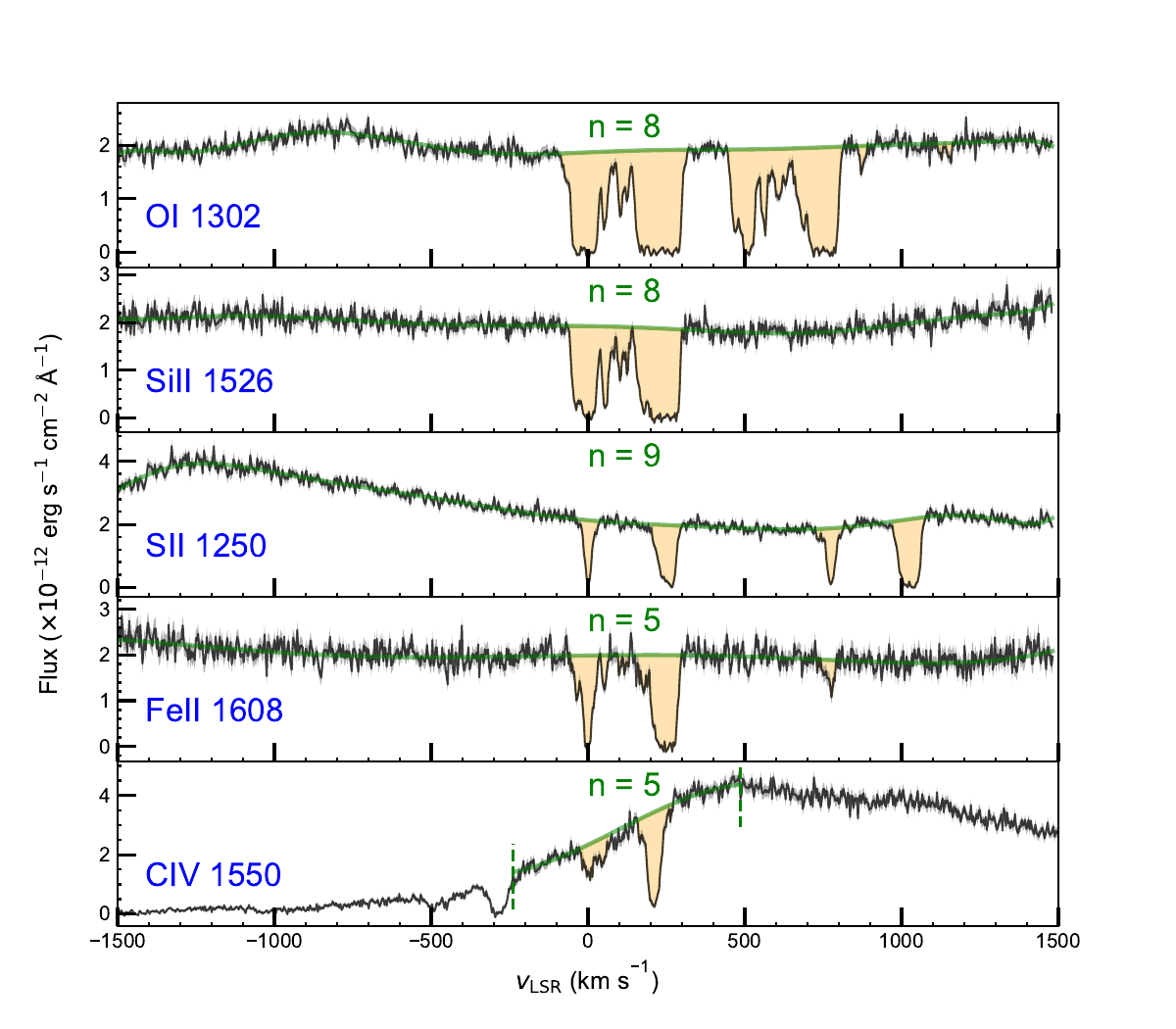}
  \caption{Non-normalized spectra and associated continuum fits for few key transitions observed toward the Sk$-$69$^\circ$246 background stellar target. In each panel, the data in black represents the observed flux, with the gray envelope indicating its uncertainty. The green lines mark the continuum fit. The absorption features shaded in yellow represent data excluded during the continuum fitting process. The order of the polynomial fit ``n'' is provided in each panel. For C\textsc{~iv}, only a small portion of the spectrum was used (see the region bounded by two vertical dashed lines in green) to fit the continuum due to the complexity of the stellar continuum.}
  \label{fig:continuum}
\end{figure*}

For each of the low-ion transitions that we explored, we created smaller spectral snippets that spanned $-1{,}500\le v_{\rm LSR}\le+1{,}500\,\kms$ for well-behaved continuum. We fit the stellar continuum using polynomials that have a typical order of $3\le n\le 8$. These polynomial orders tend to be larger than that typically used for observations with quasi-stellar objects as the background target of the stellar continuum of massive stars tends to be intricate as it is imprinted by stellar winds and rotation. These imprints tend to be more severe for the high-ion transitions. We established best-fit polynomial orders using goodness of fit tests by minimizing the reduced Chi-square ($\tilde{\chi}^2$) and used the smallest order that resulted in a marginal change in the $\tilde{\chi}^2$. The spectral regions, which exhibited absorption features inconsistent with the stellar continuum, were excluded from the fit. As an example, the continuum fits for several key transitions towards the sightline Sk$-$69$^\circ$246, along with the masked regions, are shown in Figure~\ref{fig:continuum}.

For the high-ion species C\textsc{~iv} and Si\textsc{~iv}, we used Legendre polynomials of degrees 1--8 to fit each continuum. This fitting was done within a velocity range of 500--1000 $\kms$ from the absorption line of interest. The choice of polynomial degree was determined by the complexity of the stellar continuum around the absorption lines. When the continuum was well behaved, we employed a low-degree Legendre polynomial, typically less than~5 (see \citealt{2011ApJ...727...46L} for details). The continuum for high-ions in the Sk$-$69$^\circ$175 sightline is too complex to reliably model, we therefore do not provide any measurements of Si\textsc{~iv} and C\textsc{~iv} towards that star.

\subsection{Voigt Profile Fitting}
\label{subsec:Voigt}
For all the absorption features, we fit the spectral lines with Voigt line profiles using the \textsc{VoigtFit} Python software (version 3.14.1.1, \citealt{2018arXiv180301187K}). This software utilizes analytic approximation to the Voigt profile developed by \citet{2006MNRAS.369.2025T}. We performed all of these fits using already continuum-normalized spectral segments for each ion transition. 
The \textsc{VoigtFit} program can fit various atomic lines with multiple components simultaneously and hold the Doppler parameters ($b$), line centers ($v$), and column densities ($N$) fixed or allow them to freely vary. To determine column densities, the \textsc{VoigtFit} program uses atomic data (i.e., rest frame wavelengths and oscillator strengths of transitions) from both the Vienna Atomic Line Data Base (VALD: \citealt{1995A&AS..112..525P}) and \citet{2017ApJS..230....8C}. This line-fitting software determines the best fit through a non-linear least-squares minimization. As the current version of \textsc{VoigtFit} does not have the functionality to use the tabulated line spread function (LSF) for STIS (\textsc{VoigtFit} only supports the tabulated LSF for HST/COS), we accounted for the broadening associated with the detector by using a LSF with a Gaussian shape which has a FWHM that matches the spectral resolution of the instrument and mode of the observation. For example, we used a Gaussian profile with $\rm FWHM_{\rm res}\approx6.5\,\kms$ for the STIS/E140M LSF that has a spectral resolution of $R\approx45{,}800$; we list all of the FWHM for the LSF that we used for all observing modes in the footnotes of Table~\ref{table:targets}. The potential effects of using a Gaussian approximation to the LSF on the fit parameters, particularly column density and linewidth, have been explored in previous studies. \citet{2020ApJ...897...23F} specifically examined this for HST/COS observations of HVCs and found no significant difference in the component parameters when using the tabulated LSF versus a Gaussian LSF. We have also tested this for the C\textsc{~iv} and Si\textsc{~iv} ions using a different software that utilizes the tabulated LSFs for the STIS E140M grating, and the differences in the $\log\,\rm N$ values were less than $0.1\, \rm dex$. Doppler $b$-values were also consistent within $<0.5\, \sigma$.

For the low-ions, when multiple transitions of the same ion are available, we fitted them independently. We prioritized lines based on their signal-to-noise ratio and the degree of saturation, using the weaker lines (with lower oscillator strengths) when possible to avoid saturation effects. The stronger transitions (with higher oscillator strengths) are generally chosen to analyze the weak components in the wind and the lines with weaker transitions are better for the strong components in the disk that often saturate. For example, we used the Si\textsc{~ii} $\lambda1808$ transition for very strong components in the slow-moving part of the wind and in the LMC disk; however, we often used the Si\textsc{~ii} $\lambda1526$ transition for the fast-moving wind components as they are often too weak to be detected in Si\textsc{~ii}\,$\lambda1808$. Occasionally, we also compare the $3\,\sigma$ upper limit in the column density for a non-detection of Si\textsc{~ii} $\lambda1808$ line with the column density measured from Si\textsc{~ii}\,$\lambda1526$. This comparison was helpful to check for unresolved saturation of the stronger Si\textsc{~ii}\,$\lambda1526$ line. Similarly, for Fe\textsc{~ii}, we utilized the stronger Fe\textsc{~ii}\,$\lambda1608$ and Fe\textsc{~ii}\,$\lambda2344$ lines and the weaker Fe\textsc{~ii}\,$\lambda2249$ and Fe\textsc{~ii}\,$\lambda2260$ lines. We used both the S\textsc{~ii}\,$\lambda1250$ and S\textsc{~ii} $\lambda1253$ transitions to measure the S\textsc{~ii} column density.

For neutral oxygen, we constrain its column density using O\textsc{~i} $\lambda1302$ from STIS for most of our sightlines. However, for the Sk$-$69$^\circ$175 and BI\,173 sightlines, we additionally used the O\textsc{~i}\,$\lambda1039$ line that is available through FUSE observations. 
For these two sightlines, this transition did not suffer from obvious airglow emission contamination. Although H$_2$ absorption is littered across the continuum for this transition, the LMC's ISM and outflows were not significantly impacted. However, H$_2$ contamination litters the other parts of this spectral segment, impacting our continuum fit and adding uncertainty in our normalization. We accounted for the uncertainty in continuum placement by estimating the column densities based on the highest plausible continuum positions and propagated the corresponding errors. For the Sk$-$69$^\circ$175 sightline, we identified absorption from the MW molecular line 4--0 R(2) H$_2$ $\lambda$1051.5 at $v_{\rm LSR}\approx+320\,\kms$. However, since we were not interested in this velocity region, we masked the molecular hydrogen absorption components at this velocity when performing the Voigt profile fitting. The LMC component of the 5--0 R(2) H$_2$ $\lambda$1038.7 line may be contaminated with the O\textsc{~i} $\lambda1039$ absorption at $v_{\rm LSR}\approx+120\,\kms$. To estimate the potential contribution to the O\textsc{~i}\,HVC profile from this molecular line, we examined the absorption of the 4--0 R(2) H$_2$ $\lambda$1051.5 transition. This line was chosen because it is free from contamination by other lines and the stellar continuum near this region is clean. Moreover, the strengths of both of these transitions are comparable. We found that the contribution from the molecular hydrogen line was insignificant. While we generally allowed the velocity centroids for each ions to vary freely, there were some instances where we opted to fix them. In cases of saturation and blending (which is generally the case in the disk region), we fixed the velocity centroids for all blended or saturated lines to match lines produced by elements in similar ionization states that had the least blending and saturation.

In the case of high-ions, we fitted the C\textsc{~iv} and Si\textsc{~iv} lines independently. However, the C\textsc{~iv} doublets were fitted simultaneously, as were the doublets of Si\textsc{~iv}. Moreover, we allowed for all of their fit parameters to typically vary freely. We list the Voigt profile fitting results used for our analysis for both the low- and high-ions for sightline Sk$-$69$^\circ$246 in Table~\ref{tab:Voigt_results}. We list the remainder of our fit results for the other sightlines explored and analysed in this study in Table~\ref{tab:Voigt_results_cont}, located in the Appendix \ref{section:voigt_all}. While Tables~\ref{tab:Voigt_results} and \ref{tab:Voigt_results_cont} list the column densities and uncertainties produced by \textsc{VoigtFit}, we treat them as a lower limit in our analysis for cases of saturation. Such components are provided with a footnote (`c' for a component which might be saturated and `s' for a component which is saturated) in Tables~\ref{tab:Voigt_results} and \ref{tab:Voigt_results_cont}. In situations where a line is not detected, we provide $3\,\sigma$ upper limit on the column density. We graphically illustrate the fits for each component along sightline Sk$-$69$^\circ$246 in Figures~\ref{fig:width_method} and~\ref{fig:SK-69d246_lines} and in the Appendix \ref{section:voigt_all} for all other sightlines.

We also rely on the apparent optical depth (AOD) method \citep{1991ApJ...379..245S} to obtain total column densities within set velocity ranges that we use to explore trends across all sightlines (for example, see Section~\ref{subsection:radial_dist}). This method involves converting the observed flux ($F\left( v\right)$) into an apparent column density profile ($N_{a} = \int N_{a}\left( v\right) dv$): 
\begin{equation}
\label{eq:colden}
 \frac{N_{a}}{\rm cm^{-2}} = \frac{3.768 \times 10^{14}}{f\lambda_o}\int \tau_{a}(v) \frac{\cm^{-2}}{\kms} dv,
 \end{equation}
 where $\lambda_o$ is the rest wavelength of the transition in $\AA$ and $f$ is the corresponding oscillator strength. The optical depth $(\tau)$ is related to the observed flux and continuum flux ($F_{c}\left( v\right)$) by:
\begin{equation}
\tau_{a}\left( v\right) = \ln{\left(\frac{F_{c}\left( v\right)}{F\left( v\right)}\right)}
\end{equation}
For lines with clear saturation (when the flux reaches zero), we adopt the AOD column densities as lower limits. We additionally compare the apparent column densities for different transitions of the same ion to check and correct for the unresolved saturation using the methods described in \citet{1996ApJ...471..292J} and \citet{1991ApJ...379..245S}. For unsaturated lines, the AOD column densities obtained from both the strong and weak transitions should be similar. Voigt profile fitting may not always provide a robust estimate of a lower limit on column density as it depends on the modelled $b$-value. Therefore, for the saturated components, we report only those values from the profile fits which are also consistent with the values obtained from the AOD method (see the components with footnotes `s' and `c' in Tables~\ref{tab:Voigt_results} and \ref{tab:Voigt_results_cont}).

\begin{table*}
\centering
\caption{Results of Voigt profile-fitting analysis} 
\label{tab:Voigt_results}
\begin{tabular}{cccccc}
\hline
\hline
Ion & $v_{\rm LSR}$ & $v_{\rm LMCSR}$  &  $\log{\left(N_x/\cm^{-2}\right)}$ & $b$  \\
    & (\kms) & (\kms) &   & (\kms)  \\
\hline
\multicolumn{5}{c}{\textbf{Sk$-$69$^\circ$246}} \\ \hline
O I &  +103.7 $\pm$ 0.6\tablenotemark{} & $-165.7$ & 13.94 $\pm$ 0.03 &  7.7 $\pm$ 0.9 \\
    &  +124.5 $\pm$ 0.9\tablenotemark{} & $-144.9$ & 13.67 $\pm$ 0.06 &  6.7 $\pm$ 1.5 \\

Si II &  +102.2 $\pm$ 0.9\tablenotemark{} & $-167.2$ & 13.37 $\pm$ 0.05 &  6.9 $\pm$ 1.3 \\
      &  +123.6 $\pm$ 0.9\tablenotemark{} & $-145.8$ & 13.30 $\pm$ 0.06 &  7.9 $\pm$ 1.6 \\
      &  +177.6 $\pm$ 0.0\tablenotemark{c} &  $-91.8$ & 14.17 $\pm$ 0.02 & 18.5 $\pm$ 1.0 \\
      &  +209.5 $\pm$ 0.0\tablenotemark{} &  $-59.9$ & 14.82 $\pm$ 0.09 &  9.4 $\pm$ 2.7 \\    
Fe II &  +103.1 $\pm$ 0.6\tablenotemark{} & $-167.9$ & 13.03 $\pm$ 0.04 &  5.4 $\pm$ 1.4 \\
      &  +123.0 $\pm$ 0.6\tablenotemark{} & $-146.9$ & 12.93 $\pm$ 0.04 &  4.8 $\pm$ 1.7 \\
      &  +175.7 $\pm$ 0.9\tablenotemark{} &  $-93.7$ & 13.83 $\pm$ 0.02 & 19.6 $\pm$ 1.1 \\
      &  +206.5 $\pm$ 0.0\tablenotemark{} &  $-62.9$ & 13.78 $\pm$ 0.04 &  7.8 $\pm$ 1.0 \\

S II  & +102.2  & $-167.2$ & $<13.76$ &  \nodata \\
      & +123.6  & $-145.8$ & $<13.60$ &  \nodata \\
     &  +177.6  & $-91.8$  &  $<13.64$ & \nodata \\
     &  +214.5 $\pm$ 0.0\tablenotemark{} &  $-54.9$ & 14.56 $\pm$ 0.18 &  10.4 $\pm$ 3.1 \\
     &  +241.5 $\pm$ 3.9\tablenotemark{c} &  $-27.9$ & 15.55 $\pm$ 0.12 &  16.7 $\pm$ 4.1 \\
  
Al II &   +95.4 $\pm$ 1.8\tablenotemark{} & $-174.0$ & 12.11 $\pm$ 0.13 & 8.3 $\pm$ 2.9 \\
      &  +117.0 $\pm$ 1.8\tablenotemark{} & $-152.4$ & 12.31 $\pm$ 0.09 &  10.7 $\pm$ 2.7 \\
      &  +168.4 $\pm$ 4.2\tablenotemark{c} &  $-101.0$ & 12.74 $\pm$ 0.15 & 16.7 $\pm$ 3.6 \\
      &  +206.5 $\pm$ 0.0\tablenotemark{s} &  $-62.9$ & 13.03 $\pm$ 0.11 & 19.5 $\pm$ 9.9 \\
      &  +235.4 $\pm$ 0.0\tablenotemark{s} &  $-34.0$ & 14.76 $\pm$ 0.36 & 15.7 $\pm$ 0.8 \\
 
Al III &  +103.8 $\pm$ 1.2\tablenotemark{} & $-165.6$ & 11.65 $\pm$ 0.24 &  5.5 $\pm$ 3.2 \\
       &  +124.3 $\pm$ 2.2\tablenotemark{} & $-145.1$ & 11.79 $\pm$ 0.19 &  6.8 $\pm$ 1.5 \\
       &  +177.6 $\pm$ 0.0\tablenotemark{} &  $-91.8$ & 11.51 $\pm$ 0.39 & 10.5 $\pm$ 2.2 \\
       &  +206.5 $\pm$ 0.0\tablenotemark{c} &  $-62.9$ & 13.31 $\pm$ 0.02 & 14.5 $\pm$ 3.1 \\
       &  +235.4 $\pm$ 0.0\tablenotemark{} &  $-34.0$ & 12.96 $\pm$ 0.02 & 15.4 $\pm$ 2.7 \\

C IV   &  +100.2 $\pm$ 0.0 & $-169.2$ & 13.23 $\pm$ 0.07 & 28.2 $\pm$ 6.1  \\
       &  +210.6 $\pm$ 0.3\tablenotemark{s} & $-58.8$ & 14.37 $\pm$ 0.01 & 22.5 $\pm$ 0.5  \\
Si IV  &  +98.1 $\pm$ 3.0 & $-171.3$ & 12.34 $\pm$ 0.01 & 14.4 $\pm$ 4.3  \\
       &  +212.9 $\pm$ 0.3\tablenotemark{s} & $-56.5$ & 14.06 $\pm$ 0.01 & 22.6 $\pm$ 0.4  \\
\hline     
\end{tabular}
\tablenotetext{c}{These components could potentially be saturated.}
\tablenotetext{s}{These components are saturated.}
\tablecomments{A summary of the Voigt profile fitting outcomes for $+90\, \kms \lesssim v_{\rm LSR} \lesssim \rm Disk~Boundary$ across various ion species along sightline Sk$-$69$^\circ$246; we list the outcomes for all other sightlines in Table~\ref{tab:Voigt_results_cont} located in the Appendix \ref{section:voigt_all}. We exclude the results for the components which are saturated and blended together as they may not represent the correct values. While the Tables list the errors in column densities produced by \textsc{VoigtFit}, we treat them as a lower limit in cases of saturation and such components are marked with a corresponding footnote. Since the Voigt profile fitting may not always provide a robust estimate of a lower limit on column density, we report only those values which are also consistent with the values obtained from the AOD method (for the components represented by the footnotes `c' and `s'). In situations where a line is not detected, we provide $3\,\sigma$ upper limit on the column density. In some cases, when we need to fix the velocity centroids, the errors for those particular components are set to $0.0\,\kms$. The errors in $v_{\rm LMCSR}$ are not listed in the table, as they are the same as the errors in $v_{\rm LSR}$.}
\end{table*}

\begin{figure}
  \centering
  \includegraphics[width=0.45\textwidth, trim={6.5 25 0 0}, clip]{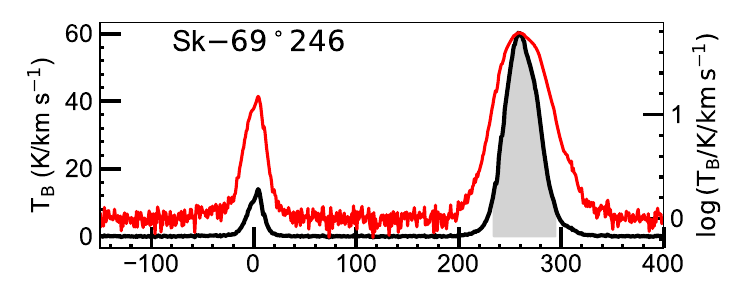}\vspace{0pt}
  \includegraphics[width=0.43\textwidth, trim={0 0 0 72}, clip]{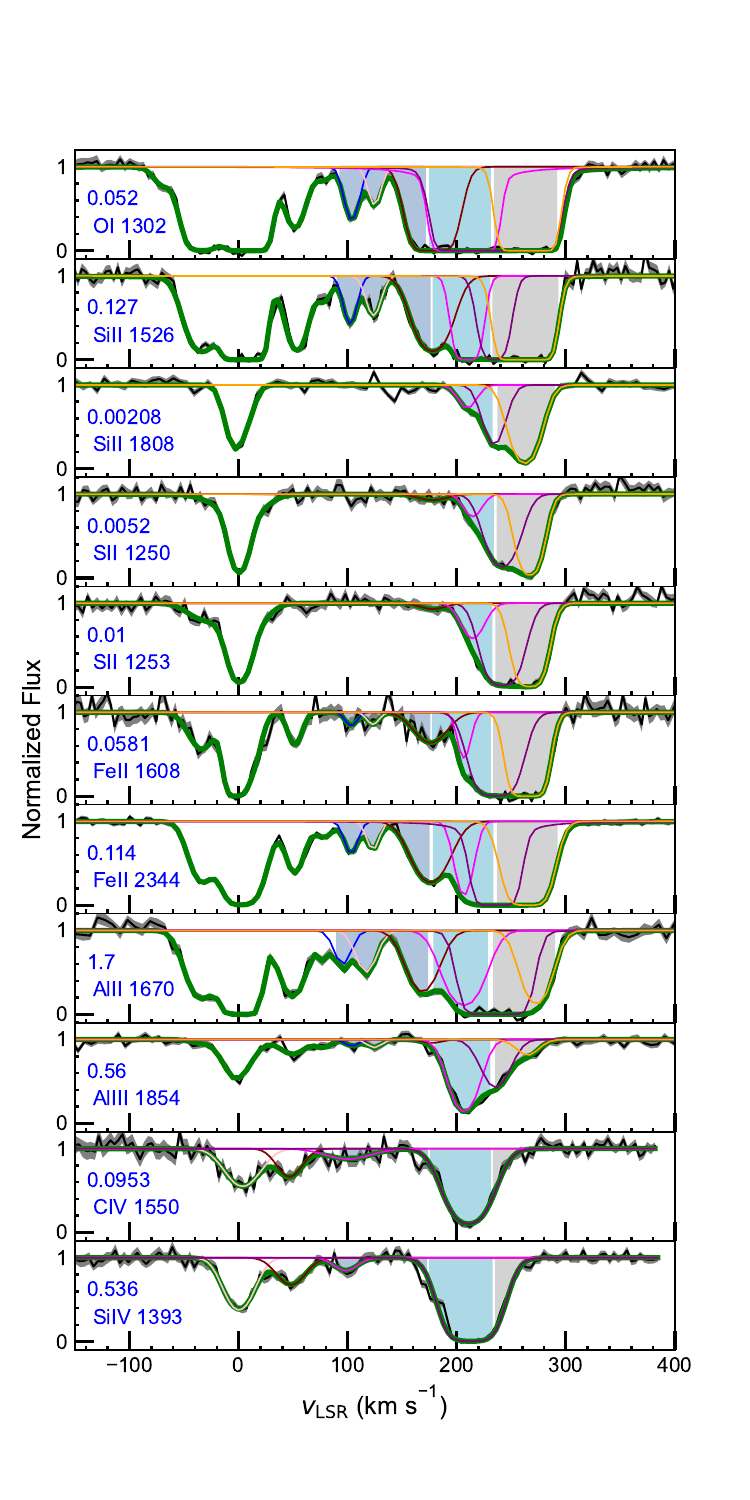}
  \caption{Upper Panel: H\textsc{~i} 21-cm emission-line data using the Parkes Galactic All-sky Survey (GASS; \citealt{2010AA...521A..17K}) for the Sk$-$69$^\circ$246 sightline (shown in black). The spectrum is also presented in a logarithmic scale (shown in red) to facilitate the visual magnification of possible weaker emission lines outside the disk. The shaded region marks the boundaries for the LMC's H\textsc{i} disk (see Section~\ref{subsection:wind_disc} for details). Lower Panel: Voigt profile fits to various metal absorption lines. The oscillator strength values for the corresponding transitions are provided in each panel. The normalized flux is depicted as a black curve, and the 1-$\sigma$ error is represented by the gray envelope around the normalized flux. The multi-colored lines correspond to Voigt profile fits for individual components, while the green line represents the total co-added fit for all these components. As in the upper panel, the gray shaded region indicates LMC disk absorption. The different shades of blue correspond to the HVC ($+90\, \kms \lesssim v_{\rm LSR} \lesssim +175\, \kms$) and the LMC wind ($+175\, \kms \lesssim v_{\rm LSR} \lesssim \rm Disk~Boundary$). Similar plotstacks for the sightlines explored in this study are included in the Appendix \ref{section:voigt_all}.} 
  \label{fig:SK-69d246_lines}
\end{figure}

\section{Gas Distribution}\label{sec:gas_distribution}
To investigate the origin of the gas absorption towards 30~Doradus, we explore the kinematic and radial distribution of the gas using our 8 stellar sightlines.

\subsection{Radial distribution of the gas}
\label{subsection:radial_dist}
\begin{figure}
  \centering
  \includegraphics[width=0.44\textwidth, trim={0 30 0 0}, clip]{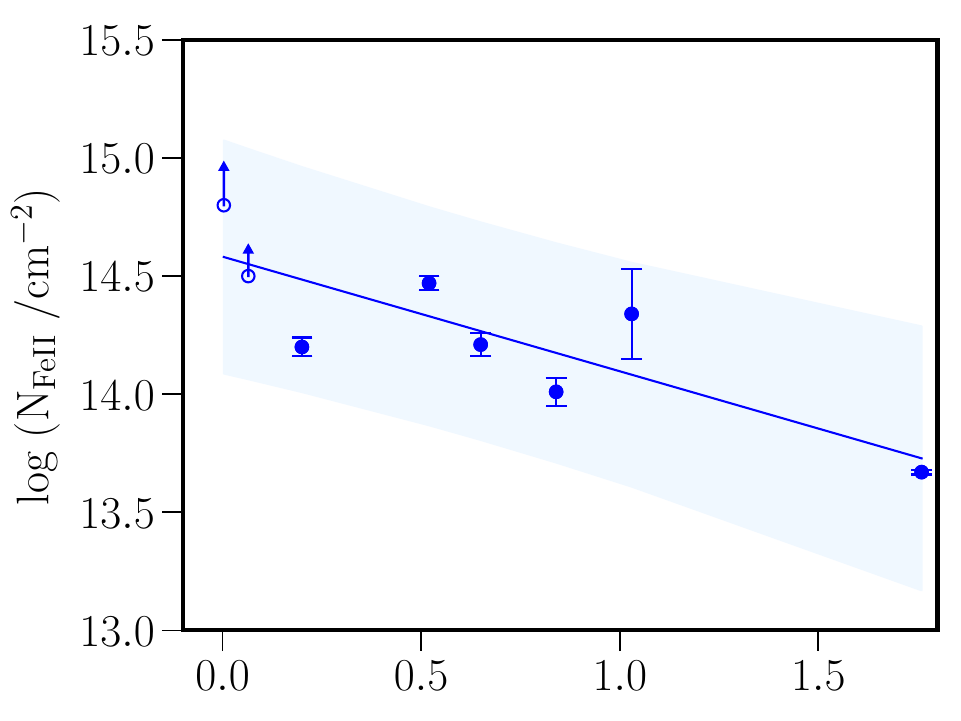}
  \includegraphics[width=0.44\textwidth, trim={0 30 0 10}, clip]{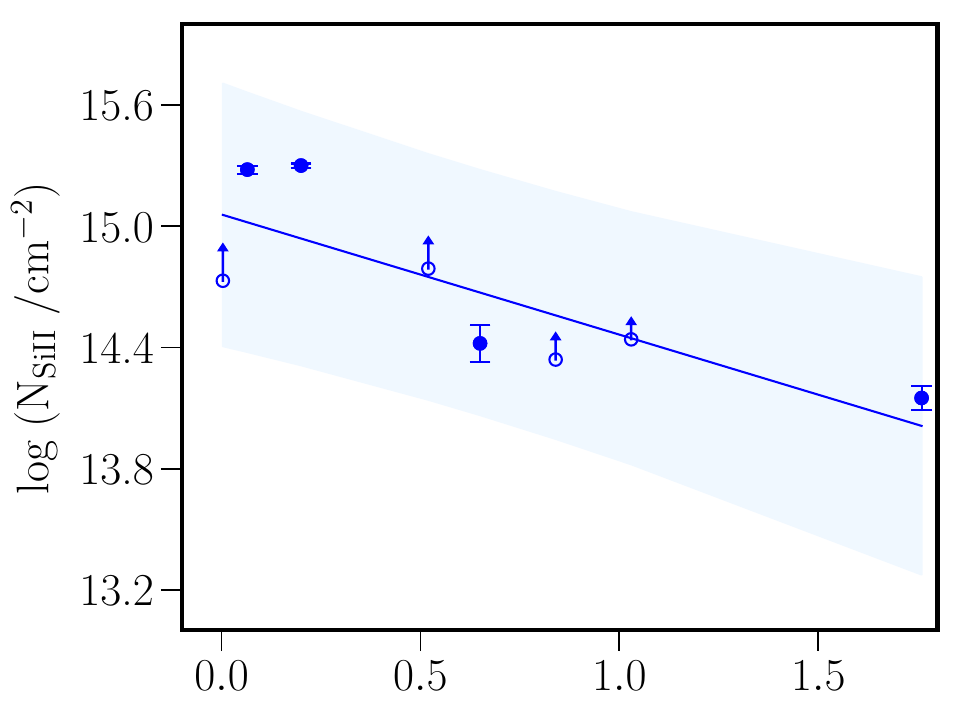}
  \includegraphics[width=0.44\textwidth, trim={0 0 0 10}, clip]{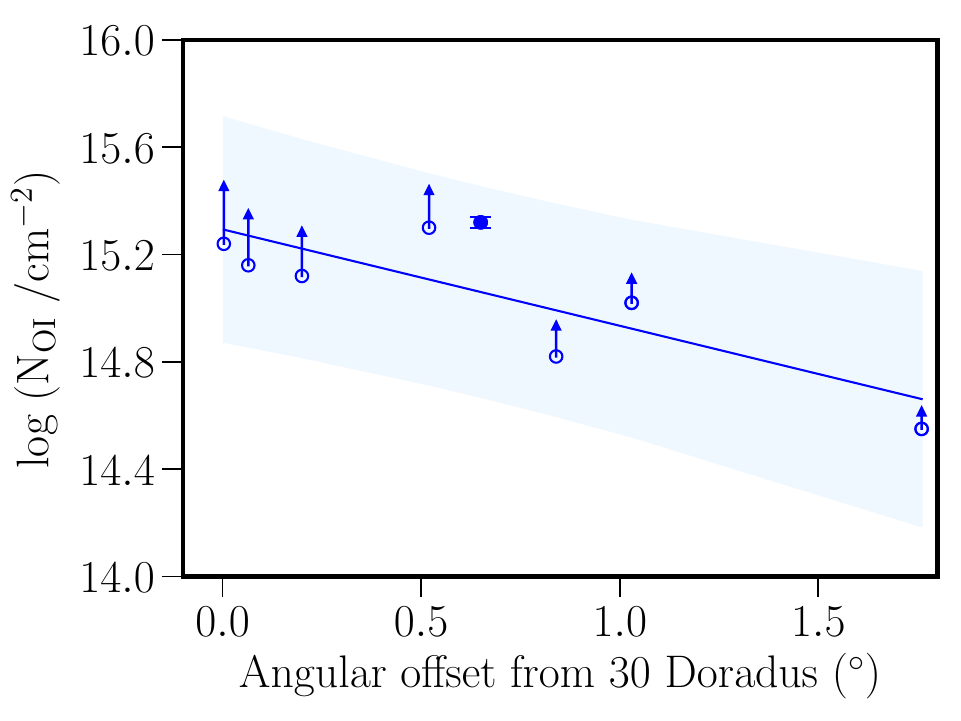}
  \caption{Variations in integrated column densities of Fe\textsc{~ii}, Si\textsc{~ii}, and O\textsc{~i} as a function of angular separation from 30~Doradus in a velocity range of $+150\, \kms <  v_{\rm LSR} < \rm LMC's~disk~boundary$, where the disk boundaries are defined in Table~\ref{tab:boundaries}. The solid lines represent linear regression fits applied to the observed data points depicted by the circles. The open circles are presented without the error bars as these values are saturated. The upward arrows represent the extent of saturation with larger arrows signifying a higher level of saturation. The shaded region in blue shows the uncertainty in the fit.}
  \label{fig:colden}
\end{figure}

We analyze the distribution of the gas absorption as a function of the angular offset from the center of 30~Doradus. For this, we calculated the total column densities for various atomic species in different velocity regions. First, we estimate the integrated column densities for Fe\textsc{~ii}, Si\textsc{~ii}, and O\textsc{~i} for velocities that range from $v_{\rm LSR}=+150\,\kms$ to the kinematic boundary of the LMC's galactic disk (see Table~\ref{tab:aod_N}). We prefer the AOD method to adding individual components from the Voigt profile fitting to determine these column densities as we want to compare the sightlines over consistent velocity binning. Second, the kinematically neighboring components to the LMC disk boundary in the closest sightlines to 30~Doradus are mostly saturated, especially for O\textsc{~i}. In cases of saturation, we treat the column densities as lower limits. We find that the integrated column densities of the blueshifted absorbers relative to the LMC's disk with $\vlsr>+150\,\kms$ decrease linearly as the angular offset from center of the 30~Doradus starburst region increases (see Figure~\ref{fig:colden}). This characteristic holds for all three of the aforementioned atomic species. However, the most notable trend is observed in the Fe\textsc{~ii} column density distribution, which experiences less saturation compared to Si\textsc{~ii} and O\textsc{~i}. 

To test the robustness of the declining column densities with angular offsets, we perform a Kendall's Tau test which provides a reliable assessment of the monotonic association between two variables. Specifically, we implement a censored Kendall’s Tau analysis to account for lower limits in column density by treating censored values as greater than or equal to their observed limits. The Kendall rank correlation coefficient $\tau$ and the significance level of the correlation $\rm p$ for all the three ions are listed in Table~\ref{tab:k_tau}. We adopt a $\rm p$-value threshold of 0.05 to compare the level of significance. We get a large negative value of $\tau=-0.64$ with a value of $\rm p=0.03$, signifying a negative correlation between the Fe\textsc{~ii} column density and the angular offset from 30~Doradus. Although there is a negative correlation between the column density and the angular offset from 30~Doradus for Si\textsc{~ii} and O\textsc{~i}, this correlation is not quite significant and is likely attributed to the elevated level of saturation in the sightlines closest to this starburst region (see Figure~\ref{fig:SK-69d246_lines} and Appendix plotstacks). The closer the sightlines are to 30~Doradus, the larger is the radial velocity range in which the profile goes all the way to zero flux. Accounting for the saturation, the radial decrease of the column densities with projected distance from 30~Doradus would be stronger than suggested by the slopes in Figure~\ref{fig:colden}.

\begin{figure}
  \centering
  \includegraphics[width=\columnwidth]{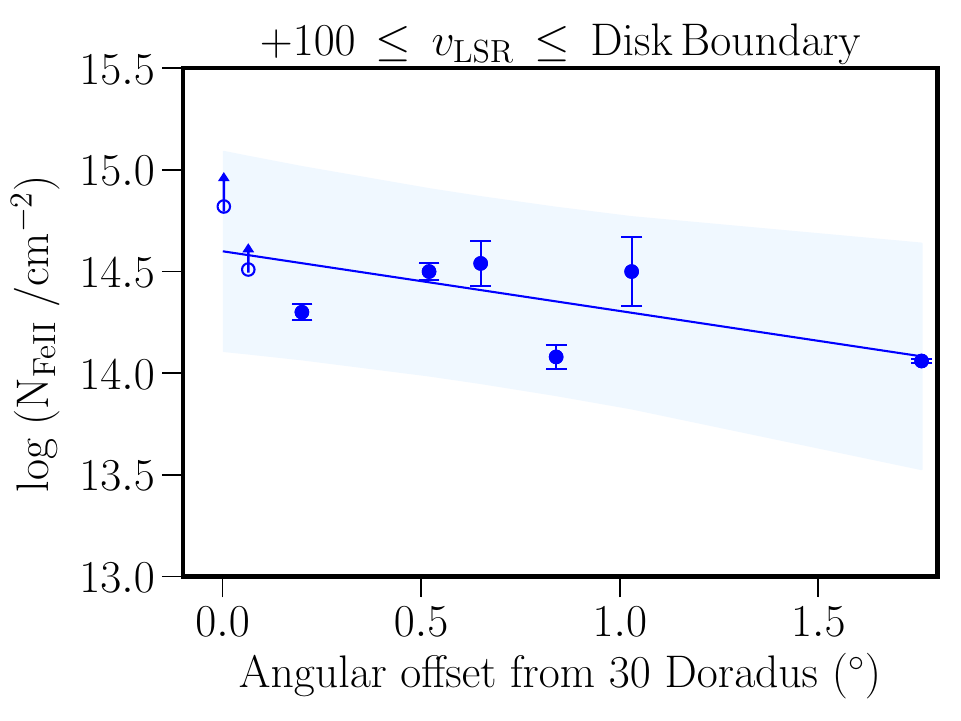}
  \caption{Similar to Figure \ref{fig:colden} but for Fe\textsc{~ii} in a velocity range of $+100\, \kms <  v_{\rm LSR} < \rm LMC's~disk~boundary$.}
  \label{fig:colden_Fe}
\end{figure}

\begin{figure}
  \centering
  \includegraphics[width=0.44\textwidth, trim={0 60 0 0}, clip]{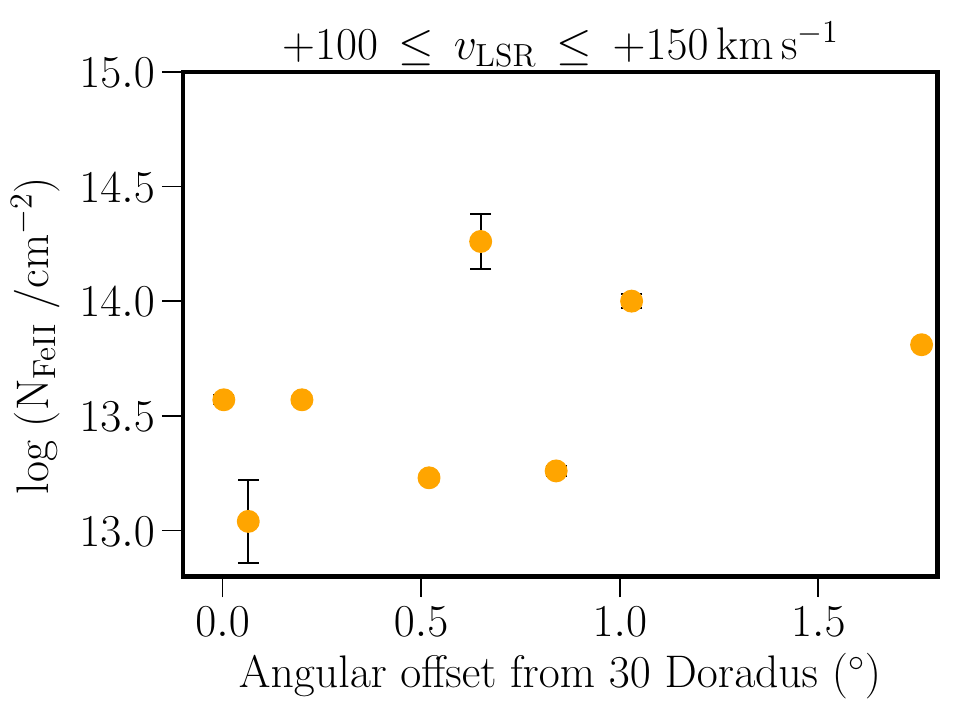}
  \includegraphics[width=0.44\textwidth, trim={0 60 0 0}, clip]{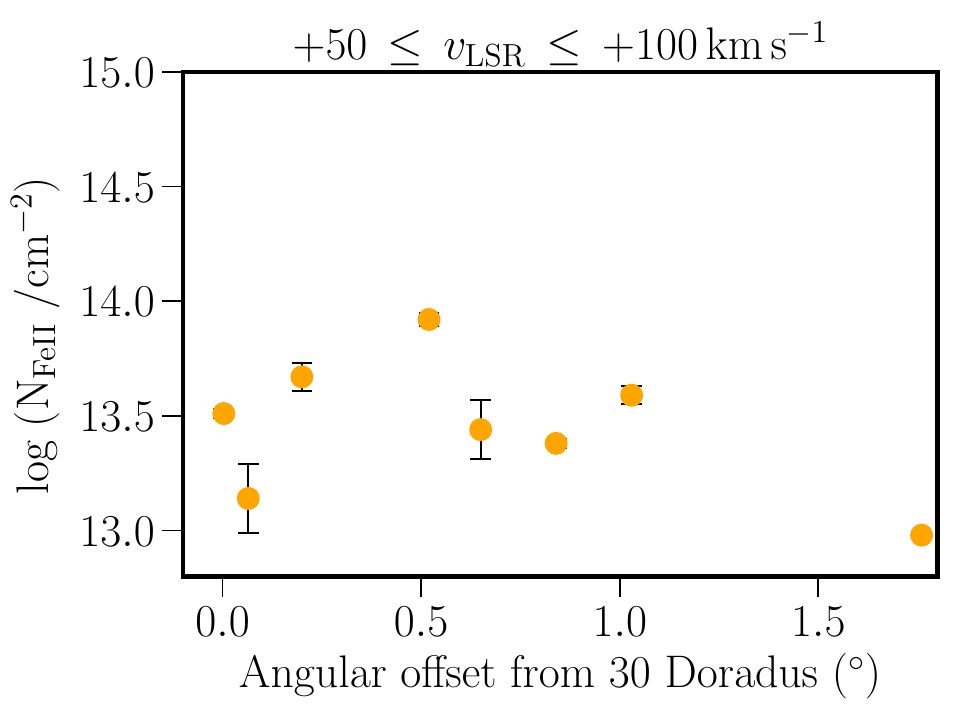}
  \includegraphics[width=0.44\textwidth, trim={0 0 0 0}, clip]{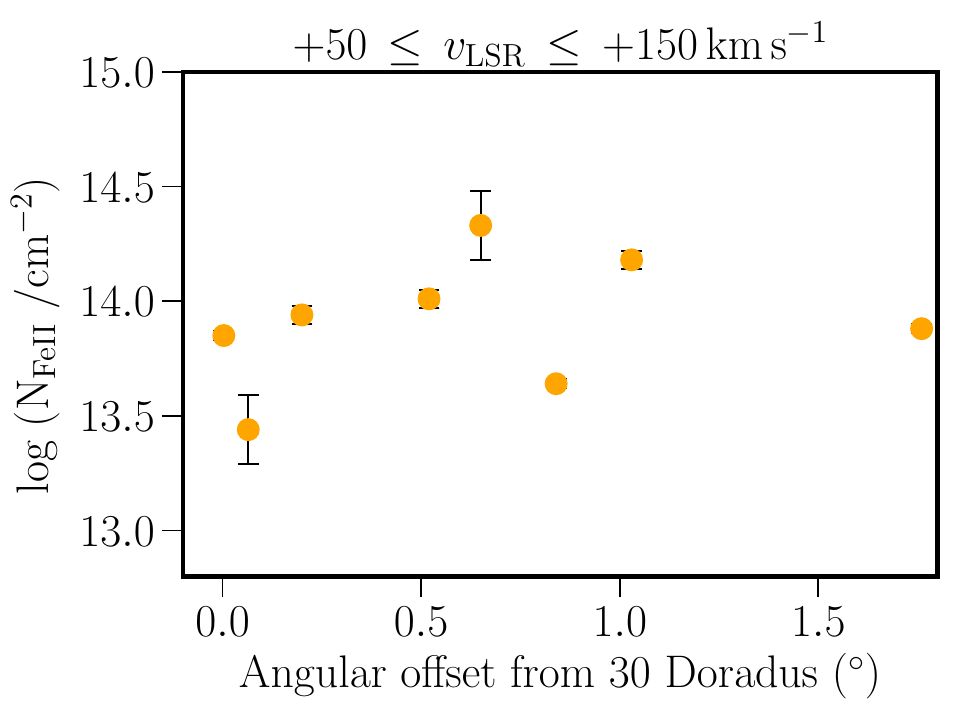}
  \caption{Variations in integrated column densities of Fe\textsc{~ii} as a function of angular separation from 30~Doradus in various velocity ranges as shown on the top of each panels.}
  \label{fig:colden_multi}
\end{figure}

We tested how varying the kinematic boundary affected how the column densities varied with increase projection distance from 30~Doradus. When we expanded the integration range by a small amount of 10--20\, \kms, this did not have any significant change in the trend we obtained earlier. This is because there is a minimal contribution to the total integrated column densities from the components that lie between $+130\,\le v_{\rm LSR}\le +150\, \kms$. If we increased this integration range by $50\,\kms$ to $v_{\rm LSR}=+100\,\kms$, we still observe a radial decrease in the column densities, though including the absorbers in this region would soften the decline (see Figure~\ref{fig:colden_Fe}).  Indeed, examining the total column densities distribution spanning the region $+100\,\le v_{\rm LSR}\le +150\, \kms$, we do not observe any specific trend in the data (see first panel in Figure~\ref{fig:colden_multi}). Moreover, we find no specific pattern in the integrated column density with respect to angular distance both in the velocity range of $v_{\rm LSR}=+50$ to $+100\, \kms$ (see middle panel in Figure~\ref{fig:colden_multi}) and in $v_{\rm LSR}=+50$ to $+150\, \kms$ (see third panel in Figure~\ref{fig:colden_multi}). This suggests that the absorption at the $v_{\rm LSR}\gtrsim +150\, \kms$ is more likely associated with the LMC wind originated from the 30~Doradus. The absorption at $v_{\rm LSR}=+100$ to $+150\, \kms$ may be associated both with the MW HVC and the LMC wind.

\begin{table*}[htbp]
\centering
\caption{The total AOD column densities integrated between $v_{\rm LSR}=+150\, \kms$ to the left boundary of the LMC's \hi\, disk (see Table~\ref{tab:boundaries}).}
\label{tab:aod_N}
\begin{tabular}{ccccc}
\hline
\hline
Sightline & Angular offset ($^\circ$) & $\log N_{\rm Fe\textsc{~ii}}$ & $\log N_{\rm Si\textsc{~ii}}$ & $\log N_{\rm O\textsc{~i}}$ \\
\hline
BAT99\,105 & 0.003 & $>14.80$ & $>14.73$ & $>15.24$ \\
Sk$-$69$^\circ$246 & 0.06 & $>14.50$ & 15.28 $\pm$ 0.01 & $>15.16$ \\
Sk$-$68$^\circ$135 & 0.20 &14.20 $\pm$ 0.04 & 15.30 $\pm$ 0.01 & $>15.12$ \\
BI\,214 & 0.52 & 14.47 $\pm$ 0.03 & $>14.79$ & $>15.30$ \\
Sk$-$69$^\circ$175 & 0.65 & 14.21 $\pm$ 0.05 & 14.42 $\pm$ 0.09 & 15.32 $\pm$ 0.02 \\
Sk$-$68$^\circ$112 & 0.84 & 14.01 $\pm$ 0.06 & $>14.34$ & $>14.82$ \\
BI\,173 & 1.03 & 14.34 $\pm$ 0.19 & $>14.44$ & $>15.02$ \\
Sk$-$69$^\circ$104 & 1.76 & 13.67 $\pm$ 0.01 & 14.15 $\pm$ 0.06 & $>14.55$ \\ \hline
\end{tabular}
\end{table*}

\begin{table}
\centering
\caption{Kendall's Tau tests between the angular offsets from 30~Doradus and the column densities.}
\label{tab:k_tau}
\begin{tabular}{cccc}
\hline
\hline
Parameters & \underline{Fe\textsc{~ii}} & \underline{Si\textsc{~ii}} & \underline{O\textsc{~i}}\\
\hline
$\tau$ & $-$0.64 & $-$0.57 & $-$0.43\\
$\rm p$ & 0.03 & 0.06 & 0.18\\
\hline
\end{tabular}
\tablecomments{Kendall rank correlation coefficient $\tau$ and the significance level $\rm p$ to measure the association between the angular offsets from 30~Doradus and the column densities.}
\end{table}

\subsection{Kinematic distribution of the gas}
\label{subsection:MWvsLMC}

\begin{figure*}
  \centering
  \includegraphics[width=0.45\textwidth]{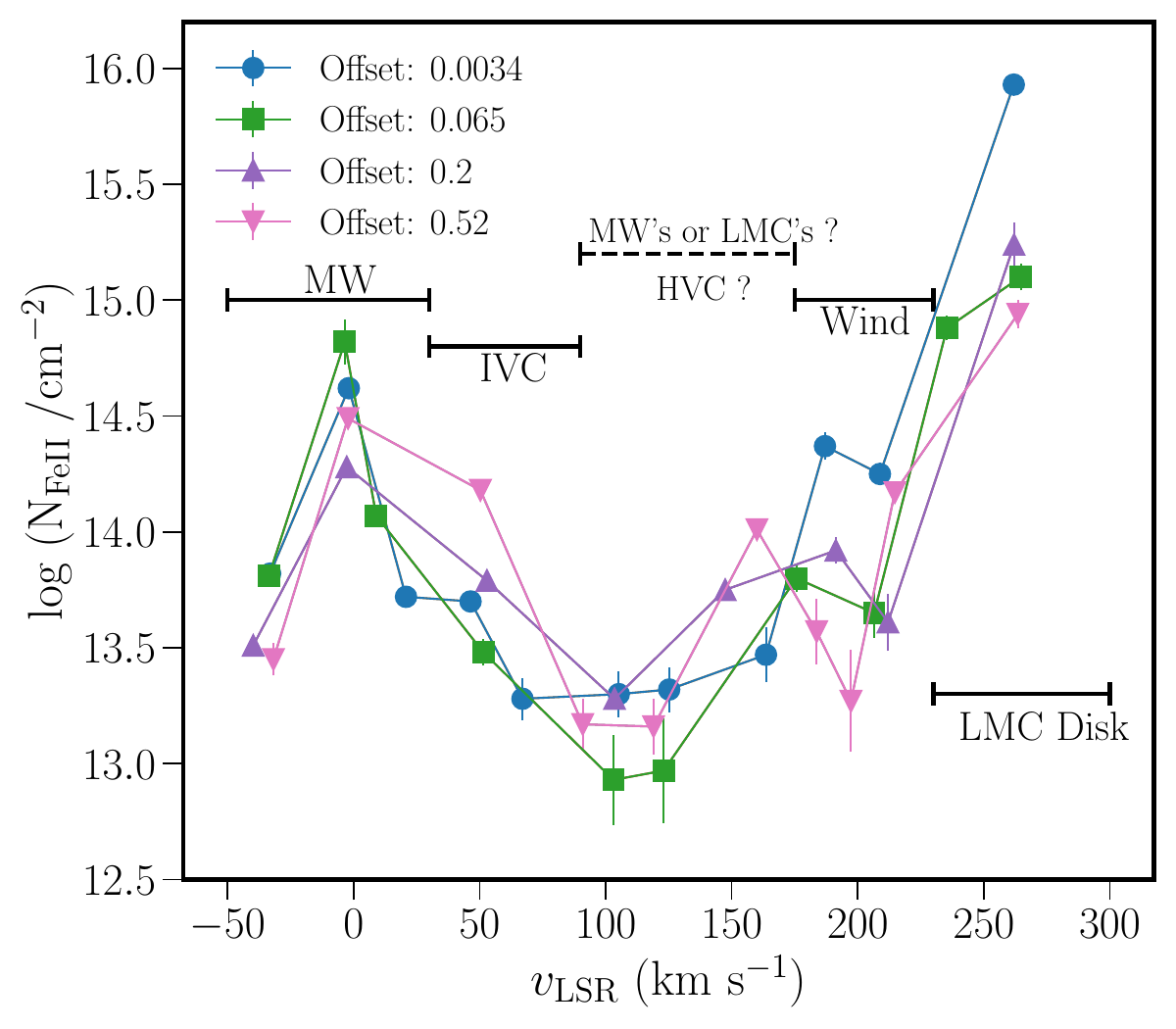}
  \includegraphics[width=0.45\textwidth]{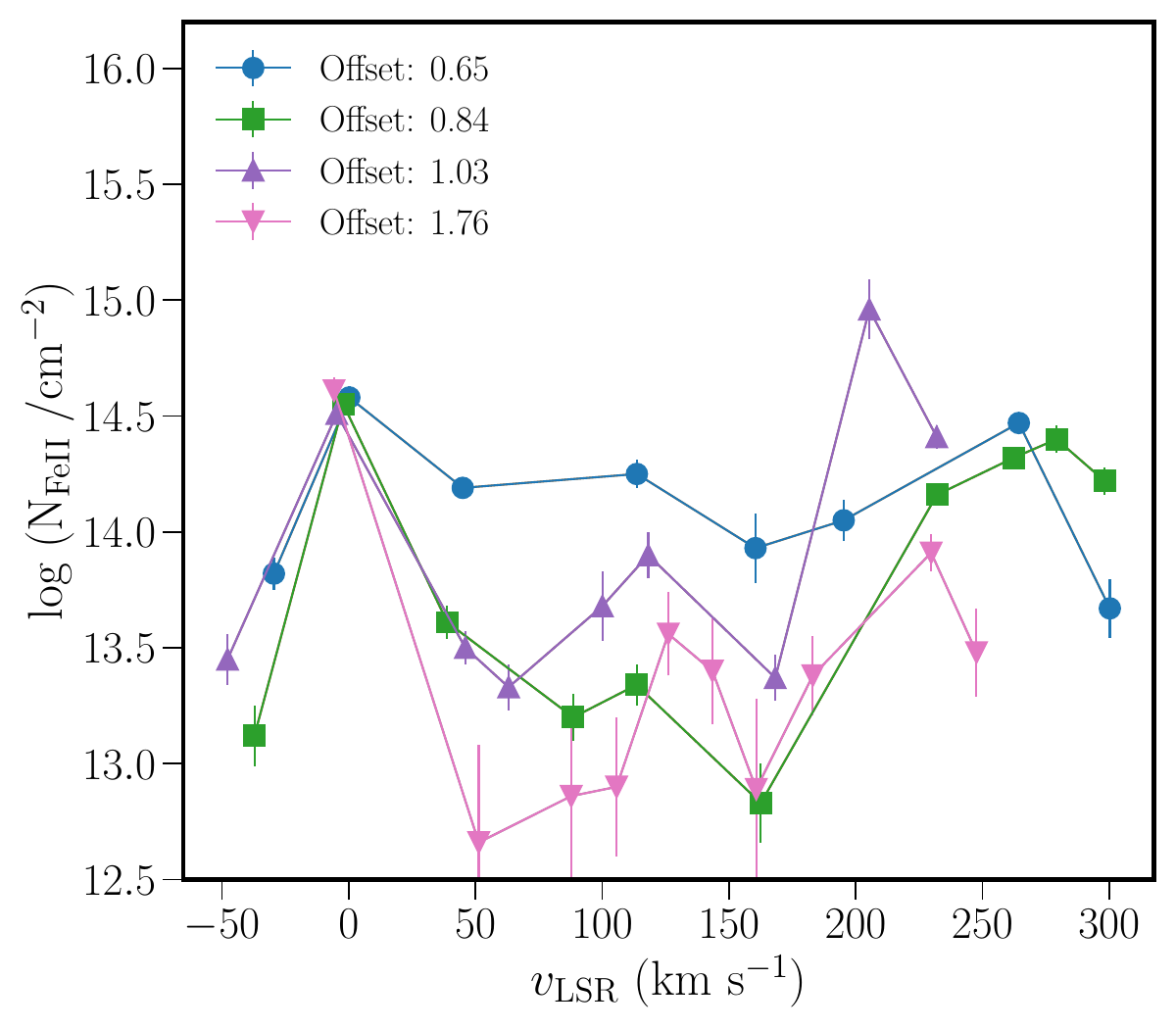}
  \caption{Variations in component column densities derived from Voigt profile fitting for Fe\textsc{~ii} with respect to $v_{\rm LSR}$. The left panel includes only the first four sightlines which are within an angular separation $\lesssim$ 0.52 degrees from the center of 30~Doradus and the right panel includes the sightlines with an offset angle $\gtrsim$ 0.65 degrees. In the left panel, the different horizontal solid lines in black show the expected velocity ranges for the MW, IVC, Wind, and the LMC disk regions. The dashed line in black shows the velocity range which may be associated with the MW HVC or the LMC wind.}
  \label{fig:NvsV}
\end{figure*}

\begin{figure*}
  \centering
  \includegraphics[width=0.45\textwidth]{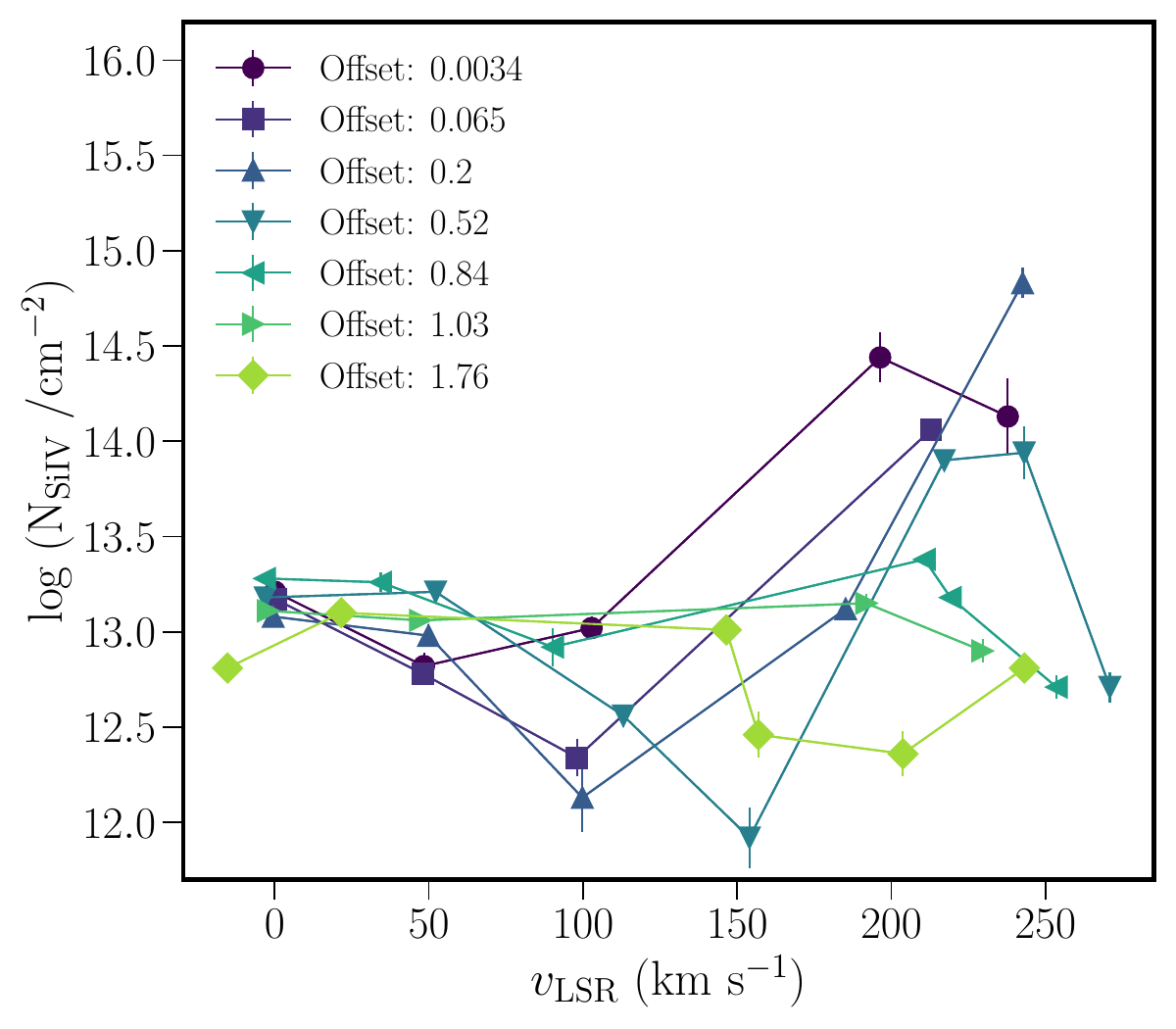}
  \includegraphics[width=0.45\textwidth]{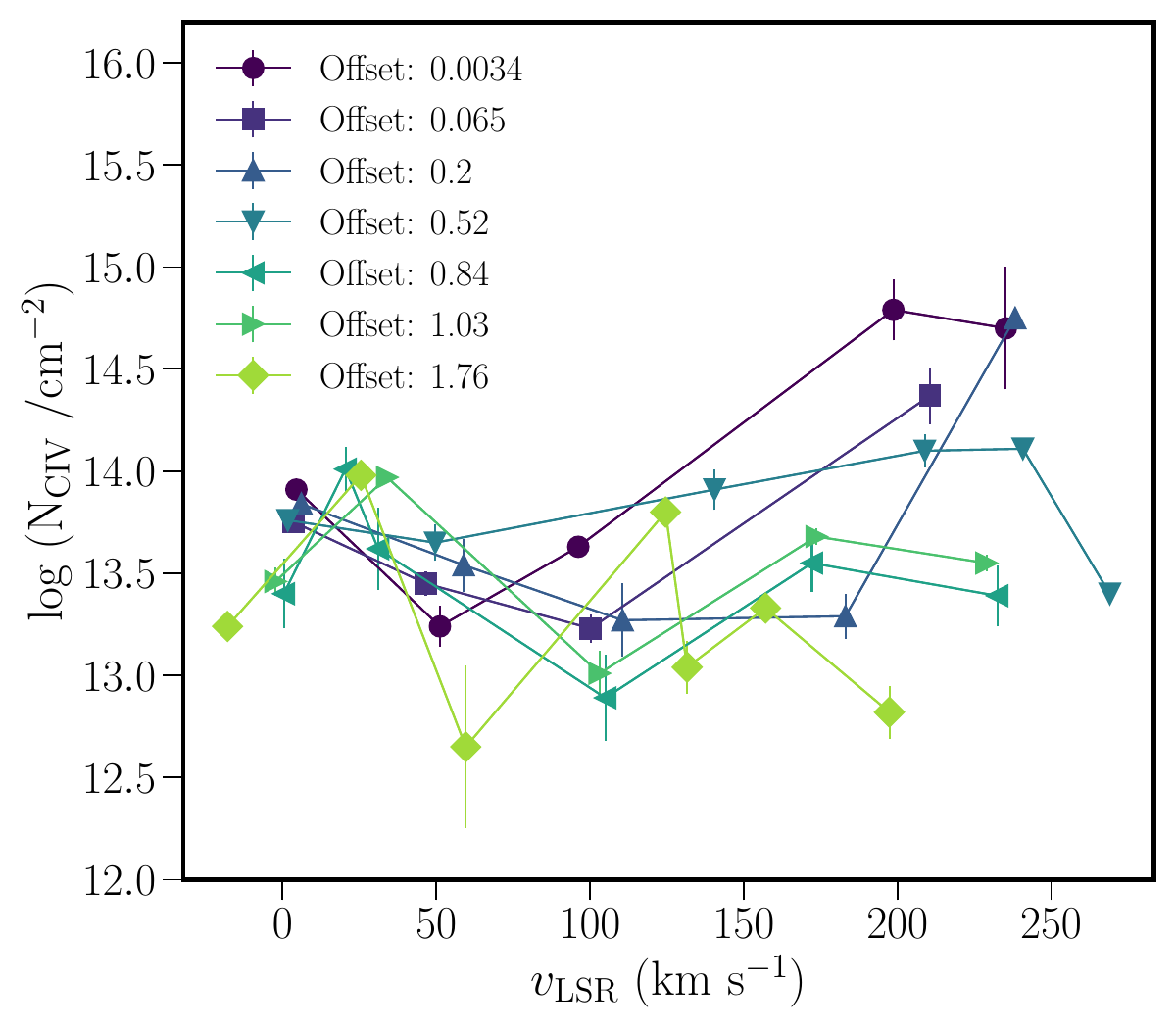}
  \caption{Variations in component column densities of Si\textsc{~iv} (Left) and C\textsc{~iv} (Right) with respect to $v_{\rm LSR}$ for all the sightlines for which we have the Si\textsc{~iv} and C\textsc{~iv} measurements.}
  \label{fig:NvsV_high}
\end{figure*}

We analyze the variations in component column densities derived from Voigt profile fitting for Fe\textsc{~ii} with respect to $v_{\rm LSR}$ in the entire velocity spectrum. We choose Fe\textsc{~ii} due to the availability of multiple transitions (Fe\textsc{~ii}\,$\lambda1608$, 2249, 2260, 2344). The components associated with these transitions exhibit fewer issues with blending and saturation when compared to other ions, which  enabled us to identify and characterize the largest number of components. 

Our examination reveals a pattern in the column densities and velocities for absorbers within 0.52~degrees from the center of 30~Doradus (see left panel of Figure \ref{fig:NvsV}). As we kinematically progress from the LMC's disk region toward the MW, the column densities steadily decrease, reaching a minimum at approximately $v_{\rm LSR}\approx+100\, \kms$ before starting to rise again as we approach the MW. For the $\rm offset=0.52^\circ$ curve, there appears to be a local minimum at $v_{\rm LSR}\approx+200\, \kms$, which seems consistent with the column densities observed near $v_{\rm LSR}\approx+100\, \kms$. However, the component at $v_{\rm LSR}\approx+200\, \kms$ is less reliable due to its very narrow line width of $2.8\, \kms$ and a larger error in column density of $0.21\, \rm dex$. Additionally, this component is not observed in other ions along this sightline, suggesting it may not be a robust feature. Therefore, we consider the more significant minimum to be around $v_{\rm LSR}\approx+100\, \kms$ for most of these first four sightlines, indicating a likely transition velocity. While some of the Fe\textsc{~ii} components in the LMC disk and the MW disk are potentially saturated, this does not affect the region where the column density is reaching towards the minimum value.

This trend is not obvious for the four farthest sightlines located from the center (see right panel of Figure~\ref{fig:NvsV}). For the outer sightlines, one can speculate the minimum column density may shift to $v_{\rm LSR}\approx+150\, \kms$ as most of them show a local minima in the column density close to that velocity. We conducted a similar analysis for the high-ions C\textsc{~iv} and Si\textsc{~iv} (see Figure~\ref{fig:NvsV_high}) and find no distinct minimum in their column densities, though two sightlines near 30~Doradus exhibit minimum column density at around $v_{\rm LSR}\approx+100\, \kms$.

If the lighter clouds ejected from a galaxy are more easily accelerated by momentum-driven forces, they should reach higher velocities. Additionally, if faster clouds traveling through an ambient medium disperse more rapidly due to stronger ram-pressure affects, then we would expect higher-velocity material to be more diffuse. Under these assumptions, the observed trend---where the material moving at higher velocities relative to the LMC’s disk exhibits lower column densities---suggests that higher-speed gas is more diffuse. This pattern aligns with the inverse relationship between cloud speed ($v_{c}$) and column density ($v_{c} \propto \frac{1}{\sqrt{N}}$) observed in galactic winds (e.g., \citealt{2002ASPC..254..292H}, see also Appendix~\ref{section:N_v}). Within this framework, we interpret the velocity at which the column density reaches a minimum as the maximum speed at which detectable diffuse gas can escape the LMC’s gravitational pull. Beyond this minimum, the subsequent rise in column densities at even higher velocities may indicate a transition to material associated with the MW’s HVCs.

Under the assumption described above, the lack of a clear trend for the outer four sightlines (see right panel of Figure~\ref{fig:NvsV}) may result from contamination of the LMC wind by the MW's HVC as well as LMC's CGM in the velocity range of $\vlsr=+100\,\kms$ to $\vlsr=+150\,\kms$. A recent study by \citet{2024arXiv241011960M} detected the LMC's CGM in several low-ionization species, including Fe\textsc{~ii}, out to an impact parameter of $17\, \kpc$. While their column densities for Fe\textsc{~ii} are generally similar to the values we obtained for the wind components, the population of Doppler b-values they observed differs significantly from our results. However, we cannot entirely rule out the possibility that the LMC wind may also be contaminated by the LMC's CGM. For the high-ions, we believe that the LMC wind may be polluted by the LMC's coronal gas \citep{2022Natur.609..915K} at this high velocity resulting into an absence of the trend of column density vs. velocity obtained for the low-ions. As the most active star-forming region in the LMC, 30~Doradus is expected to drive high-speed winds. \citet{Ciampa2020} have found evidence of similar velocity of outflowing gas from this region reaching a maximum of $175\, \kms$ relative to the LMC disk (or $v_{\rm LSR}\approx 100\, \kms$) from \ha\, emission-line observations. 

Recent study by \citet{2024arXiv240204313Z} has only considered the absorption at $v_{\rm helio}>+175\, \kms$, assuming that the material at greater speeds relative to the LMC is less likely to be associated with the galaxy's outflows. Their interpretation was based on the results from \citet{2015A&A...584L...6R}, which reported an HVC associated with the MW at $v_{\rm LSR}\approx +115\, \kms$. However, the sightline from \citet{2015A&A...584L...6R} was several degrees away from 30~Doradus and even outside the $\log(N_{\rm H\textsc{~i}}/\cm^{{-2}})\approx18.5$ contour for the LMC's disk. 

Our primary focus of this study is the LMC's outflows, which are blueshifted in absorption compared to the systemic velocity of the LMC using stars embedded within that galaxy's disk. Using the same procedure for defining the kinematic bounds of the LMC's gaseous disk that we used to identify material that is outflowing from this galaxy, we additionally find redshifted material that is presumably flowing toward it. Specifically, for three sightlines---Sk$-$68$^\circ$135, BI\,214, and Sk$-$68$^\circ$112---we observe redshifted absorption in the low-ions beyond the disk that is consistent with the weaker H\textsc{~i} components from the GASS data. \citet{2002ApJ...569..214H} have also reported a possible inflow towards a sightline, based on redshifted O\textsc{~vi}, just within ${\sim} 0.2^\circ$ of Sk$-$68$^\circ$112. We have shaded the redshifted absorption associated with these sightlines in red in Figure~\ref{fig:plotstacks} in the Appendix. This redshifted absorbers might represent gas that is inflowing onto the LMC that could be recycling back from the outflows. 

\section{Kinematics}\label{sec:gas_kinematics}
The Doppler $b$-parameter characterizes the velocity spread of an absorption line. The $b$-values obtained through Voigt profile fitting represent the combined effects of both the thermal and non-thermal broadening in an absorption line. By comparing the $b$-values for different ions and for various sightlines, we can gain insight into the mechanisms that are responsible for the broadening of the absorption lines. Additionally, comparing the velocity centroids across different sightlines and various ions within the same sightline can offer crucial information about the potential origin of the gas and the distinct gas phases. We investigate these aspects throughout the velocity spectrum, extending from the MW to the LMC. Nevertheless, our main emphasis is on the region associated with the LMC wind.

\begin{table}
\centering
\caption{$b$-value Distributions}
\label{tab:b-stat}
\begin{tabular}{cccc}
\hline
\hline
Region & $\langle b_{\mathrm{Fe\textsc{~ii}}}/\mathrm{km\,s^{-1}}\rangle$ & $\langle b_{\mathrm{C\textsc{~iv}}}/\mathrm{km\,s^{-1}}\rangle$ & $\langle b_{\mathrm{Si\textsc{~iv}}}/\mathrm{km\,s^{-1}}\rangle$\\
\hline
(I)  & $11.1 \pm 3.9$ &  &\\
(II)  & $8.3 \pm 3.4$ &  &\\
(III) & $11.6 \pm 7.7$ & $24.9 \pm 15.2$ & $18.4 \pm 11.9$\\
(IV) & $11.8 \pm 5.1$ &  &\\
\hline
\end{tabular}
\tablecomments{The values given in the second, third, and fourth columns are (average $\pm$ standard deviation) of the $b$-values for different regions based on the expected LMC wind region deduced from Section~\ref{sec:gas_distribution}: Milky Way (Region I: $v_{\rm LSR} < 100\, \mathrm{km\,s^{-1}}$), Region II: $100\, \mathrm{km\,s^{-1}} < v_{\mathrm{LSR}} < 150\, \mathrm{km\,s^{-1}}$, LMC winds (Region III: $150\, \mathrm{km\,s^{-1}} < v_{\mathrm{LSR}} < 210\, \mathrm{km\,s^{-1}}$), and LMC disk (Region IV: $v_{\mathrm{LSR}} > 210\, \mathrm{km\,s^{-1}}$) for various ions.}
\end{table}

\subsection{Doppler $b$-parameter distributions}\label{subsec:b_dist}
We compare the magnitude of $b$-values across the entire velocity range from the MW to the LMC using Fe\textsc{~ii} (see details in the Appendix \ref{section:b-val}). The average $b$-value and the standard deviation from the mean for the LMC wind region ($+150\, \lesssim v_{\rm LSR} < +210\, \kms$) is found to be $\langle b\rangle_{\rm Fe\textsc{~ii}}=11.6 \pm 7.7\,\kms$ (see Table~\ref{tab:b-stat}). These values for $+100\, \lesssim v_{\rm LSR} < +150\, \kms$ are $\langle b\rangle_{\rm Fe\textsc{~ii}}=8.3 \pm 3.4\,\kms$. We do not find any significant differences in the average $b$-values between any of these regions. Moreover, we also do not observe any specific trend in the magnitude of $b$-values with respect to the angular distance from the 30~Doradus.

We also estimate the average $b$-values and the standard deviations from the mean for the high ions to be $\langle b\rangle_{\rm C\textsc{~iv}}=24.9 \pm 15.2\,\kms$ and  $\langle b\rangle_{\rm Si\textsc{~iv}}=18.4 \pm 11.9\,\kms$ for the absorbers within the velocity range of $+150\, \lesssim v_{\rm LSR} < +210\, \kms$. These values significantly exceeded the average values observed for their low-ion counterparts. The increased $b$-values for the high-ions is mostly because they typically probe a distinct phase of the warm ionized gas compared to the low-ions. Despite the generally elevated average $b$-values observed for high-ions, a subset of components exhibits $b$-values $< 10\, \kms$. Such weaker narrow Si\textsc{~iv} and C\textsc{~iv} components could potentially indicate absorbers that are photoionized. However, there is also a possibility that they may trace non-equilibrium collision ionization (NECI) (see \citealt{2011ApJ...727...46L}). Additionally, the $b\gtrsim50\, \kms$ for some of the C\textsc{~iv} components may be an indication of more non-thermal broadening, i.e., collisions and possibly tracing a warmer gas than Si\textsc{~iv}. However, since the ionization potential of C\textsc{~iv} is higher than that of Si\textsc{~iv}, it is not surprising that they do not trace a multi-phase gas in the same way. For three sightlines, namely Sk$-$69$^\circ$246, Sk$-$68$^\circ$135, and BI\,173, we also possess measurements for the $b$-values of the intermediate-ion Al\textsc{~iii}. The average $b$-value and the standard deviation from the mean for the Al\textsc{~iii} ions are found to be $\langle b\rangle_{\rm Al\textsc{~iii}}=13.9 \pm 8.0\,\kms$. The average $b$-value for the Al\textsc{~iii} ions are found to lie between the values for the low-ions (Fe\textsc{~ii}) and high-ions (Si\textsc{~iv} and C\textsc{~iv}) as one would expect from the intermediate ionization potential of Al\textsc{~iii} at $\rm IP_{\rm Al^{+2}}=18.8\,\rm eV$.

\begin{figure*}
  \centering
  \includegraphics[scale=0.45]{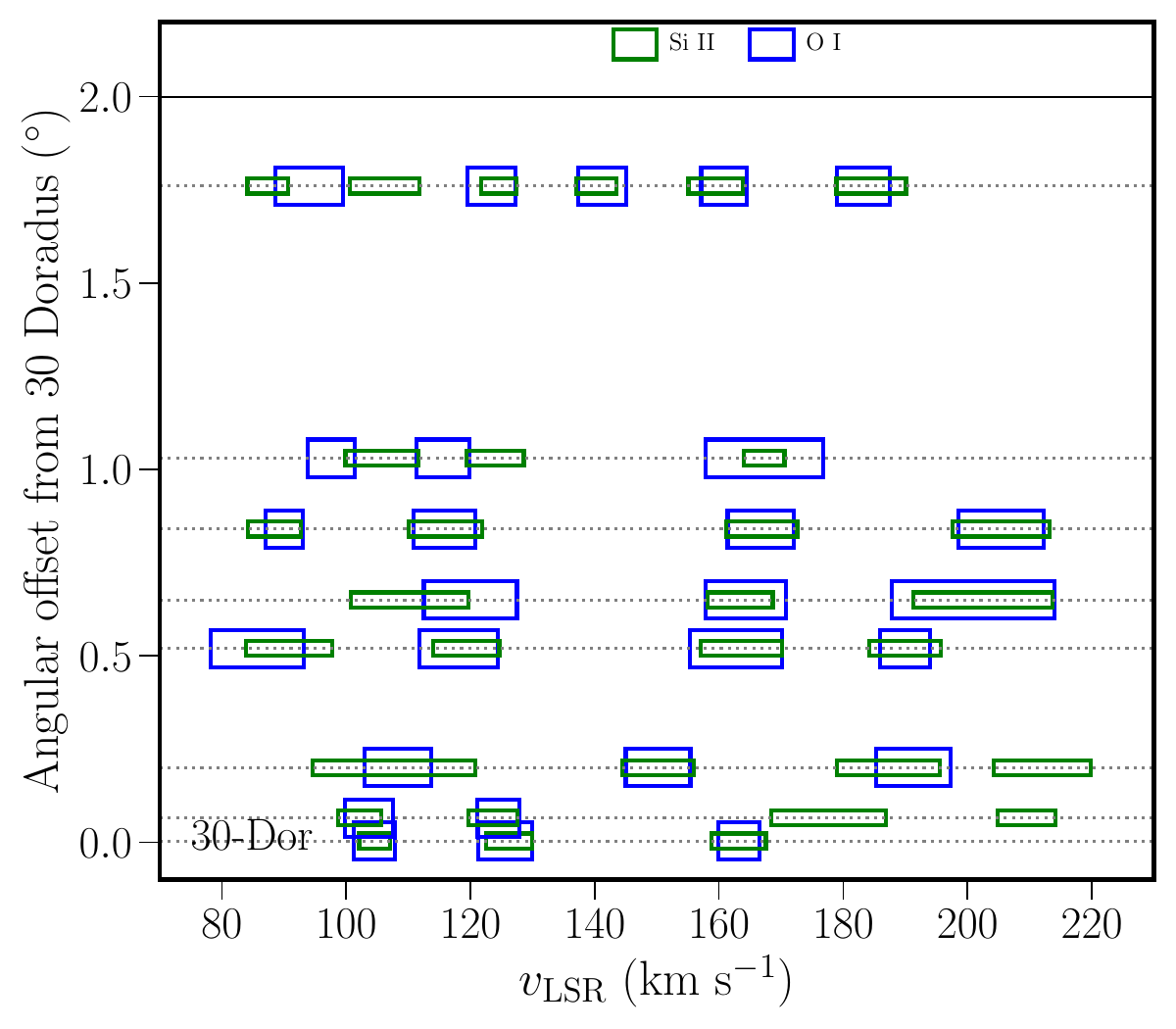}
  \includegraphics[trim=77 0 0 0,clip,scale=0.45]{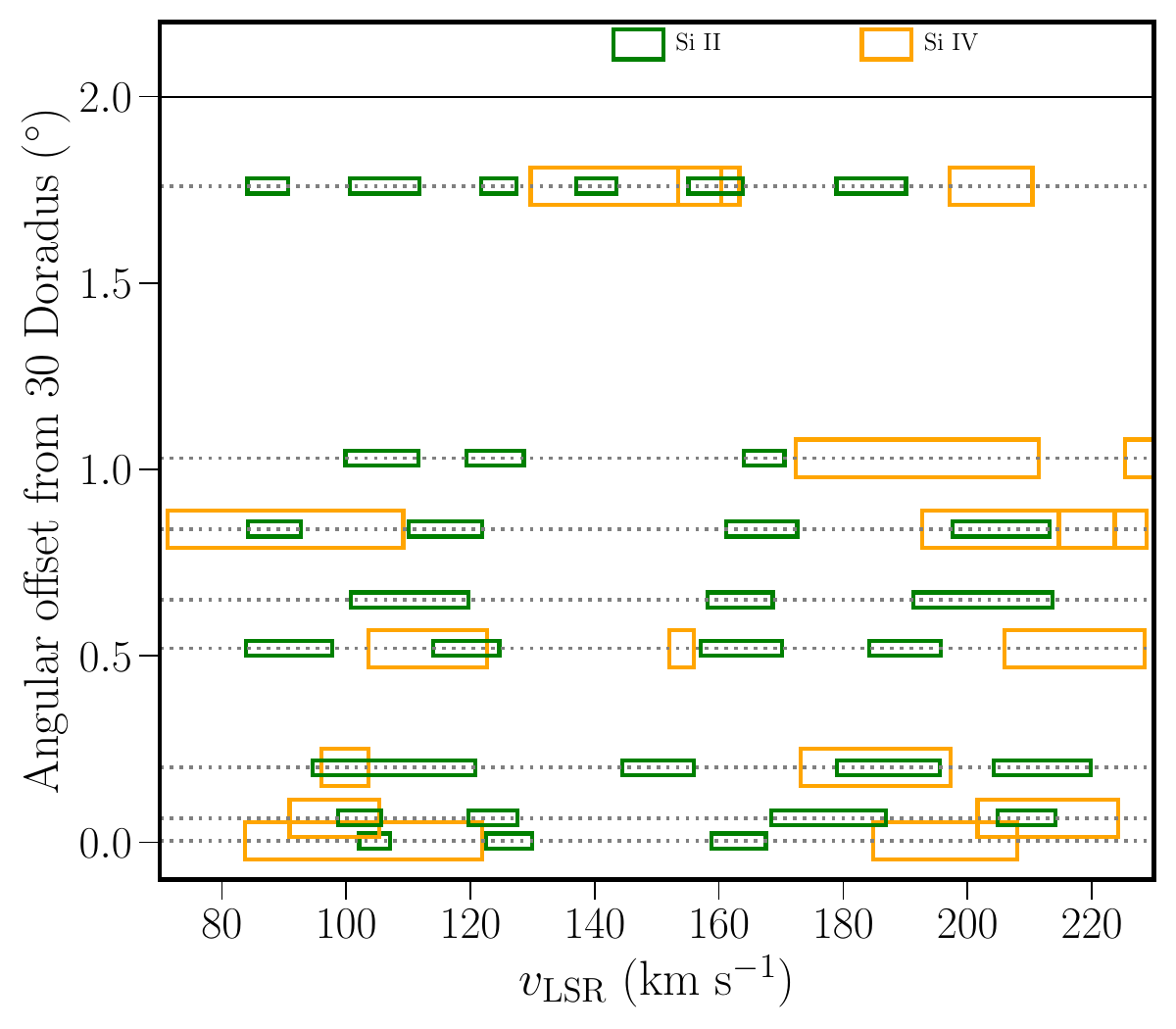}
  \caption{Comparisons of Doppler $b$-parameters distributions across various sightlines at varying angular separations from 30~Doradus: (Left) between the low ions Si\textsc{~ii} and O\textsc{~i} in the range of $+90 \lesssim v_{\rm LSR} \lesssim +220\, \kms$. (Right) between the high-ions Si\textsc{~iv} and the low-ions Si\textsc{~ii}. In both figures, the width of the rectangular boxes is equal to the magnitudes of the Doppler $b$-parameters, while the varying heights of the boxes are intentionally adjusted to enhance visibility. The center of these boxes represents the velocity centroid of the individual components. The horizontal dotted lines indicate the angular separations of these sightlines from the center of 30~Doradus.}
\label{fig:doppler_wind}
\end{figure*}

\subsection{Alignment of the absorption profiles}
We compare the magnitudes of the $b$-values and the velocity centroids of the components between Si\textsc{~ii} and O\textsc{~i} within the velocity range of $+100\,\lesssim v_{\rm LSR} \lesssim+220\, \kms$ which corresponds to the LMC wind region in Figure~\ref{fig:doppler_wind} (left). For the purpose of comparing with the low-ions with high-ions, we also display the $b$-values distributions for Si\textsc{~iv} and Si\textsc{~ii} within the wind velocities, in Figure \ref{fig:doppler_wind} (Right). In cases of complex profiles for high ions, the application of Voigt fitting can occasionally yield excessively large $b$-values in order to minimize the $\chi^2$. The components with $b$-values exceeding ${\sim}50\, \kms$ may be less reliable due to potential blending with other components or too low S/N and a smaller optical depth (see \citealt{2011ApJ...727...46L} for details). Therefore, we need to be careful when interpreting these broad components, as they could represent either a single broad component or a composite of numerous narrower components. 

\begin{figure}
  \centering
  \includegraphics[width=0.48\textwidth]{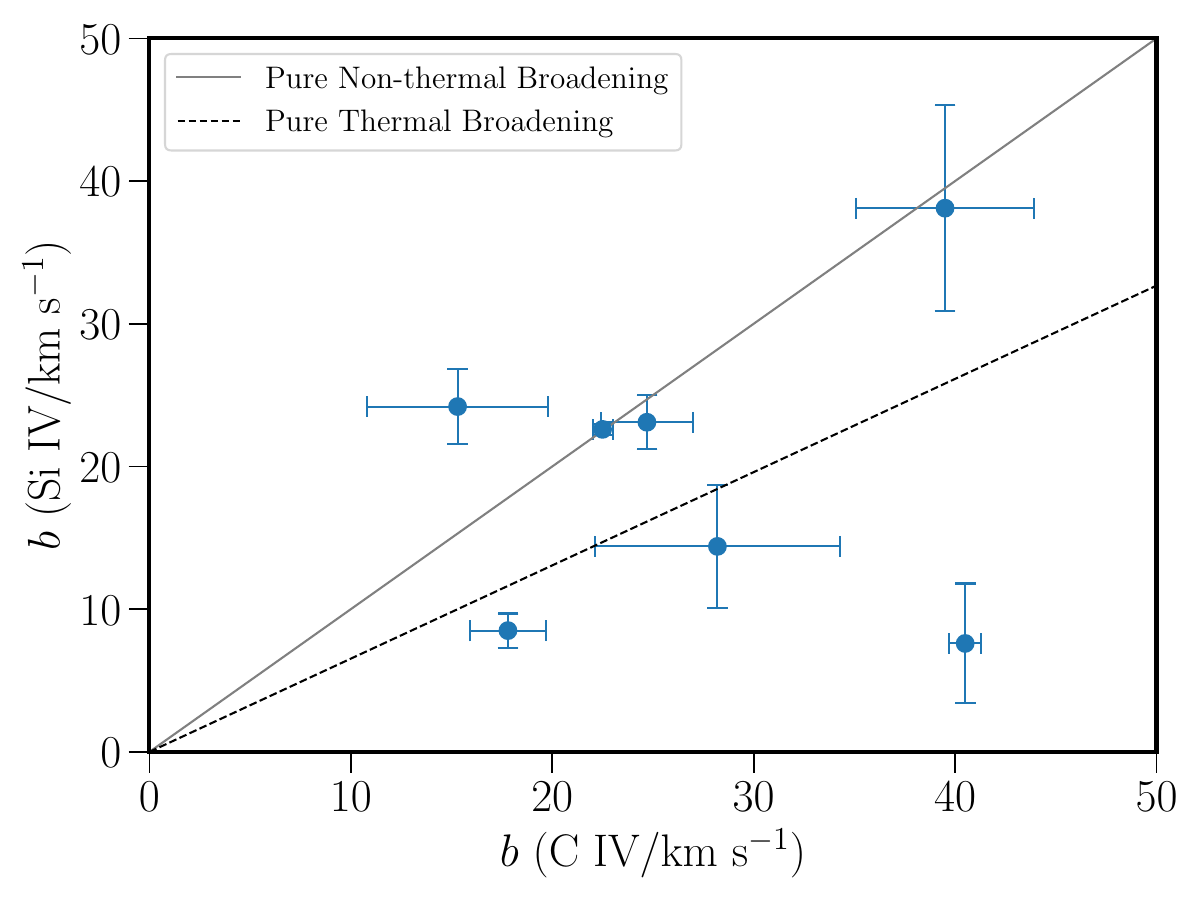}
  \caption{Comparison of the $b$-parameters for the velocity matching components between Si\textsc{~iv} and C\textsc{~iv} that falls within $90\, \kms \lesssim v_{\rm LSR} \lesssim 220\, \kms$. The solid line in gray is the broadening due to pure nonthermal motion and the dashed line in black is the broadening assuming a pure thermal motion.}
  \label{fig:b_CIV_SiIV}
\end{figure}

Comparing the velocity centroid offsets between O\textsc{~i} and Si\textsc{~ii} can be informative, as O\textsc{~i} serves as a proxy for neutral hydrogen, while Si\textsc{~ii} acts as a proxy for both neutral and ionized hydrogen. However, no significant offsets were observed between O\textsc{~i} and Si\textsc{~ii} within the velocity range where we anticipated the presence of LMC winds. The similar $b$-values and the absence of significant offsets is likely due to our detection of predominantly single-phase gas in the winds probed by the low-ions, which is photoionized. When comparing the positions of the components for Si\textsc{~ii} and Si\textsc{~iv}, there are instances where they may share similar velocity centroids. However, due to the significantly higher number of components in the low-ions compared to the high-ions in the velocity range of the wind probed by low-ions, they predominantly probe different phases of the gas. When comparing the Doppler $b$-values associated with matching components in Si\textsc{~iv} and C\textsc{~iv}, we observe that in 4 cases the data fall near to the nonthermal line, and in 3 cases, they are closer to the pure thermal broadening line (see Figure \ref{fig:b_CIV_SiIV}). This implies that both thermal and nonthermal motions are important in producing the high-ions.

Interestingly, we observe a consistent difference in the absorption profiles redward to the disk between low-ions and high-ions. In most cases, the absorption in low ions is more redshifted compared to C\textsc{~iv} and Si\textsc{~iv}, and also in comparison to Al\textsc{~iii} for the two systems where Al\textsc{~iii} data is available (see Figure \ref{fig:SK-69d246_lines} and the Figures \ref{fig:plotstacks} in the Appendix \ref{section:voigt_all}). This effect is particularly pronounced in the closest sightlines to 30~Doradus, where we observe the most concentrated and highest speed wind probed by low-ions (see Section \ref{sec:gas_distribution}). For instance, in the BAT99\,105 and Sk$-$69$^\circ$246 sightlines, most of the low-ion absorption is detected beyond $v_{\rm LSR}\gtrsim+300\, \kms$. However, for the high-ions, the absorption is limited at below $v_{\rm LSR}<+300\, \kms$.

\section{Gas ionization}\label{sec:ionization}
To evaluate the ionization state of the gas, we analyze the observed ion ratios for various low-ions and the line ratios for the high-ions. Additionally, we also incorporate multidimensional photoionization models using the radiative transfer code \textsc{Cloudy} (version 22.01, \citealt{chatzikos2023}) to characterize the physical and ionization properties of the gas.
\subsection{Low-ions}\label{subsec:low_ion}
As we will discuss in Section~\ref{subsection:Cloudy}, we lack direct measurements of H\textsc{~i} values for most of the wind components; therefore, it is not feasible to conduct photoionization modeling for all these components. However, we can assess the the spatial and kinematic variation in ionization and dust by comparing multiple low ionization species ratios and checking for any associated trends (see \citealt{2009ApJ...702..940L}). For this exploration, we used the [Si\textsc{~ii}/O\textsc{~i}], [Fe\textsc{~ii}/O\textsc{~i}], and [Si\textsc{~ii}/S\textsc{~ii}] ion ratios, which are logarithmic with respect to the solar abundances. We adopt the solar abundances from the \citet{2009ARAA..47..481A} catalog. We present those results in Table \ref{tab:ion_ratio} for different sightlines across a range of velocities. When we observe saturation or non detection of an ion, we report only the corresponding limits. For S\textsc{~ii}, we rely on either S\textsc{~ii}\,$\lambda1250$ or S\textsc{~ii}\,$\lambda1253$, while excluding S\textsc{~ii}\,$\lambda1259$ due to significant blending with the Si\textsc{~ii}\,$\lambda1260$ line. It is important to note that S\textsc{~ii}\,$\lambda1250$ and S\textsc{~ii}\,$\lambda1253$ lines are weaker transitions and are primarily detected in the slower-moving portions of the wind, which are kinematically closer to the LMC's disk ($v_{\rm LMCSR} > -160\,\kms$). The mean and dispersion values for [Si\textsc{~ii}/O\textsc{~i}] and [Fe\textsc{~ii}/O\textsc{~i}] for the components at $v_{\rm LSR} \gtrsim +100\,\kms$ are $0.73\pm0.29$ and $0.43\pm0.25\,\rm dex$, respectively. These dispersions substantially exceed the measurement errors for individual components, indicating the presence of genuine variations in ionization or dust depletion conditions across different sightlines and components. However, we do not find a clear trend in the [Si\textsc{~ii}/O\textsc{~i}] and [Fe\textsc{~ii}/O\textsc{~i}] values concerning their proximity to the 30~Doradus region. Also, there is not a discernible pattern of these values kinematically. Below, we discuss potential factors, such as ionization or dust depletion effects, which may be contributing to these variation in the ion ratios.

\begin{figure}
  \centering
  \includegraphics[width=\columnwidth]{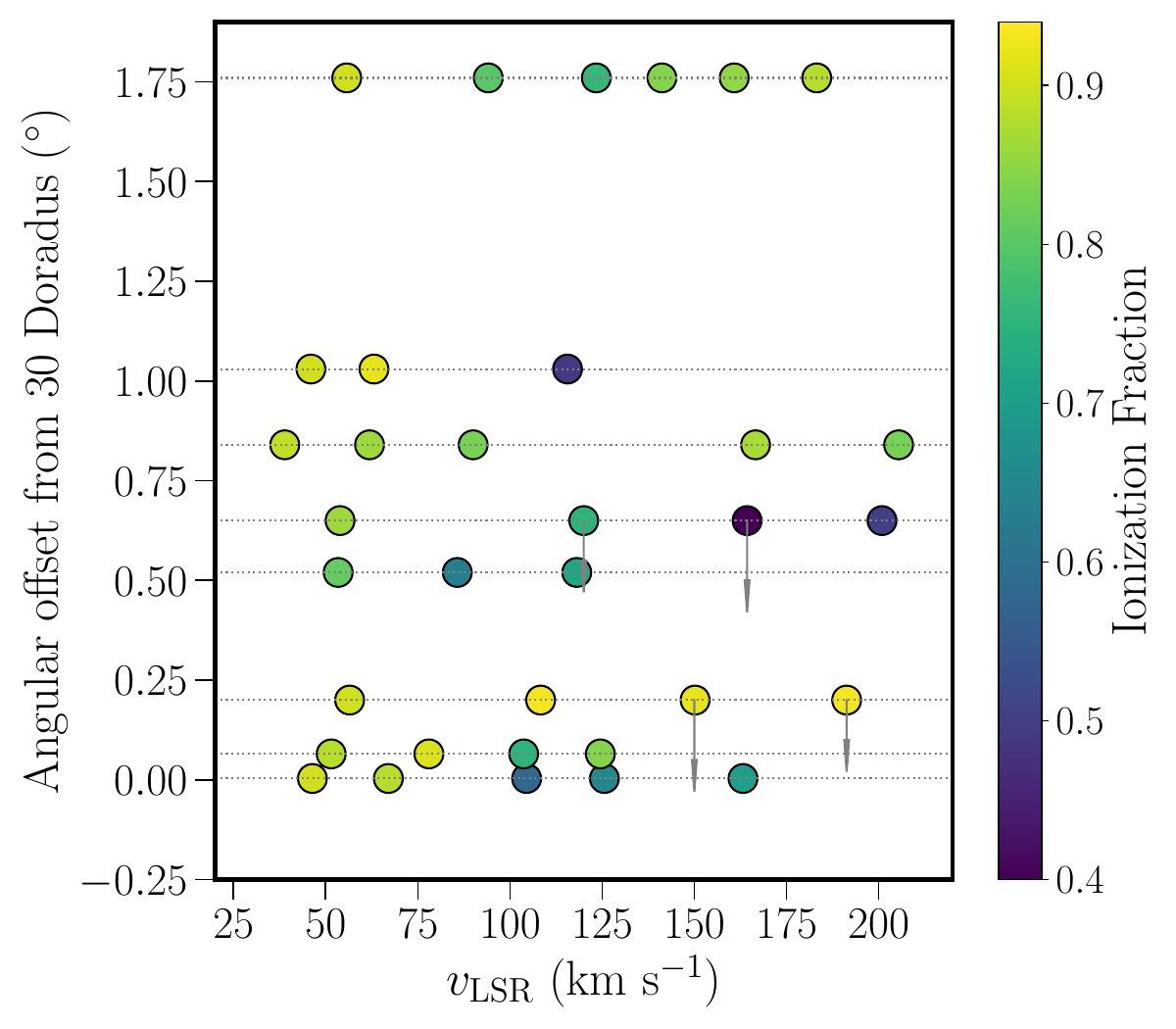}
  \caption{Ionization fraction as a function of angular separation from the 30~Doradus region and $v_{\rm LSR}$ as determined by employing O\textsc{~i} as a proxy for neutral hydrogen and Si\textsc{~ii} as a substitute for neutral and ionized hydrogen. Only components where both O\textsc{~i} and Si\textsc{~ii} were detected are included in the analysis. In few cases, the upper limits in the values are depicted by the downward directed arrows. The horizontal dotted lines are drawn at the corresponding angular separations of
these sightlines from the center of 30~Doradus.}
  
  \label{fig:ionmap}
\end{figure}

To accurately interpret ion ratios, we consider three fundamental properties of atomic species: chemical evolution, ionization, and dust depletion. Iron, for instance, is primarily produced during type-I supernova explosions \citep{1997NuPhA.621..467N}, while type-II supernovae produce $\alpha$-elements along with iron \citep{2006NuPhA.777..424N}. These distinct phenomena may cause variations in the quantities of atomic species they disperse into the ISM. For example, an $\alpha$-element enhancement of $\sim0.26\, \rm dex$ is observed in the neutral ISM in the LMC \citep{2024A&A...683A.216D}. Therefore, given the disparate time scales of these phenomena and the variations they produce in the abundances of atomic species, it is crucial to focus on elements with similar chemical evolution when interpreting ion ratios. Moreover, the dust depletion level depends on different density environments and may also correlate with the metallicity (\citealt{2024A&A...683A.216D} and references therein). Furthermore, dust depletion patterns differ for refractory elements like silicon and iron compared to volatile elements like sulfur, oxygen, and phosphorus \citep{ 2017ApJ...838...85J,2009ApJ...700.1299J,1996ApJ...470..893S}. Silicon and iron are subject to more extensive depletion than oxygen and sulfur in the MW's ISM \citep{2009ApJ...700.1299J}. Since silicon and oxygen are both $\alpha$-elements, the [Si\textsc{~ii}/O\textsc{~i}] ratio should remain largely unaffected by chemical evolution. However, this ratio is influenced by dust and ionization effects. Oxygen being a volatile element, we do not anticipate significant dust depletion \citep{2009ApJ...700.1299J} and the ionization correction for oxygen is also minimal \citep{viegas1995}. Therefore, to interpret [Si\textsc{~ii}/O\textsc{~i}], we must weigh the relative impact of dust depletion versus ionization on Si\textsc{~ii}. In general, if moderate dust depletion is expected in HVC gas, we might anticipate $-0.3 \lesssim [{\rm Si\textsc{~ii}}/{\rm O\textsc{~i}}] \lesssim 0.0\,\rm dex$ (see also \citealt{2009ApJ...702..940L}). For example, \citet{2023ApJ...946L..48F} found the dust depletion for HVC Complex~C to be $\delta(\rm Si)\approx0.29\, \rm dex$ suggesting moderate dust depletion. Therefore, our observed average value of $[\langle{\rm Si\textsc{~ii}}/{\rm O\textsc{~i}}\rangle] = 0.73\pm0.29\,\rm dex$ is primarily driven by the ionization of Si\textsc{~ii}, with the considerable dispersion in values stemming from spatial and kinematic ionization variations (see \citealt{2009ApJ...702..940L}). We also estimate the ionization fraction by using O\textsc{~i} as a proxy for neutral hydrogen and Si\textsc{~ii} as a substitute for both neutral and ionized hydrogen (H\textsc{~i}+H\textsc{~ii}) (see \citealt{2016ApJ...817...91B, 2009ApJ...702..940L, 2001MNRAS.323..904L}). The hydrogen ionization fraction can be approximated using the following equation:
\begin{align}
\chi_{\rm H\textsc{~ii}} &= 1 - \chi_{\rm H\textsc{~i}} \\
    &> 1 - \frac{\rm N_{O\textsc{~i}}}{\rm N_{Si\textsc{~ii}}} \cdot \frac{\rm (Si/H)_{\odot}}{\rm (O/H)_{\odot}},
\end{align}
where the hydrogen ionization fraction is $\rm \chi_{H\textsc{~ii}}=N_{\rm H\textsc{~ii}}/(N_{\rm H\textsc{~i}}+N_{\rm H\textsc{~ii}})$, the fraction of neutral hydrogen is $\rm \chi_{H\textsc{~i}}$, and $\rm (Si/H)_{\odot}$ and $\rm (O/H)_{\odot}$ represent the solar abundance of silicon and oxygen, respectively. Using this method, we estimated the hydrogen ionization fraction of the absorbers along our \nsightlines\,sightlines, which we display in Figure~\ref{fig:ionmap}. We include only components which have both the O\textsc{~i} and Si\textsc{~ii} detections. Generally, we observe that the wind components are ionized to varying degrees, ranging from at least 40$\%$  to as high as 90$\%$ based on the observation. It is important to note that if Si is depleted relative to O, the estimated [Si\textsc{~ii}/O\textsc{~i}] would provide a lower limit, and the gas could be even more ionized. The data uncover both spatial and kinematic anisotropy in the ionization fractions. However, no clear trend in the ionization fraction concerning the distance from 30~Doradus is evident.

To examine how the hydrogen ionization fraction varies across different velocity regimes, we compute their weighted averages in three distinct velocity bins. For the absorbers across all 8~sightlines, we find that $\langle \chi_{\rm H\textsc{~ii}}\rangle = 0.87 \pm 0.03$ at $v_{\rm MW,\,disk}< v_{\text{LSR}} \le +100\,\kms$, $\langle \chi_{\rm H\textsc{~ii}}\rangle=0.72 \pm 0.11$ at \(+100 \leq v_{\text{LSR}} < +150\, \kms\), and $\langle \chi_{\rm H\textsc{~ii}}\rangle= 0.76 \pm 0.14$ at $+150\,\kms \le v_{\text{LSR}} < v_{\rm LMC,\,disk}$. Notably, the ionization fraction of the gas at \(v_{\text{LSR}} < +100\, \kms\) that likely traces MW material is higher than and does not overlap with the gas moving at \(+100 \leq v_{\text{LSR}} < +150\,\kms\). However, the ionization fraction of the gas moving at this intermediate velocity bin does overlap with the highest velocity bin material that is moving at speeds more closer with the LMC, albeit the latter has a much larger spread. The overlap between these two may indicate a causal connection between the gas in these two velocity bins that would be speculated if they shared the same physical origin. One might expect the LMC wind highly ionized due to the stellar feedback that heats the gas through a combination of photoionization and shock heating. However, other factors---such as mixing with the surrounding media---could promote cooling. Combined, this has led to a large spread in ionization conditions in this region.

\begin{figure}
  \centering
  \includegraphics[width=\columnwidth]{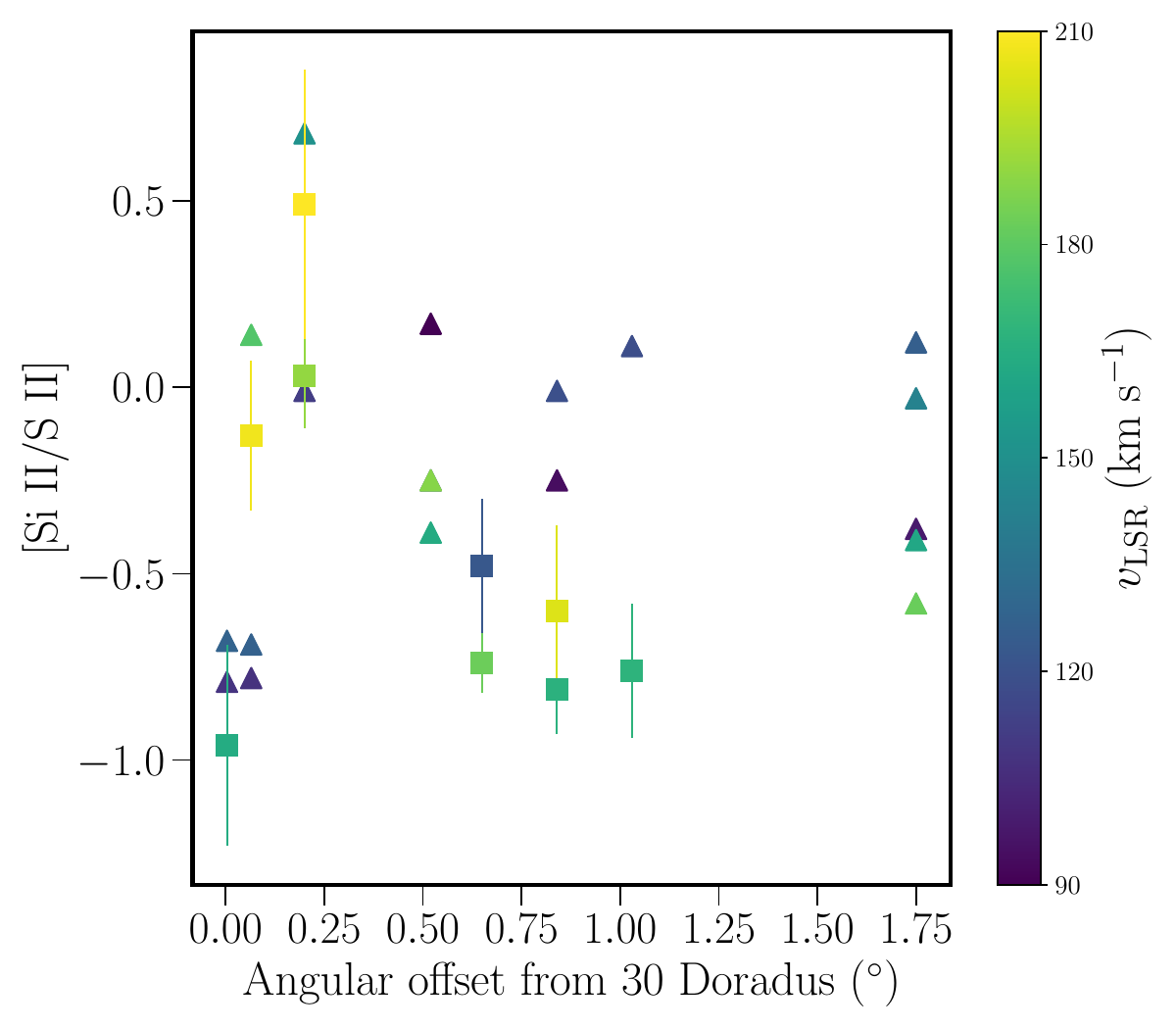}
  \caption{Variation of [Si\textsc{~ii}/S\textsc{~ii}] with offset angle from 30~Doradus and $v_{\rm LSR}$. The upward directed triangles without the error bars are only the lower limits.}
  
  \label{fig:dustmap}
\end{figure}

Another crucial ion ratio to consider is [Si\textsc{~ii}/S\textsc{~ii}]. Since both silicon and sulfur are $\alpha$-elements, this ratio depends on the relative effects of dust and ionization. We compare the distribution of this ratio across various offsets from 30~Doradus and a wide range of velocities (see Figure \ref{fig:dustmap}). In the cases of saturation of the Si\textsc{~ii} lines and non detection of S\textsc{~ii} lines, we present the lower limit of the ratio shown by the upward triangle. While there is not a distinct dependence of [Si\textsc{~ii}/S\textsc{~ii}] on velocity, a noticeable trend emerges regarding projected distance from 30~Doradus, especially for slow-moving components. Sightlines in close proximity to 30~Doradus tend to exhibit solar or occasionally supersolar values, while those farther away predominantly have subsolar values. If we consider sulfur as an undepleted element due to its lower condensation temperature, the ratio might suggest a depletion pattern that resulted from more dust destruction in the central region of 30~Doradus by the higher number of supernova remnants (see \citealt{2015ApJ...799...50L}). However, we cannot draw a conclusion from this ratio alone without considering the ionization state of the gas, as the ionization potential for S\textsc{~ii} is higher than that of Si\textsc{~ii}. Additionally, recent studies indicate that there is evidence of sulfur depletion in the ISM \citep{2022ApJ...928...90R}.

\begin{table*}
\centering
\caption{Observed ion ratios}
\label{tab:ion_ratio}
\begin{tabular}{ccccccc}
\hline
\hline
Sightline & Angular offset ($^\circ$) & $v_{\rm LSR}$(\kms) & [Fe\textsc{~ii}/O\textsc{~i}] & [Si\textsc{~ii}/O\textsc{~i}] & [Si\textsc{~ii}/S\textsc{~ii}] & Comments \\
\hline
BAT99\,105 & 0.003 & $+$104.5 & $0.37 \pm 0.10$  & $0.38 \pm 0.06$ & $>-0.79$ & HVC/Wind \\
          && $+$125.6 & $0.45 \pm 0.10$   & $0.46 \pm 0.07$ & $>-0.68$ & HVC/Wind\\
          && $+$163.2 & $0.63 \pm 0.17$   & $0.52 \pm 0.18$ & $-0.96 \pm 0.27$ & LMC wind\\

Sk$-$69$^\circ$246 & 0.06 & $+$103.7 & $0.28 \pm 0.05$ & $0.61 \pm 0.06$ & $>-0.78$ & HVC/Wind\\
                   && $+$124.5 & $0.45 \pm 0.07$ & $0.81 \pm 0.08$ & $>-0.69$ & HVC/Wind\\
                   && $+$177.6 &     ...         &      ...        & $>0.14$ & LMC wind \\
                   && $+$209.5 &    ...          &     ...         & $-0.13 \pm 0.20$ & LMC wind \\

Sk$-$68$^\circ$135 & 0.20 & $+$108.3 & $0.39 \pm 0.04$ & $1.14 \pm 0.04$ & $>-0.01$ & HVC/Wind\\
                   && $+$150.2 & $<0.21$         & $<1.12$         & $>0.68$  & LMC wind \\
                   && $+$191.3 & $<0.19$         & $<1.18$         & $0.03 \pm 0.14$ & LMC wind\\
                   && $+$212.0 &      ...        &       ...       & $0.49 \pm 0.36$ & LMC wind \\

BI\,214    & 0.52 & $+$85.7  & $-0.09 \pm 0.12$ & $0.43 \pm 0.08$ & $>0.17$ & HVC/Wind\\
          && $+$118.1 & $0.07 \pm 0.13$   & $0.55 \pm 0.06$ & $>-0.25$ & HVC/Wind\\
          && $+$162.7 & $<0.21$           & ...             & $>-0.39$ & LMC wind \\
          && $+$189.9 &  ...              &    ...          & $>-0.25$ & LMC wind\\
          
Sk$-$69$^\circ$175 & 0.65 & $+$120.0   & $0.24 \pm 0.07$ & $0.60 \pm 0.17$  & $-0.48 \pm 0.18$ & LMC Wind\tablenotemark{P} \\
                   && $+$164.3   & $0.52 \pm 0.23$ & $0.17 \pm 0.27$  & ...  & LMC wind\\
                   && $+$200.9   & $-0.02 \pm 0.11$ & $0.30 \pm 0.07$ & $-0.74 \pm 0.08$ & LMC wind \\

Sk$-$68$^\circ$112 & 0.84 & $+$90.0  & $0.35 \pm 0.12$ & $0.77 \pm 0.07$ & $>-0.25$ & HVC/Wind \\
                   && $+$116.6 & $<0.13$         &   ...           & $>-0.01$ & HVC/Wind \\
                   && $+$166.6 & $0.17 \pm 0.17$ & $0.90 \pm 0.04$ & $-0.81 \pm 0.12$ & LMC wind \\
                   && $+$205.4 & $<-0.21$        & $0.76 \pm 0.06$ & $-0.60 \pm 0.23$ & LMC wind \\

BI\,173 & 1.03 & $+$115.6 & $0.08 \pm 0.09$ & $>-0.04$ & $>0.11$ & MW HVC\tablenotemark{P}\\
       && $+$167.3 &   ...           &       ...        & $-0.76 \pm 0.18$ & LMC wind \\

Sk$-$69$^\circ$104 & 1.76 & $+$94.1 & $0.07 \pm 0.30$ & $0.70 \pm 0.19$ & $>-0.38$ & HVC/Wind \\
                   && $+$123.4 & $0.48 \pm 0.18$ & $0.62 \pm 0.11$ & $>0.12$ & HVC/Wind \\
                   && $+$141.2 & $0.35 \pm 0.23$ & $0.79 \pm 0.08$ & $>-0.03$ & HVC/Wind \\
                   && $+$160.8 & $0.15 \pm 0.39$ & $0.83 \pm 0.09$ & $>-0.41$ & LMC wind \\
                   && $+$183.2 & $0.74 \pm 0.08$ & $0.91 \pm 0.06$ & $>-0.58$ & LMC wind \\
\hline
\end{tabular}
\tablenotetext{P}{Based on the photoionization modelling from Section \ref{subsection:Cloudy}.}
\tablecomments{This table shows the ion ratios for the matching components in a velocity range of $+90\, \kms \lesssim  v_{\rm LSR} < \rm LMC's~disk~boundary$. For most of the components, the $v_{\rm LSR}$ values are provided based on the velocity centroids for O\textsc{~i}. However, when a reliable  O\textsc{~i} measurement is not available, we provide this value based on Si\textsc{~ii}. Sightlines are ordered in the increasing angular offsets from the center of 30~Doradus. The comments in the final column indicate the association of these absorption components with either the MW HVC, the LMC wind, or a combination of HVC and wind when a distinct connection with the MW HVC and/or LMC wind is not evident. For all of the sightlines, the absorption at $v_{\rm LSR} > 150\, \kms$ is most likely associated with the LMC wind based on the discussion in Section~\ref{sec:gas_distribution}. The absorption at $+100\, \kms \lesssim  v_{\rm LSR} \lesssim +150\, \kms$ does not have a clear association with the MW HVC or the LMC wind, except for two components in two sightlines where the photoionization modelling is available.}
\end{table*}

\subsection{High-ions}\label{subsec:high_ions}
\begin{figure*}
  \centering
  \includegraphics[width=0.49\textwidth]{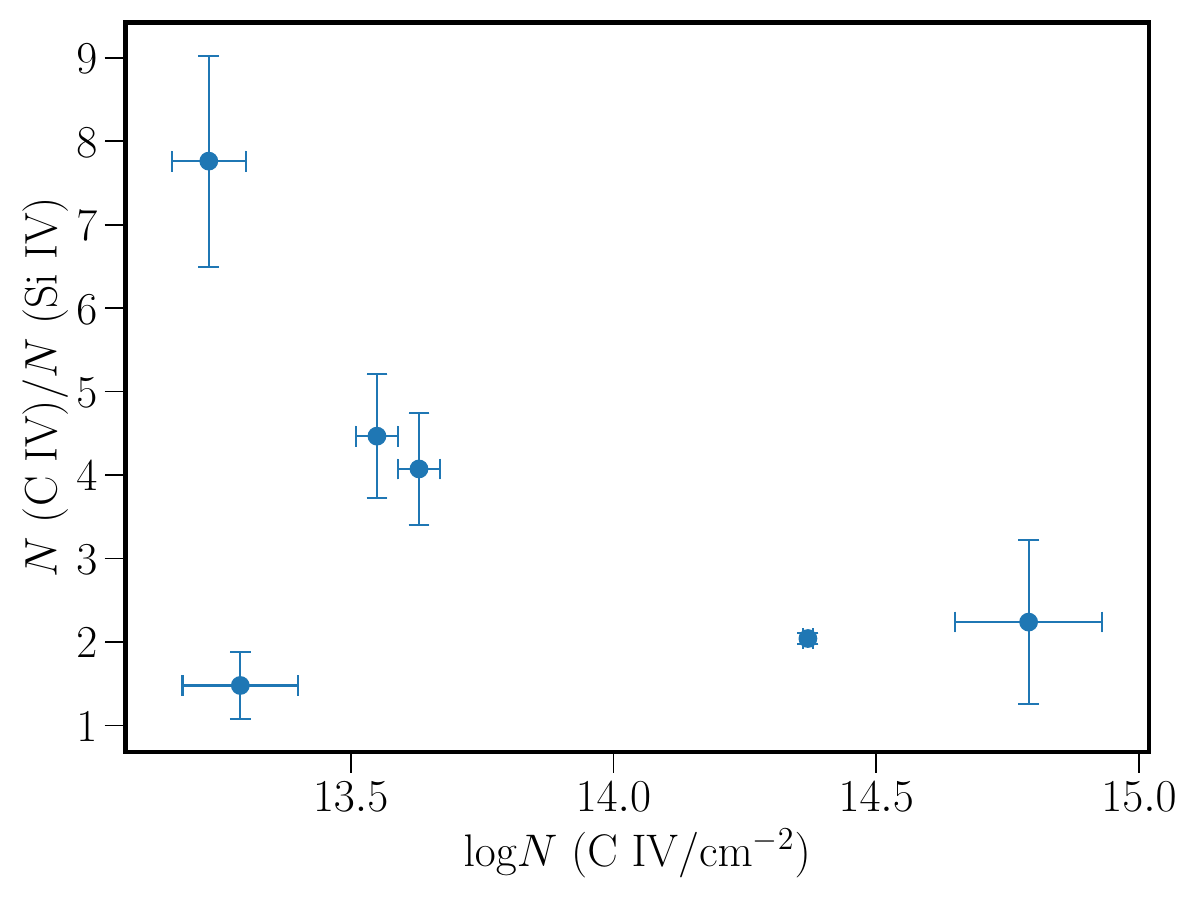}
  \includegraphics[width=0.49\textwidth]{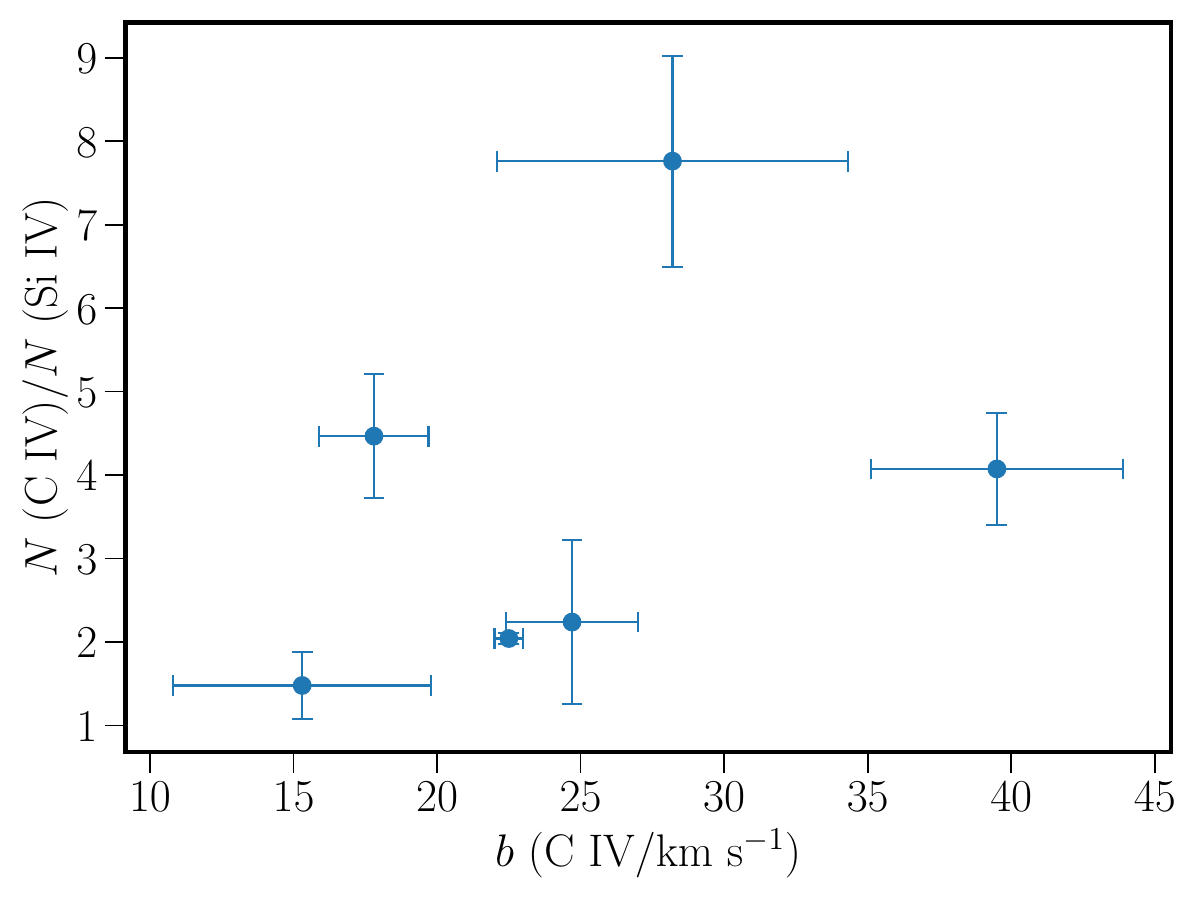}
  \caption{Ratio of C\textsc{~iv} and Si\textsc{~iv} column densities for velocity matching components that fall within $90\, \kms \lesssim v_{\rm LSR} \lesssim 220\, \kms$ a) as a function of logN C\textsc{~iv} (Left) and b) as a function of $b$-values for C\textsc{~iv} (Right).}
  \label{fig:Ratio_high}
\end{figure*}

While the neutral and low-ionization atomic species trace the neutral and photoionized outflows from the LMC, the high-ions, C\textsc{~iv} and Si\textsc{~iv}, may probe both the photoionized and collisionally ionized gas. The theoretical models describing the ionization mechanisms responsible for generating the high-ions present varying predictions for their ratios \citep{1996ApJ...458L..29S}. The column density ratios of C\textsc{~iv} and Si\textsc{~iv} have been widely used to understand the physical and ionization conditions in both the MW and the LMC \citep{2011ApJ...727...46L,2016ApJ...817...91B,2019ApJ...887...89W,2022Natur.609..915K}. To accurately estimate the ratios, it is imperative that the Si\textsc{~iv} and C\textsc{~iv} ions exhibit similar velocity structures and velocity centroids. Across the first three sightlines, two components each have matching counterparts between Si\textsc{~iv} and C\textsc{~iv} within the velocity range of $+90 \lesssim v_{\rm LSR} \lesssim +220\, \kms$. However, the outer sightlines generally lack matching pairs for both ions. We compare the column density ratios for these matching components as a function of C\textsc{~iv} column density and $b_{\rm C\textsc{~iv}}$ (see Figure \ref{fig:Ratio_high}). We observed that the higher values of $N_{\rm C\textsc{~iv}}/N_{\rm Si\textsc{~iv}}$ are typically associated with higher values of $b_{\rm C\textsc{~iv}}$. We estimate that the C\textsc{~iv} and Si\textsc{~iv} ions sharing identical velocities, exhibit the mean and dispersion of $\langle N_{\rm C\textsc{~iv}}/N_{\rm Si\textsc{~iv}}\rangle = 3.7\pm2.3$, with values spanning from 1.5 to 7.8. Our $N_{\rm C\textsc{~iv}}/N_{\rm Si\textsc{~iv}}$ ratios and values of $b_{\rm C\textsc{~iv}}$ do not suggest that photoionization alone is influencing this gas as incident ionizing radiation is anticipated to produce more $N_{\rm Si\textsc{~iv}}$ than $N_{\rm C\textsc{~iv}}$ due to the much lower ionization potential of the silicon ion at $\rm IP_{\rm Si^{+3}}=33.5\,\rm eV$ compared to $\rm IP_{\rm C^{+3}}=47.9\,\rm eV$. Instead, this average ratio indicates a preference for collisional ionization, with turbulent mixing layers (TMLs; \citealt{2015ApJ...812..111K,2018ApJ...866...34S}) which exist in the wind appearing to be the best candidates for the production of these high ions (see \citealt{2011ApJ...727...46L}).

\begin{figure}
    \centering
    \includegraphics[width=\columnwidth]{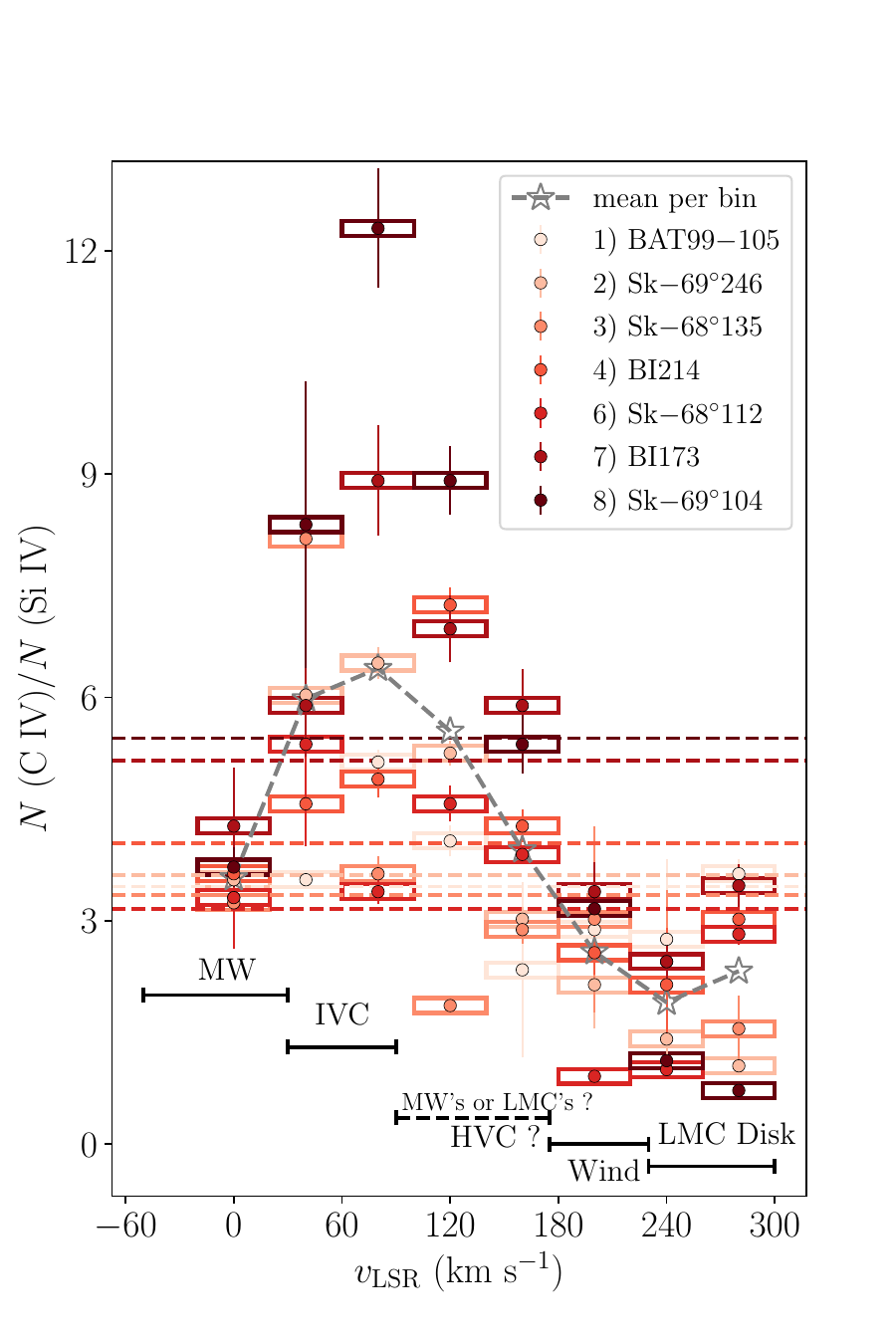}
    \caption{Ratio of C\textsc{~iv} and Si\textsc{~iv} AOD column densities for eight different bins between $v_{\rm LSR}=-20$ to $+300\, \kms$, with each bin size equal to $40\, \kms$ that is indicated by the span of the horizontal bars. The mean values for these ratios for each sightline over the entire sampled velocity range are represented with a horizontal dotted line. Darker red colors represent sightlines farther from the center of 30~Doradus, while lighter shades correspond to closer sightlines. Additionally, we plot the mean values across all sightlines for each velocity bin, represented by gray stars, and connect these points with dashed gray line. Similar to Figure~\ref{fig:NvsV}, the different horizontal solid lines in black mark the expected velocity ranges for the MW, IVC, Wind, and the LMC disk regions and the dashed black line marks the velocity range which may be associated with the either the MW's HVCs or the LMC wind.}
    \label{fig:AOD_high}
\end{figure}

We also analyzed the line ratios for all sightlines that have well behaved continuum (7 out of 8) over velocities that span from the MW to the LMC ($v_{\rm LSR}=-20$ to $+300\,\kms$) by estimating the AOD column density for Si\textsc{~iv} and C\textsc{~iv} in discrete $40\,\kms$ bins (see Figure~\ref{fig:AOD_high}). While these ratios vary among different sightlines, we generally observe that they are larger for the MW region than for the LMC. To compare them further, we plotted the mean values for these ratios for each sightline with a dotted line. For many of the sightlines, the ratios for the bins located at $v_{\rm LSR}>+150$ $\kms$ are smaller than the mean value, possibly suggesting their association with the LMC. In \citet{2007MNRAS.377..687L}, they observed that the line ratios of O\textsc{~VI} and C\textsc{~iv} vary from sightline to sightline. However, they also found that the MW and IVC components had similar ratios and that the HVC and LMC components ratios were similar, suggesting two different associations. If we compare our global mean $N_{C\textsc{~iv}}/N_{Si\textsc{~iv}}$ ratios across all sightlines for each velocity bin, the ratio reaches its maximum at the bin center of $v_{\rm LSR}=+80$ $\kms$ and then decreases toward the LMC, suggesting a transition region somewhere between $+80 \lesssim v_{\rm LSR} \lesssim +120\, \kms$. While the physical meaning of this trend is unclear, it might indicate a transition similar to what we observed in Section~\ref{subsection:MWvsLMC} for the column density versus velocity of Fe\textsc{~II}.

Our photoionization models (which we discuss in more detail in Section~\ref{subsection:Cloudy}) for the $v_{\rm LSR}=+115\, \kms$ component along slightline BI\,173 and $v_{\rm LSR}=+120\, \kms$ component along sightline Sk$-$69$^\circ$175 also predict the column densities of Si\textsc{~iv} and C\textsc{~iv} in the single-phase photoionized gas in which the low and high ions are well mixed. For the Sk$-$69$^\circ$175 sightline, the modelled photoionized column densities are \(\log (N_{\rm Si\textsc{~iv}}/\text{cm}^{-2}) = 14.24\) and \(\log (N_{\rm C\textsc{~iv}}/\text{cm}^{-2}) = 14.77\). For this component, the model predicts that the dominant phases for Si and C are in the form of Si\textsc{~iii} and C\textsc{~iii}. However, we lack the observed column densities of Si\textsc{~iv} and C\textsc{~iv} for this sightline to compare with the predictions. For the BI\,173 sightline, the predicted photoionized Si\textsc{~iv} and C\textsc{~iv} column densities are extremely small, at \(\log (N_{\rm Si\textsc{~iv}}/\text{cm}^{-2}) = 8.86\) and \(\log (N_{\rm C\textsc{~iv}}/\text{cm}^{-2}) = 7.37\). The much larger observed value of \(\log (N_{\rm Si\textsc{~iv}}/\text{cm}^{-2}) = 13.02\) for a component at a similar velocity suggests that photoionization alone cannot account for such a high column density for Si\textsc{~iv} and therefore, requires a higher contribution from the non-thermal motion of the gas. From the average Doppler parameter values for the ions Si\textsc{~iv} and C\textsc{~iv}, we estimate that the gas probed by these ions can reflect temperatures of $T<8.5 \times 10^5\,\rm{K}$ and $<6.9 \times 10^5 \, \rm{K}$, respectively, suggesting TMLs being the most likely candidates of these high ions. However, these ions may also be indicative of probing the corona of the LMC or an interface region where the wind feeds into the LMC's halo.

\subsection{Photoionization Modeling}\label{subsection:Cloudy}
While the observed ion ratios in Section~\ref{subsec:low_ion} provide some insight into the ionization conditions, we further assess the physical and ionization properties through photoionization modeling. The neutral- and low-ionization species, i.e., O\textsc{~i}, Si\textsc{~ii}, S\textsc{~ii}, Al\textsc{~ii}, and Fe\textsc{~ii}, with narrow component structure should trace the cool photoionized phase of the LMC's outflow. 
We observe multiple components of neutral, low, and intermediate ions in each of our sightlines between the MW and the LMC; however, the location of these components is not known for certain. 
Components observed at $ v_\mathrm{LSR}\geq +100\,\kms$ can be associated with MW halo HVCs as well as with the LMC wind. Given the discussion in Sections~\ref{subsection:wind_disc}, \ref{subsection:HVC_wind}, and \ref{subsection:MWvsLMC} above and that we are looking in the direction of an active star-forming region, we explore if blue-shifted absorbers that are kinematically offset from the LMC's \hi\ disk by more than $100\,\kms$ could be part of a high velocity outflow from the LMC. For example, this would include the $\vlsr\approx+120\,\kms$ ($v_\mathrm{LMCSR}\approx-152\,\kms$) components along the Sk$-$69$^\circ$175 sightline. 

To assess the role of photoionization in these components, as well as to characterize their physical conditions and element abundances, we include multidimensional photoionization models using the radiative transfer code \textsc{Cloudy} (version 22.01, \citealt{chatzikos2023}) in our analysis. These models aim to 1) provide measurements of unseen ion stages in order to determine an elemental abundance, 2) quantify the amount of dust depletion, and 3) provide a constraint on the size and location of the cloud. All models assume that the outflow component is a plane-parallel slab of uniform density exposed to radiation fields from MW's disk, Magellanic Clouds, and extragalactic background. We did not use the direct radiation field from the 30~Doradus star-forming region as exposing these fast-moving clouds with a velocity of $\vlsr\approx+120\,\kms$ to such field does not reproduce the observed \hi\ and O\textsc{~i} column densities as all of the oxygen and hydrogen will be ionized under such exposure. The MW disk-halo model that we used is a nonisotropic composite hard and soft radiation field composed from the escaping UV flux of O--B stars \citep{1999ApJ...510L..33B, 2005ApJ...630..332F,2014ApJ...787..147F, bland2019}. The Magellanic radiation fields from \citet{2013ApJ...771..132B} are rotated versions of the Milky Way's model that have been re-scaled so they match the observation of \ha\ emission towards the Magellanic Bridge. We adopt the extragalactic UV background from \citet{khaire2019}. With this co-added and 3D reconstructed radiation field, we were able to interpolate its strength at any distance between the Milky Way and LMC\footnote{Milky Way Radiation Field and Ionization Photon Flux Tool, which includes contributions from the Magellanic Clouds: \href{https://github.com/Deech08/galrad}{https://github.com/Deech08/galrad}}. 

As we do not know the precise position of the gas along the line of sight, we run our models over a large grid of flux values that correspond to different physical positions between the Milky Way and LMC. Following the presentation of the disk-halo model in \citet{bland2019}, an absorber along the line of sight to the LMC could be exposed to an escaping hydrogen ionizing flux in the range of $4.65\lesssim\log{\left(\Phi_{\rm H}/{\rm photons\,\cm^{-2}\,\s^{-1}}\right)}\lesssim7.05$. We therefore ran all of our \textsc{Cloudy} photoionization models over this flux range in step sizes of 0.15 dex to match the contours of the model (see \citealt{bland2019}). Thus, this modeling approach will ascertain what range of flux values is permitted by the observations to help narrow down the physical location of the cloud components along the line of sight. We ran each of these models over a range of hydrogen number densities from $-3\le \log{\left(n_{\rm{H}}/\cm^{-3}\right)} \le 0$ in steps of $0.1\,\rm dex$.  

We are only able to perform photoionization modeling for high-velocity components toward two LMC stars, Sk$-$69$^\circ$175 and BI\,173, as they have suitable H\textsc{~i} emission data, which serves as the model stopping criteria since \ion{H}{1} should be co-spatial with our most robust metallicity tracers \ion{O}{1} and \ion{S}{2}. For our purposes, suitable H\textsc{~i} data are detections of at least $3\sigma$ that are similar velocities to the observed neutral and low ion species. Details and results for each sightline are described in the following subsections. 

\subsubsection{The $+$115 km s$^{-1}$ component toward BI\,173}\label{subsubsec:cloudy_bi173}

We observe an absorption component of O\textsc{~i} at $\vlsr=+115.6\,\kms$ in the STIS UV spectrum toward BI\,173 that kinematically overlaps with the H\textsc{~i} 21-cm emission centered at $\vlsr=+117.9\,\kms$ (see Figure~\ref{fig:plotstacks}). Corresponding components for Si\textsc{~ii}, Fe\textsc{~ii}, and Al\textsc{~ii} are also observed within the velocity range of $+118\le\vlsr\le+124\,\kms$. However, this component does not have an adjacent ion ratio to constrain the ionization parameter, therefore we designed a \textsc{Cloudy} optimization model to find model parameters that are consistent with the observed column densities in this component (see Table~\ref{tab:Voigt_results_cont}). 

We use \textsc{Cloudy}'s built-in \texttt{optimize} command with two free parameters: the hydrogen-ionizing photon flux $\Phi_{\rm H}$ and the total hydrogen number density $n_{\rm H}$. As described above, these initial broad range requirements ensure that \textsc{Cloudy} does not settle at a local minima during the optimization process. The neutral hydrogen column density measured in this component ($\log{\left( N_{\rm H\textsc{~i}}/\cm^{-2}\right)} = 19.01 \pm 0.10$ from the GASS survey) serves as our stopping criterion. Because H\textsc{~i} and O\textsc{~i} are known in this component, we use the $[{\rm O\textsc{~i}}/{\rm H\textsc{~i}}] = -0.69\pm0.13\,\rm dex$ abundance as an additional input parameter. We want to emphasize that this value of $[{\rm O\textsc{~i}}/{\rm H\textsc{~i}}]$ is in excellent agreement with the value estimated by \citet{2009ApJ...702..940L}. To fine tune the optimization model even further, we permit this metallicity to vary within a margin of error of $\pm 0.15\,\rm dex$ in $0.05\,\rm dex$ increments. 
After the optimization model returns the optimal $\log{\Phi_{\rm H}}$, $\log{\left(n_{\rm H}\right)}$, and metallicity, we run a final \textsc{Cloudy} model at those parameters to determine the best-matched model predicted ion column densities, ionization corrections, and potential dust depletion. 

The results of the optimization model describe a cloud that is predominately neutral, with $\log{\left( N_{\rm H\textsc{~ii}}/\cm^{-2} \right)}= 18.31$ and hydrogen ionization fraction of  $\chi_{\rm H\textsc{~ii}}= 0.17$. We find an optimal hydrogen number density and hydrogen-ionizing photon flux of $\log{\left( n_{\rm H}/\cm^{-3} \right)}= -0.29$ and $\log{\left( \Phi_{\rm H}/{\rm photons\,\cm^{-2}\,\s^{-1}} \right)}= 5.16$. The characteristic thickness of the absorbing neutral gas layer is $6.5\,\rm pc$, calculated from the hydrogen number density and the neutral gas column density, i.e., $r = N_{\rm H\textsc{~i}}/n_{\rm H}$. Thus, the model predicts a small, mostly neutral cloudlet with $\log{\left( T/\rm K \right)}\approx 3.8$. Since this absorber lies somewhere along the line of sight between the star BI\,173 in the LMC and the Sun, exposure to this ionizing photon flux places the receding cloud at a radial distance of either $7.5\,\kpc$ from the center of the LMC or $4.3\,\kpc$ from our position in the disk of the MW. 

\begin{figure}
    \centering
    \includegraphics[width=\columnwidth]{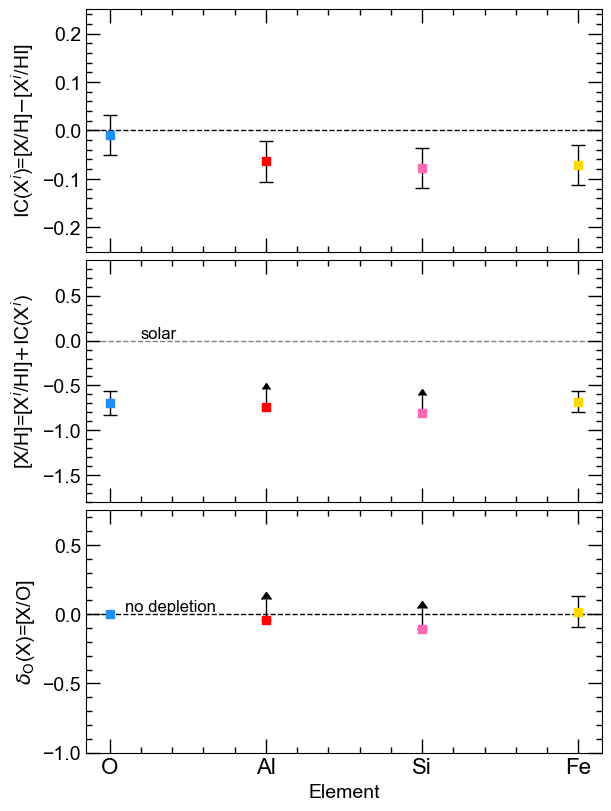}
    \caption{Photoionization modeling results for the $ v_\mathrm{LSR} \approx +115\,\kms$ component toward BI\,173 with optimal hydrogen number density $\log{\left(n_{\rm H}/\cm^{-3}\right)} = -0.29$ and hydrogen ionizing photon flux of $\log{\left(\Phi_{\rm H}/{\rm photons\,\cm^{-2}\,\s^{-1}}\right)} = 5.16$. Top panel: ionization corrections for the low ions O\textsc{~i}, Al\textsc{~ii}, Si\textsc{~ii}, and Fe\textsc{~ii}. The dashed gray horizontal line at zero indicates no ionization correction. Middle panel: comparison of the elemental abundances after correcting the ionization. The dashed gray horizontal line at zero marks the solar abundance.  Bottom: comparison of the levels of depletion among different elements $\delta_\mathrm{O}$(X) = [X/O]. The dashed gray horizontal line at zero indicates no depletion relative to oxygen.}
    \label{fig:BI173_cloudy}
\end{figure}

Because the ionic column density we observe may not be the dominant ion stage present in the gas, we utilize the model results to calculate ionization corrections for the observed ion stages. The ionization correction can be expressed as a difference between the true and observed ion abundance, i.e.,
\begin{equation}
    \mathrm{IC}(\mathrm{X}^i) = [\mathrm{X/H}] - [\mathrm{X}^i/\mathrm{H\;I}].
\end{equation}
We find a low ionization correction of IC(O) = $-$0.01, which results in a low gas-phase oxygen abundance of $[{\rm O}/{\rm H}] = -0.70 \pm 0.13\,\rm dex$. We calculate ionization-corrected gas-phase abundances of $[{\rm Al}/{\rm H}] \geq -0.74$, $[{\rm Si}/{\rm H}] \geq -0.81$, and $[{\rm Fe}/{\rm H}] = -0.68 \pm 0.12\,\rm dex$ (see Table \ref{tab:cloudy_bi173} and the top and middle panels of Figure~\ref{fig:BI173_cloudy}). The error bars on the ionization corrections in the top panel represent the highest and lowest possible correction to the ion abundance if the model column density (i.e., O\textsc{~i}, Al\textsc{~ii}, Si\textsc{~ii}, or Fe\textsc{~ii}) were 10\% higher or lower. We include this range to demonstrate the low sensitivity this change has on the ionization correction. This range of sensitivity is also propagated through to the calculated element abundance and the associated error. 

We check for dust depletion effects in this component following the convention that the depletion, $\delta_\mathrm{O}$(X), of a refractory element X is measured as a difference between ionization-corrected abundances of X and oxygen:
\begin{equation}
    \delta_\mathrm{O}(\mathrm{X}) \equiv [\mathrm{X/H}] - [\mathrm{O/H}].
\end{equation}
A negative value for $\delta_\mathrm{O}$(X) indicates depletion of X with respect to the volatile element oxygen, under the assumption that the total (gas$+$dust) abundances are solar. The error for the depletion depends only on the observed ion column density errors and the IC sensitivity values for the refractory and volatile elements. Overall, we find a lower limit for the  depletion level for both silicon and aluminum of $\delta_\mathrm{O}({\rm Si}) \geq -0.11$ and $\delta_\mathrm{O}({\rm Al}) \geq -0.04,\rm dex$, respectively. We find $\delta_\mathrm{O}({\rm Fe}) = -0.02 \pm 0.11,\rm dex$ (see Table \ref{tab:cloudy_bi173} and the bottom panel of Figure \ref{fig:BI173_cloudy}). The dust depletion measurements suggest there is no dust in this cloud. Overall, these physical characteristics (i.e., mostly neutral, cool, low-metallicity, dust-free) suggest the cloud seen near +115 \kms\ is a MW halo HVC. 

\begin{table}
\centering
\caption{Summary of photoionization modeling results for $ v_\mathrm{LSR} \approx +115\,\kms$ component toward BI\,173}
\label{tab:cloudy_bi173}
\begin{tabular}{cccc}
\hline
\hline
Element (X) & IC(X$^{i}$) & [X/H] & $\delta_\mathrm{O}$(X)\\
\hline
O  & $-0.01\pm0.04$ & $-0.70\pm0.13$ & ...  \\
Al & $-0.06\pm0.04$ & $\geq -0.74$   & $\geq -0.04$ \\
Si & $-0.08\pm0.04$ & $\geq -0.81$   & $\geq -0.11$ \\
Fe & $-0.07\pm0.04$ & $-0.68\pm0.12$ & $-0.02\pm0.11$ \\
\hline
\end{tabular}
\end{table}

\subsubsection{The $+$120 km s$^{-1}$ component toward Sk$-$69$^\circ$175}\label{subsubsec:cloudy_sk69}
Along the Sk$-$69$^\circ$175 sightline, we observe an absorption component of O\textsc{~i} that is centered at $\vlsr=+120.0\,\kms$ in the FUSE FUV spectrum that overlaps with the H\textsc{~i} 21-cm emission line at $\vlsr = +116.4\,\kms$ (see Figure~\ref{fig:plotstacks}). \citet{2009ApJ...702..940L} measured  $\log{\left(N_{\rm H\textsc{~i}}/\cm^{-2}\right)} = 19.18 \pm 0.10$ for this component using Parkes radio 21\, cm data from the LMC H\textsc{~i} survey \citep{2003MNRAS.339...87S}.
We also observe multiple singly ionized species Al\textsc{~ii}, Si\textsc{~ii}, S\textsc{~ii}, Fe\textsc{~ii}, and Ni\textsc{~ii} that overlap and with centroids within the velocity range of $+108\le\vlsr\le+116\,\kms$. We measure a lower limit of $\log{\left(N_{\rm Si\textsc{~iii}}/N_{\rm Si\textsc{~ii}}\right)} \geq -1.16\,\rm dex$ (Si\textsc{~iii} is saturated) for this absorber, which we can use to bound the ionization parameter. 

We designed a \textsc{Cloudy} optimization model to find the optimal parameters that explain the observed column densities for this absorber (see Table~\ref{tab:Voigt_results_cont}). We once again use \textsc{Cloudy}'s built-in \texttt{optimize} command with free parameters $\Phi_{\rm H}$ and $n_{\rm H}$, where we used the neutral hydrogen column density measured in this component of $\log{\left(N_{\rm H\textsc{~i}}/\cm^{-2}\right)} = 19.18\,\rm dex$ as a model stopping criterion. 
The model optimizes over the range of flux values from $4.65 \leq \log{\left( \Phi_{\rm H}/\rm{cm}^{-2}\rm{s}^{-1} \right)}\leq 7.05$ in steps of 0.15\,dex and number density $-3.0 \leq \log{\left(n_{\rm H}/\cm^{-3}\right)} \leq 0$ in steps of 0.10\,dex, as we describe above in Section~\ref{subsubsec:cloudy_bi173}. 

\begin{figure}
    \centering
    \includegraphics[width=\columnwidth]{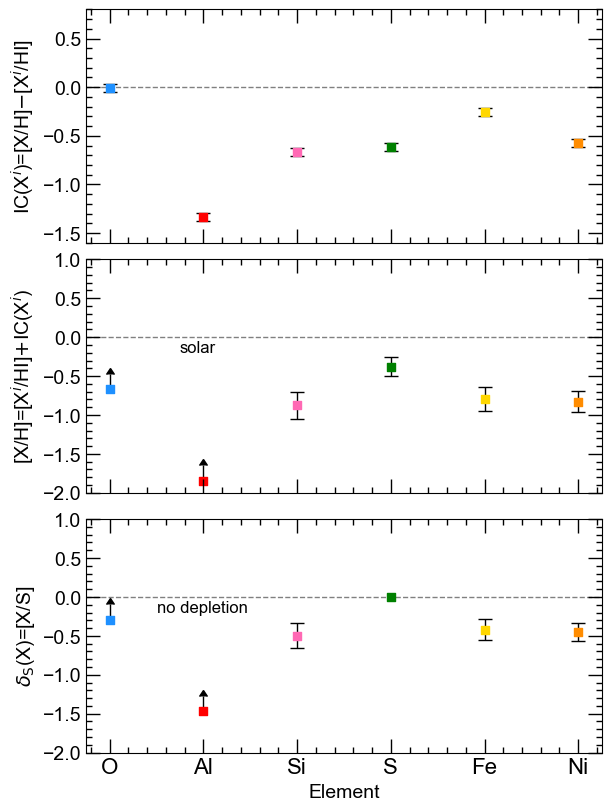}
    \caption{Photoionization modeling results for the $\vlsr\approx+120\,\kms$ component toward Sk$-$69$^\circ$175 with optimal hydrogen number density $\log{\left(n_{\rm H}/\cm^{-3}\right)} = -2.13$ and hydrogen ionizing photon flux of $\log{\left(\Phi_{\rm H}/{\rm photons\,\cm^{-2}\,\s^{-1}}\right)} = 5.85$. Top panel: ionization corrections for the six low ions (O\textsc{~i}, Al\textsc{~ii}, Si\textsc{~ii}, S\textsc{~ii}, Fe\textsc{~ii}, and Ni\textsc{~ii}). Middle panel: comparison of the ionization-corrected elemental abundances. The gray dashed horizontal line marks solar abundance. Bottom panel: comparison of the levels of depletion among different elements $\delta_\mathrm{S}$(X)=[X/S]. The gray horizontal line marks no depletion of the element relative to sulfur, assuming a solar abundance.}
    \label{fig:SK-69d175_cloudy}
\end{figure}

We looked to estimate the metallicity based on O\textsc{~i} as this element depletes very little in warm ionized gas and has a negligible ionization correction for clouds with $\log{\left(N_{\rm H\textsc{~i}}/\cm^{-2}\right)}\geq 19$ due to charge exchange between hydrogen and oxygen \citep{viegas1995}, which is consistent with what we found in our photoionization models (see Figures~\ref{fig:BI173_cloudy} and~\ref{fig:SK-69d175_cloudy}). However, given the FUSE resolution of $20\,\kms$, we cannot rule out that the O\textsc{~i} 1039 {\AA} component at $\vlsr = +120\,\kms$ may be saturated, therefore we give a limit on [O\textsc{~i}/H\textsc{~i}] $\geq$ $-0.67$. Additionally, adopting the O\textsc{~i} column density as it is from the profile fitting results in nonphysical outcomes from the photoionization modeling. Fortunately, we have a good unsaturated measurement for S\textsc{~ii} which yields a sulfur ion abundance of [S\textsc{~ii}/H\textsc{~i}] = $+0.24\pm0.11$. Sulfur is also known to deplete very little onto dust, but does require an ionization correction to the ion abundance to account for the amount of sulfur present in other ion stages. \citet{collins2003} conducted a study to determine the range of ionization corrections for S\textsc{~ii} as a function of $\log{N_{\rm H\textsc{~i}}}$ in different density regimes, where the highest and lowest ionization corrections corresponding to our \hi\ column densities range between $-0.7\lesssim {\rm IC}_{\rm S\textsc{~ii}}\lesssim-0.3$, respectively. 
This range of correction values could result in an elemental abundance for [S/H] from approximately $-$0.5 to nearly solar, therefore we permitted the optimize model to vary over a larger range of metallicity values to fully explore the effect of metallicity on the model column densities. We chose a range from $-0.76$ to $+0.06$ in steps of $0.10\,\rm dex$, which conservatively contains within it both the observed [O\textsc{~i}/H\textsc{~i}] lower limit and the possibility of a solar abundance.

The results of the optimization model describe a mostly ionized cloud with a hydrogen ionization fraction of $\chi_{\rm H\textsc{~ii}} = 0.97$ and $\log{\left(T/\rm K\right)} \approx 4$. The optimal hydrogen number density and hydrogen-ionizing photon flux are $\log{\left(n_{\rm H}/\cm^{-3}\right)} = -2.13$ and $\log{\left(\Phi_{\rm H}/{\rm photons\,\cm^{-2}\,\s^{-1}}\right)} = 5.85$, respectively. The characteristic thickness of the absorbing neutral gas layer is 0.67\,kpc, calculated from the hydrogen number density and the neutral gas column density, i.e., $r = N_{\rm H\textsc{~i}}/n_{\rm H}$. Since this component is seen somewhere along the line of sight to Sk$-$69$^\circ$175 in the LMC, exposure to this ionizing photon flux places the receding cloud at a radial distance of either 3.6\,kpc from the center of the LMC or 6.6\,kpc from our position in the disk of the MW.

We find an ionization correction of $\rm IC(S) = -0.62\,dex$ and a gas-phase element abundance of [S/H] = $-$0.38 $\pm$ 0.12. This is consistent with the lower limit to the gas-phase oxygen element abundance [O/H] $\geq$ $-$0.67. We calculate ionization-corrected gas-phase abundances of [Al/H] $\geq$ $-$1.85, [Si/H] = $-$0.88 $\pm$ 0.18, [Fe/H] = $-$0.80 $\pm$ 0.15, and [Ni/H] = $-$0.83 $\pm$ 0.13 (see Table \ref{tab:cloudy_sk69d175} and the top and middle panels of Figure \ref{fig:SK-69d175_cloudy}). We also checked for UV dust depletion effects in this component using the same procedure described above in Section~\ref{subsubsec:cloudy_bi173}. In this case, we calculate the depletion of the refractory elements relative to the unsaturated volatile element sulfur. For this component, we find evidence of moderate dust depletion for silicon, iron, and nickel of $\delta_\mathrm{S}$(Si) = $-$0.50 $\pm$ 0.16, $\delta_\mathrm{S}$(Fe) = $-$0.42 $\pm$ 0.13, and $\delta_\mathrm{S}$(Ni) = $-$0.45 $\pm$ 0.11.

Overall, these physical characteristics (i.e., mostly photoionized, warm, sub-solar metallicity, moderate dust depletion) suggest the cloud seen near $+120$ \kms\ originates from a galactic disk. We note that the sub-solar element abundance [S/H] = $-0.38\pm0.12$ (42$^{+13}_{-10}$\% solar) is consistent with the current day metallicity of the LMC of 46\% \citep{russell1992}. For the cloud to have originated from the MW Galactic disk, we expect that the cloud should have a solar or above-solar metallicity unless the cloud has undergone significant mixing \citep{2014ApJ...795...99G} over the predicted distance from the MW of 6.6 kpc. 

\begin{table}
\centering
\caption{Summary of photoionization modeling results for $\vlsr\approx+120\,\kms$ component toward Sk$-$69$^\circ$175}
\label{tab:cloudy_sk69d175}
\begin{tabular}{cccc}
\hline
\hline
Element (X) & IC(X$^{i}$) & [X/H] & $\delta_\mathrm{S}$(X)\\
\hline
O  & $-0.01\pm0.04$ & $\geq -0.67$   & $\geq -0.29$  \\
Al & $-1.34\pm0.04$ & $\geq -1.85$   & $\geq -1.47$ \\
Si & $-0.67\pm0.04$ & $-0.88\pm0.18$ & $-0.50\pm0.16$ \\
S  & $-0.62\pm0.04$ & $-0.38\pm0.12$ & ... \\
Fe & $-0.26\pm0.04$ & $-0.80\pm0.15$ & $-0.42\pm0.13$ \\
Ni & $-0.58\pm0.04$ & $-0.83\pm0.13$ & $-0.45\pm0.11$ \\
\hline
\end{tabular}
\end{table}

\section{Galaxy Simulation}
\label{section:simulation}

\begin{figure*}
  \centering
  \includegraphics[width=0.90\textwidth]{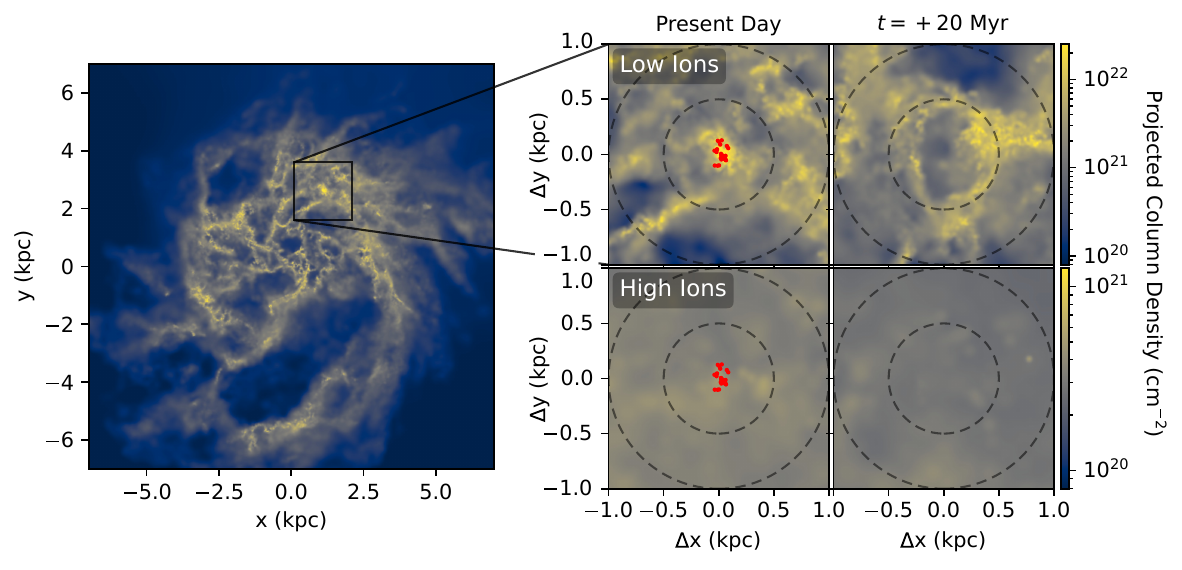}
  \caption{Face-on projection of the simulated LMC highlighting the 30~Doradus analog region. Color bars represent projected column density in cm$^{-2}$. The left panel depicts the entire face of the LMC disk with the four panels on the right showing a zoomed region around the 30~Doradus analog. Of the four panels on the right, the top two panels highlight the distribution of H\textsc{~i} corresponding to the low-ions, and the bottom panels highlight the H\textsc{~ii} corresponding to the high-ions. The left two panels feature the gas at the present day, having formed $1.2\times10^5$~$M_\odot$ of stars within 2 Myr, and the right panels emphasize the appearance of the region at 20 Myr of evolution. A supernova bubble is visible extending $\sim$1~kpc in diameter. The grey dashed lines mark circles with radii of 0.5 and 1~kpc as in Figure~\ref{fig:hi_target_map}.}
\label{fig:sim}
\end{figure*}

To aid in the interpretation of our observations, we incorporate high-resolution hydrodynamic simulations of the LMC, SMC, and MW galaxies. These GIZMO simulations utilized a ``meshless finite-mass" hydrodynamics scheme \citep{gizmo,gadget}. These simulated galaxies consist of live dark matter halos that include coronal gas at their respective virial temperatures. The LMC and SMC have stellar and gaseous disks that are in agreement with observations and follow the orbital history of the MCs. This simulation follows the \citet {springel03} star formation, mechanical stellar feedback, and the \citet{wiersma09} and \citet{hopkins18} metal-dependent heating and cooling prescriptions. We outline the details of these simulations in \citet{2024ApJ...967...16L}.

Our simulation produces a starburst region that is similar to 30~Doradus, but one that has a less active history. This simulated region formed $1.3\times 10^{5}\, M_{\odot}$ of stars in the past 2\,Myr and $2.4\times10^{5}\, M_\odot$ within the past 20\,Myr. We include snapshots of the low and high ion gas distribution of the simulated 30~Doradus analog at the present day and 20\,Myr in the future in Figure \ref{fig:sim}. Due to the high present-day star formation rate, a large ($\sim$1~kpc in diameter) bubble in the ISM forms in 20~Myr as a consequence of the supernovae feedback. The present-day morphology of the low-ionization species in this simulation is similar to that of the \hi\ gas distribution of 30~Doradus (see Figure~\ref{fig:hi_target_map}), where the star-formation activity is being fed by high column density gas that is surrounded by pockets of relatively lower column density gas in which ejected material can more easily traverse.

\begin{figure*}
  \centering
  \includegraphics[width=0.75\textwidth]{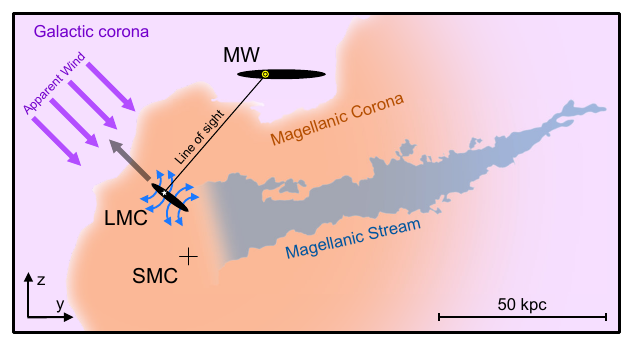}
  \caption{Schematic of the present-day, warped Magellanic Corona (in orange) in the $y-z$ plane and the predicted position of the Magellanic Stream (in blue)  \citep{2020Natur.585..203L,2021ApJ...921L..36L}. The present-day LMC velocity is indicated with a gray arrow and the opposing direction of the apparent wind through the MW's coronal gas (in purple) is marked with purple arrows. The blue arrows that emanate from the LMC's disk represent the LMC's galactic wind, which is encased within the Magellanic corona. The black line that extends between the Sun and a star embedded within the LMC's disk illustrates the line of sights that we are able to probe in this absorption-line study.}
\label{fig:corona}
\end{figure*}

As the LMC and SMC approach the MW, the  LMC's coronal gas and wind interact with MW's halo and they become distorted due to a combination of ram-pressure and tidal forces. We include a schematic of the resultant geometry of the mangled Magellanic Corona \citep{2020Natur.585..203L,2021ApJ...921L..36L} relative to the MW and SMC in Figure~\ref{fig:corona}. The present-day warped Magellanic Corona is shaded orange and the MW's coronal gas is depicted in purple. The arrows in blue emanating from the LMC's disk represent the LMC's galactic wind, which is encapsulated within the Magellanic Corona. While the Magellanic Corona is significantly warped from its initially spherical shape, it still surrounds the Clouds today and may shield the LMC's wind from the ram-pressure effects associated with the MW's gaseous halo. 

\section{Discussion} \label{section:Discussion}
Below, we discuss the origin of the wind in detail and compare our findings with the existing literature on 30~Doradus and the global LMC wind in general. 
As previously detailed in Section~\ref{sec:gas_distribution}, our analysis of the radial distribution of the blueshifted gas forground to the 30~Doradus, with velocities reaching up to $v_{\rm LSR}\approx+150\, \kms$, suggests that this material is primarily associated with the galactic wind driven by 30~Doradus. However, additional evidence---such as the relationship between cloud speed and column density described in Section~\ref{subsection:MWvsLMC}, trends in ion ratios, and insights from photoionization modeling---indicates that some of the absorption within the range $+100 \lesssim v_{\rm LSR} \lesssim +150\, \kms$ may also be attributed to the LMC wind, along with contributions from MW's HVCs. In particular, the highest wind speeds based on our lighter cloud assumption as discussed in Section~\ref{subsection:MWvsLMC} are observed along the first four sightlines, which are located near the most highly ionized central region of 30~Doradus, where the massive star cluster R136 resides (see \citealt{1991IAUS..148..145W,2018A&A...614A.147C} and references therein). Given the extreme conditions in this region, the presence of concentrated, high-velocity outflows along these sightlines is expected.

We estimate the outflow mass and the outflow rate from the low ions for the 30~Doradus region encompassing these inner four sightlines that lie out to a projected radial distance of $R=0.45\,\kpc$ from the center of 30\,Doradus. For this, we utilize the average Si\textsc{~ii} AOD column density of \(\log (N_{\rm Si\textsc{~ii}}/\text{cm}^{-2}) \approx 15.36\) for the sightlines Sk$-$69$^\circ$246 and Sk$-$68$^\circ$135 integrated in the velocity range of $+100\, \kms < v_{\rm LSR} < \rm Disk~Boundary$. Both of these sightlines have Si\textsc{~ii}\,$\lambda1808$ measurements available which offer unsaturated column densities. The total column density measured this way is also consistent with the results from the Voigt profile fitting. The outflowing mass ($M_{\rm outflow}$) is directly proportional to the product of the column density of the ion (N) and the surface area of the outflow (A).

\begin{equation}
M_{\rm outflow} = \mu\, \langle N_{H} \rangle\, A\, \cos i
\end{equation}
where we multiply by a factor of \(\cos i\) to account for the LMC's inclination angle of $i \approx 23.4\arcdeg$ \citep{2022ApJ...927..153C} and we adopt an average particle mass of \(\mu = 1.3 m_{p}\). The surface area of the outflow can be written as \(A = c_{f}\, \pi\,R^{2}\), where we consider a covering fraction of \(c_{f} = 1.0\) for the low ionization species as this region of 30~Doradus remains active across its entire area and we detect the outflows in all the sightlines we studied. Assuming the metallicity of LMC ($Z = 0.5\,Z_{\odot}$) from \citet{1992ApJ...384..508R}, the mass equation for the nearside outflow can be written as follows (see \citealt{2016ApJ...817...91B}):

\begin{equation}
M_{\rm outflow} = \frac{1.3 m_{\rm P} \langle N_{\rm Si\textsc{~ii}} \rangle\, \pi \, R^{2}\, \cos i}{Z/Z_{\odot}\, ({\rm Si/H})_{\odot}}
\end{equation}

We estimate a low-ion probed baryon mass outflow of $M_{\rm baryon}\approx8.6 \times 10^{5} M_{\odot}$ within 0.5 degrees of 30~Doradus. If we restrict the maximum wind speed to $\vlsr=150\, \kms$ instead of $\vlsr=100\, \kms$,  the average Si\textsc{~ii} AOD column density will only decrease by $-0.06\, \rm dex$ to a value of \(\log (N_{\rm Si\textsc{~ii}}/\text{cm}^{-2}) \approx 15.29\), which does not result in any significant reduction in the outflow mass. Based on the discussion in Section~\ref{subsection:wind_disc}, extending the LMC disk width by $\sim10\, \kms$ in the blueshifted direction would reduce the total average column density to \(\log (N_{\rm Si\textsc{~ii}}/\text{cm}^{-2}) \approx 15.18\) and the corresponding outflow mass would be $M_{\rm baryon}\approx5.7 \times 10^{5} M_{\odot}$. Comparing with the outflow mass estimated by \citet{2016ApJ...817...91B}, 30~Doradus alone is found to be contributing a significant fraction of the total outflowing low ions from the entire LMC.

Considering the oldest star cluster, the age of 30~Doradus has been determined to be between 26.5 and 31.5\,Myr old \citep{2016ApJ...833..154C}. Studies also indicate that the rate of star formation in the 30~Doradus region surpassed the average rate in the LMC around 20\,Myr ago \citep{2015ApJ...811...76C}. Assuming an average line-of-sight speed of $100\le\langle|v|\rangle \le150\, \kms$, the wind originating from 30~Doradus could have traveled a distance $2-4.6\,\kpc$ over the time period of $20-30\,\rm Myr$. Remarkably, this distance range aligns with one of the solutions obtained from photoionization modeling for a component at $v_{\text{LSR}} = +120\, \kms$ toward Sk$-$69$^\circ$175. This particular component is found to be at a radial distance of $3.6\, \kpc$ from the LMC, consistent with the observation. Moreover, assuming a maximum outflow timescale of t$_{\rm outflow} = 31.5\, \rm Myr$ based on the oldest star cluster in the 30~Doradus \citep{2016ApJ...833..154C}, the outflow rate would be $\dot{M}_{\rm outflow}\gtrsim~0.02 M_{\odot} \, \text{yr}^{-1}$. Comparing our measurement for the outflow mass rate with the star formation rate (SFR) of 30~Doradus region ($\rm SFR = 0.18\, M_{\odot}\, yr^{-1}$; \citealt{2023ApJ...944...26N}), we estimate a mass loading factor of $\eta \gtrsim 0.10$. Additionally, we also estimate the outflow rate using the cloud velocity and column density relation from \citet{2002ASPC..254..292H} (see Appendix \ref{section:N_v}) and based on our previous assumption of the lighter cloud hypothesis discussed in Section~\ref{subsection:MWvsLMC}. Assuming a 30~Doradus starburst radius of $0.15\, \kpc$ \citep{2011ApJ...731...91L} and an LMC metallicity of $Z = 0.5\,Z_{\odot}$, we calculate an outflow rate of $\dot{M}_{\rm outflow}=0.026\pm0.003 M_{\odot} \, \text{yr}^{-1}$, consistent with the outflow rate obtained from the conservative approach discussed above.

Our findings are in alignment with the results reported by \citet{Ciampa2020} based on \ha \, emission map of the LMC. They observed that the 30~Doradus region generates winds with significantly higher speeds, surpassing a velocity of $ v_{\rm LMCSR} > -175\, \kms$, in contrast to the more general galactic winds, which exhibit velocities around $ v_{\rm LMCSR} < -110\, \kms$. Additionally, \citet{Ciampa2020} noted an asymmetry in the morphology of the \ha \, emission associated with the global wind, where the wind is notably more concentrated in the quadrant of the galaxy that encompasses 30~Doradus. Based on high-resolution emission data of [O\textsc{~iii}] from the 30~Doradus region, \citet{2003MNRAS.344..741R} also reported the wind velocity that closely matched the velocities observed in the fast-moving components in our work.

Previously reported in the study by \citet{2009ApJ...702..940L}, there is evidence of higher concentrations of HVCs  in the vicinity of the 30~Doradus region based on the gradient in the average velocities of HVCs observed across the entire expanse of the LMC. Similarly, \citet{1999AJ....118.2797K} suggested a higher density of supergiant shells closer to the 30~Doradus region. These high-velocity clouds near 30~Doradus may be linked with H\textsc{~i} voids that have been generated by explosive stellar events, including supernova explosions. Indeed, \citet{2003MNRAS.339...87S} found a connection between the formation of H\textsc{~i} voids in the LMC's disk and supergiant shells (see their Figure 10 and 11) as a consequence of such explosive events.

We also compare the ionization properties of the winds between the 30~Doradus region and the entire LMC by examining the average [Si\textsc{~ii}/O\textsc{~i}]. To facilitate the comparison, we contrasted our values with the results presented in \citet{2009ApJ...702..940L}, which specifically investigated the HVCs directed toward the LMC within the velocity range of $+90 \lesssim  v_{\rm LSR} \lesssim +175\, \kms$. Although we find wind speeds reaching up to $\vlsr\approx +215\, \kms$, we focus solely on the wind components in the range of $+90 \lesssim  v_{\rm LSR} \lesssim +175\, \kms$ to match with the observations in \citet{2009ApJ...702..940L} for the comparison. Our findings for the 30~Doradus region reveal an average [Si\textsc{~ii}/O\textsc{~i}] ratio of 0.69\,dex with a dispersion of 0.29\,dex, compared to the average ratio of 0.48\,dex with a similar dispersion reported by \citet{2009ApJ...702..940L}. This relatively higher value observed in the 30~Doradus region is in line with expectations, as it is a result of the elevated radiation levels arising from the intense stellar activity in the area. In contrast, the average value for the entire LMC represents a mixture of active and passive regions, which naturally yields a lower average [Si\textsc{~ii}/O\textsc{~i}] ratio.

While there is substantial evidence supporting the notion that the fast-moving winds originate from the intense stellar activities within the 30~Doradus region, it is important to explore the other possibilities, such as the potential influence of the thick disk's motion and the gas dynamics resulting from tidal interactions. Typically, a galaxy's thick disk plays a dominant role in the overall rotational dynamics of the galaxy and may exhibit some lag compared to the less dense thin disk \citep{2007AJ....134.1019O,2007ApJ...663..933H,2007A&A...468..951K}. However, given the LMC's nearly face-on orientation ($i\approx23.4\arcdeg$, \citealt{2022ApJ...927..153C}), we may not be sensitive to absorption resulting from the thick disk's lag, as such effects are typically more pronounced in edge-on galaxies (see \citealt{2016ApJ...817...91B} for details). Another plausible source of influence is the gravitational interactions between the Magellanic Clouds, which have given rise to various structures, including the Leading Arm, Magellanic Bridge, and Magellanic Stream \citep{2013ApJ...771..132B,2014ApJ...787..147F}. While the Leading Arm and Magellanic Bridge are situated relatively farther away from the 30~Doradus region, the proximity of the Magellanic Stream to this area warrants consideration. \citet{2008ApJ...679..432N} traced one of the two filamentary structures in the Magellanic Stream back to the  quadrant of the LMC that contains 30\,Doradus.  \citet{2013ApJ...772..111R} later found that this filament also has a  metallicity consistent with an LMC origin. Nonetheless, our choice of sightlines on the opposite side of 30~Doradus to where that filament passes was deliberate with intentions of avoiding material from this structure.

Considering that some studies propose the association of HVCs toward the LMC at a velocity of $+90 \lesssim  v_{\rm LSR} \lesssim +150\, \kms$ with the Milky Way halo, rather than the LMC, as discussed in Section~\ref{subsection:HVC_wind}  \citep{1981ApJ...243..460S, 1990A&A...233..523D, 1999Natur.402..386R, 2015A&A...584L...6R}, it is crucial to explore these possibilities in the context of the 30~Doradus region. In particular, \citet{2015A&A...584L...6R} indicates that high-velocity absorption is identified in the spectrum of a Milky Way halo star located at $d_{\odot}\lesssim13.3\,\kpc$ away or within $5.6\,\kpc$ from the Galactic disk, strongly indicating an association with the Milky Way rather than the LMC for this absorber in this direction. However, our sightlines are specifically positioned either in the immediate core of the 30~Doradus or in its close vicinity. The background halo star used by \citet{2015A&A...584L...6R} lies in the opposite side of the 30~Doradus and outside the H\textsc{~i} contour of the LMC disk based on the H\textsc{~i} emission channel maps from \citet{Ciampa2020}; the fast moving absorber along this sightline may not represent the outflows from the LMC and could instead be associated with an HVC that lies in the foreground of the LMC. 

We further explore the origin and properties of the absorption at $+100 \lesssim v_{\rm LSR} \lesssim +150\, \kms$ using photoionization modeling available for two high-velocity components along two sightlines located at an offset angle $\gtrsim$ 0.65 degrees from the center of 30~Doradus. The photoionization modeling for the $v_{\rm LSR} = +120\, \kms$ absorber toward Sk$-$69$^\circ$175 and the $v_{\rm LSR} = +115\, \kms$ absorber towards BI\,173 provides some insights into the dust depletion and the metallicity, which may be traced back to a specific origin of these absorbers. For the component towards BI\,173, we find evidence of no dust depletion, mostly neutral, cool, and a lower metallicity of about ${-}0.7\,\rm dex$ (see Section \ref{subsubsec:cloudy_bi173}). For the component towards Sk$-$69$^\circ$175, we find evidence of moderate dust depletion and a sub-solar metallicity of [S/H] = $-0.38\pm0.12$ (see Section \ref{subsubsec:cloudy_sk69}), which is consistent with the current-day metallicity of the LMC \citep{russell1992}. 

However, there are several detections of dust depletion in MW HVCs too, typically with low to moderate levels. The $\delta(\rm Si)=0.50\pm0.16$, $\delta(\rm Fe)=0.42\pm0.13$, and $\delta(\rm Ni)=0.45\pm0.11\,\rm dex$ depletion patterns that we measure for the absorber along the Sk$-$69$^\circ$175 sightline are very similar to what \citet{2023ApJ...946L..48F} found for HVC Complex~C at $\delta(\rm Si)\approx0.29$, $\delta(\rm Fe)\approx0.42$, and $\delta(\rm Al)\approx0.53\,\rm dex$. For an HVC in the inner galaxy, \citet{2023ApJ...944...65C} reported a value of [Fe/O] = $-0.3\pm0.20\,\rm dex$, suggesting moderate dust depletion. While a metallicity of [S/H] = $-0.38\pm0.12$ for the $v_{\rm LSR} = +120\, \kms$ component towards Sk$-$69$^\circ$175 is close to the LMC metallicity, it is also close to several MW HVCs, such as the Smith Cloud of [S/H] = $-0.28\pm0.14\,\rm dex$ \citep{2016ApJ...816L..11F} and Complex~C of [S/H] = $-0.51\pm0.16\,\rm dex$ \citep{2023ApJ...946L..48F}. However, the speed of the cloud at $|v| \approx 145\, \kms$ relative to the LMC, the metallicity that agrees with present day of the LMC, the dust depletion is consistent with a galactic origin, and the alignment with the bursty 30~Doradus region, makes this absorber more likely to be associated with the LMC wind than with the MW. In contrast, the high velocity absorber along the BI\,173 sightline with a lower metallicity and no signs of dust suggest that this material is more likely a MW HVC. Therefore, the high velocity gas at  $+90 \lesssim v_{\rm LSR} \lesssim +150\, \kms$ along the 4~outer sightlines may represent a combination of both foreground MW HVCs and LMC wind.  

As we observe significant outflows from 30~Doradus, questions arise about how these outflows withstand the direct headwind as the LMC orbits through the MW's gaseous halo. Studies have determined that this headwind directly impacts the leading edge of the LMC, resulting in the suppression of outflows on the near side, while the effect is milder on the far side \citep{2020ApJ...893...29B, Ciampa2020}. However, these studies did not consider the existence of the Magellanic corona, which has been recently discovered both in observations \citep{2022Natur.609..915K} and in simulations \citep{2020Natur.585..203L,2021ApJ...921L..36L}. From the galactic simulation, we discussed in Section \ref{section:simulation}, although the Magellanic Corona has deviated significantly from its initially spherical shape, it still envelops the Magellanic Clouds today, providing shielding from the ram-pressure effects of the MW's gaseous halo (see Figure \ref{fig:corona}). Furthermore, recent studies utilizing hydrodynamical simulations of the LMC \citep{2023ApJ...959L..11S} have suggested that the LMC's motion may generate a bow shock ahead of it, which could also help mitigate the effects of the ram pressure (see also \citealt{2024arXiv240204313Z}). \citet{2016ApJ...817...91B} utilizing quasar as a background source to investigate the outflows from the LMC observed symmetrical absorption on both the near and far sides, indicating that nearside outflows are not significantly suppressed. This also supports our speculation that there must be a protective mechanism against ram-pressure affecting these outflows. Since the LMC might be on its first infall to the MW \citep{2007ApJ...668..949B} and has a high velocity \citep{2013ApJ...764..161K}, both the bow shock and Magellanic Corona play an important role in protecting the LMC's disk from the MW's circumgalactic gas.

\section{Summary} \label{section:Summary}
We investigate the absorption features along eight sightlines located within 1.7~degrees of 30~Doradus. We intentionally chose sightlines residing on one side of 30~Doradus to avoid potential interference from Magellanic tidal structures, such as the Magellanic Stream. Our study combines UV absorption-line observations obtained with \textit{HST/STIS} E140M and E230M gratings and \textit{FUSE}, focusing on early-type stars in the LMC. Additionally, we complement our observations with ancillary \hi \, 21-cm radio emission-line GASS and GASKAP observations associated with the LMC. Our primary findings are as follows.
\begin{enumerate}
    \item \textit{Origin and Distribution of the Gas:} The blueshifted gas in the foreground of 30~Doradus, with speeds up to $\vlsr=+150\,\kms$, is most likely associated with the galactic wind originating from 30~Doradus, based on its radial distribution (see Section~\ref{subsection:radial_dist}). However, several results---for example, those based on the cloud speed vs. column density relation (Section~\ref{subsection:MWvsLMC}), trends in ion ratios (Section~\ref{subsec:low_ion}), and photoionization modeling (Section~\ref{subsection:Cloudy}), in addition to those from other studies---suggest that some of the $+100 \lesssim v_{\rm LSR} \lesssim +150\, \kms$ absorption might also belong to the LMC wind in addition to the Milky Way HVCs. 
    
    \item \textit{Outflow Mass and Outflow Rate:} The integrated low-ion column densities across all wind velocities suggest a greater concentration of outflowing gas in the direction  of 30~Doradus that gradually declines with the projected distance from the center of this starburst region. We estimate a low-ion probed baryon mass outflow of $M_{\rm outfflow}\approx~(5.7 \text{--} 8.6) \times 10^{5} M_{\odot}$ from within $0.52^\circ$ of 30~Doradus. We further estimate a lower limit for the outflow rate to be $\dot{M}_{\rm outflow} \gtrsim~0.02 M_{\odot} \, \text{yr}^{-1}$ and a mass loading factor of $\eta \gtrsim 0.10$. We find that the outflows from this region contributes a significant fraction of the total low-ion outflow mass from the entire LMC.
    \item \textit{Ionization:} The observed ion ratios together with the photoionization modeling reveal that the wind components are multiphased with hydrogen ionization fraction ranging from $40\%$ to $97\%$. We observe both the spatial and kinematic variations in these ionization fractions. The typical smaller average $b$-values and an absence of significant offsets in the velocity centroids between O\textsc{~i} and Si\textsc{~ii} suggest that the gas probed by the low-ions is photoionized. For the high ions, Si\textsc{~iv} and C\textsc{~iv} components are broader and kinematically offset from the low-ions, suggesting that they are susceptible to collisions and are largely consistent with TMLs that exist in the wind.
    \item \textit{Metallicity and Dust Depletion:} Our photoionization modelling reveals the properties for two absorbers that lie at (1) $v_{\text{LSR}} \approx +120\, \kms$ towards Sk$-$69$^\circ$175 and (2) $v_{\text{LSR}} \approx +115\, \kms$ towards BI\, 173. We find evidence for moderate dust depletion for silicon, iron, and nickel of $\delta_\mathrm{S}({\rm Si}) = -0.50 \pm 0.16$, $\delta_\mathrm{S}({\rm Fe}) = -0.42 \pm 0.13$, and $\delta_\mathrm{S}({\rm Ni}) = -0.45 \pm 0.11\,\rm dex$ for the absorber along the Sk$-$69$^\circ$175 sightline. This cloud also has a sub-solar element abundance of [S/H] = $-0.38\pm0.12$, which is consistent with the current day metallicity of the LMC. However, for the cloud towards BI\, 173, we find no evidence of dust ($\delta_\mathrm{O}({\rm Fe}) = -0.02 \pm 0.11\,\rm dex$) and a lower metallicity ($[{\rm O}/{\rm H}] = -0.70 \pm 0.13\,\rm dex$). The mostly neutral, cool, low-metallicity, and dust-free properties suggest that the cloud seen near $\vlsr=+115\,\kms$ is a MW halo HVC. 
\end{enumerate}
Finally, our high-resolution hydrodynamical simulations of the LMC, SMC, and MW system using the GIZMO code suggest that the winds generated in our 30~Doradus-like region are only weakly affected by the direct headwind from the MW's halo. This is because the LMC's corona acts as a shield, protecting the outflows. Given that the LMC might be on its first infall to the MW (e.g., \citealt{2007ApJ...668..949B}) and moving at high velocity (e.g., \citealt{2013ApJ...764..161K}), both the bow shock and the Magellanic Corona play crucial roles in safeguarding the LMC's wind from the MW's circumgalactic gas.

This paper is the first in a series investigating the wind from the LMC and future efforts will include a comprehensive analysis of its global wind patterns, its connection with the activity occurring within this galaxy, and the LMC's halo. 

\acknowledgments

We sincerely thank the anonymous referee for their valuable feedback, which has significantly improved the quality of this paper. Support for this program was provided by NASA through the grant HST-AR-16602.001-A from the Space Telescope Science Institute, which is operated by the Association of Universities for Research in Astronomy, Inc. under NASA contract NAS5-26555. N.L. acknowledges support by NASA through grants HST-AR-16602 and HST-AR-17051 from the Space Telescope Science Institute. Additional support for Horton was provided by NSF grant 2334434. This study used ULLYSES observations obtained with the NASA/ESA Hubble Space Telescope \citep{2020RNAAS...4..205R}, retrieved from the Mikulski Archive for Space Telescopes (MAST: http://archive.stsci.edu listed under \href{https://doi.org/10.17909/t9-jzeh-xy14}{DOI: 10.17909/T9-JZEH-XY14}). STScI is operated by the Association of Universities for Research in Astronomy, Inc. under NASA contract NAS 5-26555. This work has made use of the Vienna Atomic Line Data Base (VALD) database, operated at Uppsala University, the Institute of Astronomy RAS in Moscow, and the University of Vienna. This scientific work uses data obtained from Inyarrimanha Ilgari Bundara / the Murchison Radio-astronomy Observatory. We acknowledge the Wajarri Yamaji People as the Traditional Owners and native title holders of the Observatory site. CSIRO’s ASKAP radio telescope is part of the Australia Telescope National Facility (\url{https://ror.org/05qajvd42}). Operation of ASKAP is funded by the Australian Government with support from the National Collaborative Research Infrastructure Strategy. ASKAP uses the resources of the Pawsey Supercomputing Research Centre. Establishment of ASKAP, Inyarrimanha Ilgari Bundara, the CSIRO Murchison Radio-astronomy Observatory and the Pawsey Supercomputing Research Centre are initiatives of the Australian Government, with support from the Government of Western Australia and the Science and Industry Endowment Fund.
GASKAP-HI is partially funded by the Australian Government through an Australian Research Council Australian Laureate Fellowship (project number FL210100039 awarded to NMc-G).
This study used archived \hi\ LMC data obtained through \url{https://www.astro.uni-bonn.de/hisurvey/AllSky_gauss/}.

\software{Astropy \citep{astropy:2013,astropy:2018,astropy:2022}, Cloudy \citep{ferland2017}, GIZMO \citep{gizmo,gadget},
VoigtFit \citep{2018arXiv180301187K},
galrad (\href{https://github.com/Deech08/galrad}{https://github.com/Deech08/galrad}).
}

\clearpage



\bibliographystyle{aasjournal} 

\appendix \label{section:appendix}

\section{Voigt Profile fitting results}
\label{section:voigt_all}
We present the summary of the Voigt profile fitting results and spectral plotstacks for all the sightlines analyzed in this study in Table~\ref{tab:Voigt_results_cont} and in Figure~\ref{fig:plotstacks}, except for sightline Sk$-$69$^\circ$246 that was presented in Table~\ref{tab:Voigt_results} and Figure~\ref{fig:SK-69d246_lines}. These results are presented in increasing order of the angular separation of these sightlines from the center of 30~Doradus. Although Table~\ref{tab:Voigt_results_cont} include column densities with uncertainties generated by \textsc{VoigtFit} for lines that are saturated, we consider these values to be lower limits. 
For these saturated lines, although we provide the $b$-values, we stress they are also less certain. For these saturated components, we report only values from the profile fits which are consistent with those obtained from the AOD method. Such components are annotated with a footnote (`c' indicating a component that might be saturated and `s' indicating a component that is saturated). For non-detected lines, we provide $3\,\sigma$ upper limits on the column density.

\begin{ThreePartTable}
\centering
\begin{TableNotes}
\footnotesize
  \item[c]{These components could potentially be saturated.}
  \item[s]{These components are saturated.}
  \item[*]{This component is affected by molecular absorption.}
\end{TableNotes}
\begin{longtable}{ccccc}
\caption{Results of Voigt profile-fitting analysis} 
\label{tab:Voigt_results_cont} \\
\hline
\hline
Ion & $v_{\rm LSR}$ & $v_{\rm LMCSR}$  &  $\log{\left(N_x\right)}$ & $b$  \\
    & (\kms) & (\kms) &   & (\kms)  \\
\hline
\endfirsthead
\multicolumn{4}{l}{{\tablename\ \thetable{} -- continued from previous page}} \\
\hline
\hline
Ion & $v_{\rm LSR}$ & $v_{\rm LMCSR}$  &  $\log{\left(N_x\right)}$ & $b$  \\
    & (\kms) & (\kms) &   & (\kms)  \\
\hline
\endhead
\hline \multicolumn{5}{c}{{Continued on next page}} \\ \hline
\endfoot
\insertTableNotes
\endlastfoot

\multicolumn{5}{c}{\textbf{BAT99\,105}} \\ \hline
O I &  +104.5 $\pm$ 0.3\tablenotemark{} & $-164.6$ & 14.12 $\pm$ 0.03 &  6.6 $\pm$ 0.6 \\
    &  +125.6 $\pm$ 0.6\tablenotemark{} & $-143.5$ & 14.06 $\pm$ 0.03 & 8.6 $\pm$ 0.8 \\
    &  +163.2 $\pm$ 0.0\tablenotemark{} & $-105.9$ & 14.03 $\pm$ 0.13 & 6.7 $\pm$ 1.8 \\
Fe II &  +105.2 $\pm$ 0.9\tablenotemark{} & $-163.9$ &  13.30 $\pm$ 0.10 & 2.9 $\pm$ 2.1 \\
      &  +125.2 $\pm$ 1.5\tablenotemark{} & $-143.9$ &  13.32 $\pm$ 0.10 & 7.1 $\pm$ 2.4 \\
      &  +163.6 $\pm$ 2.1\tablenotemark{} & $-105.5$ &  13.47 $\pm$ 0.11 & 9.1 $\pm$ 3.1 \\
      &  +187.0 $\pm$ 0.9\tablenotemark{c} &  $-82.1$ &  14.37 $\pm$ 0.06 & 7.7 $\pm$ 1.4 \\
      &  +208.8 $\pm$ 0.9\tablenotemark{c} &  $-60.3$ &  14.25 $\pm$ 0.05 &  9.4 $\pm$ 1.6 \\
   
Si II &  +104.6 $\pm$ 0.6  \tablenotemark{} & $-164.5$ & 13.32 $\pm$ 0.05 & 5.0 $\pm$ 1.0 \\
      &  +126.2 $\pm$ 0.9\tablenotemark{} & $-142.9$ & 13.34 $\pm$ 0.06 &  7.4 $\pm$ 1.3 \\
      &  +163.2 $\pm$ 0.0\tablenotemark{} & $-105.9$ & 13.37 $\pm$ 0.13 & 8.7 $\pm$ 2.1 \\

S II  & +104.6  & $-109.9$ & $<13.72$ & \nodata  \\
      & +126.2  & $-142.9$ & $<13.63$ & \nodata  \\
     &  +159.2 $\pm$ 2.6\tablenotemark{} & $-109.9$ & 13.94 $\pm$ 0.24 &  6.5 $\pm$ 1.2 \\
     &  +186.2 $\pm$ 1.7\tablenotemark{} &  $-82.9$ & 15.08 $\pm$ 0.04 & 8.0 $\pm$ 0.9 \\
     &  +205.9 $\pm$ 0.6\tablenotemark{} &  $-63.2$ &  15.40 $\pm$ 0.04 & 10.3 $\pm$ 0.7 \\
C IV  &  +96.2 $\pm$ 3.3 & $-172.9$ & 13.63 $\pm$ 0.04 & 39.5 $\pm$ 4.4  \\
      &  +198.7 $\pm$ 5.4\tablenotemark{s} & $-70.4$ & 14.79 $\pm$ 0.14 & 24.7 $\pm$ 2.3 \\
Si IV &  +102.8 $\pm$ 3.6 & $-166.3$ & 13.02 $\pm$ 0.06 &  38.1 $\pm$ 7.2 \\
      &  +196.4 $\pm$ 3.9\tablenotemark{s}  & $-72.7$ & 14.44 $\pm$ 0.13 & 23.1 $\pm$ 1.9 \\
       
\hline\multicolumn{5}{c}{\textbf{Sk$-$68$^\circ$135}} \\ \hline
O I &  +108.3 $\pm$ 0.6\tablenotemark{} & $-161.2$ & 14.08 $\pm$ 0.02 &  10.8 $\pm$ 0.8 \\
    &  +150.2 $\pm$ 0.9\tablenotemark{c} & $-119.3$ & 14.73 $\pm$ 0.06 &  10.4 $\pm$ 0.9 \\
    &  +191.3 $\pm$ 0.0\tablenotemark{s} &  $-78.2$ & 14.92 $\pm$ 0.13 &  12.0 $\pm$ 5.0 \\

Si II &  +107.7 $\pm$ 0.9\tablenotemark{} & $-161.8$ & 14.04 $\pm$ 0.03 & 26.3 $\pm$ 2.9 \\
      &  +150.2 $\pm$ 0.9\tablenotemark{} & $-119.3$ & 14.67 $\pm$ 0.11 & 11.6 $\pm$ 1.1 \\
      &  +187.3 $\pm$ 2.7\tablenotemark{} &  $-82.2$ & 14.92 $\pm$ 0.08 & 16.6 $\pm$ 3.3 \\
      &  +212.0 $\pm$ 0.0\tablenotemark{} &  $-57.5$ & 14.87 $\pm$ 0.09 & 15.6 $\pm$ 1.5 \\    
Fe II &  +103.5 $\pm$ 0.9\tablenotemark{} & $-166.0$ & 13.28 $\pm$ 0.03 & 14.6 $\pm$ 1.6 \\
      &  +147.4 $\pm$ 0.3\tablenotemark{} & $-122.1$ & 13.75 $\pm$ 0.02 & 11.3 $\pm$ 0.6 \\
      &  +191.3 $\pm$ 2.1\tablenotemark{} &  $-78.2$ & 13.92 $\pm$ 0.06 & 16.2 $\pm$ 2.1 \\
      &  +212.0 $\pm$ 0.9\tablenotemark{c} &  $-57.5$ & 13.61 $\pm$ 0.12 &  7.1 $\pm$ 2.4 \\

S II  & +107.7  & $-161.8$ & $<13.66$ & \nodata  \\
      & +150.2  & $-119.3$ & $<13.60$ & \nodata  \\
     &  +197.7 $\pm$ 4.2\tablenotemark{} &  $-71.8$ & 14.50 $\pm$ 0.12 & 16.6 $\pm$ 4.9 \\
     &  +215.7 $\pm$ 1.8\tablenotemark{} &  $-53.8$ & 13.99 $\pm$ 0.35 &  4.8 $\pm$ 4.6 \\

Al II &   +97.5 $\pm$ 0.0\tablenotemark{} & $-172.0$ & 13.14 $\pm$ 0.44 & 61.5 $\pm$ 71.4 \\
      &  +147.1 $\pm$ 3.0\tablenotemark{c} & $-122.4$ & 12.60 $\pm$ 0.30 & 8.1 $\pm$ 4.9 \\
      &  +194.9 $\pm$ 0.0\tablenotemark{c} &  $-74.6$ & 13.04 $\pm$ 0.28 & 23.6 $\pm$ 12.1 \\
      &  +212.0 $\pm$ 0.0\tablenotemark{s} &  $-57.5$ & 12.41 $\pm$ 0.53 &  7.1 $\pm$ 0.0 \\

Al III &  +191.3 $\pm$ 0.0\tablenotemark{} &  $-78.2$ & 12.72 $\pm$ 0.08 & 32.6 $\pm$ 6.4 \\
       &  +235.2 $\pm$ 1.7\tablenotemark{c} &  $-34.3$ & 13.07 $\pm$ 0.14 & 17.1 $\pm$ 2.1 \\
C IV  &  +110.6 $\pm$ 19.5 & $-158.9$ & 13.27 $\pm$ 0.18 & 40.5 $\pm$ 0.8  \\
      &  +183.1 $\pm$ 3.0  &  $-86.4$ & 13.29 $\pm$ 0.11 & 15.3 $\pm$ 4.5  \\
Si IV &  +99.8 $\pm$ 2.7  & $-169.7$ & 12.13 $\pm$ 0.18 & 7.6 $\pm$ 4.2  \\
      &  +185.2 $\pm$ 1.8 & $-84.3$ & 13.12 $\pm$ 0.04 &  24.2 $\pm$ 2.6  \\
  \hline\multicolumn{5}{c}{\textbf{BI\,214}} \\ \hline          
O I &   +85.7 $\pm$ 1.2\tablenotemark{} & $-180.7$ & 14.45 $\pm$ 0.06 & 15.0 $\pm$ 2.6 \\
    &  +118.1 $\pm$ 1.2\tablenotemark{} & $-148.3$ & 14.28 $\pm$ 0.05 & 12.6 $\pm$ 1.7 \\
    &  +162.7 $\pm$ 1.8\tablenotemark{s} & $-103.7$ & 14.99 $\pm$ 0.08 & 14.8 $\pm$ 1.7 \\
    &  +189.9 $\pm$ 0.0\tablenotemark{s} & $-76.5$ & 14.82 $\pm$ 0.44 & 8.0 $\pm$ 3.7 \\

Si II &   +90.8 $\pm$ 0.9\tablenotemark{} & $-175.6$ & 13.70 $\pm$ 0.05 & 13.9 $\pm$ 2.3 \\
      &  +119.3 $\pm$ 0.9\tablenotemark{} & $-147.1$ & 13.65 $\pm$ 0.04 & 10.7 $\pm$ 1.1 \\
      &  +163.6 $\pm$ 2.1\tablenotemark{s} & $-102.8$ & 14.33 $\pm$ 0.09 & 13.2 $\pm$ 1.4 \\
      &  +189.9 $\pm$ 0.0\tablenotemark{s} & $-76.5$ & 14.18 $\pm$ 0.15 & 11.5 $\pm$ 5.5 \\

Fe II &   +90.9 $\pm$ 0.0\tablenotemark{} & $-175.5$ & 13.17 $\pm$ 0.11 &  12.9 $\pm$ 0.0 \\
      &  +119.0 $\pm$ 0.0\tablenotemark{} & $-147.4$ & 13.16 $\pm$ 0.12 & 9.5 $\pm$ 3.5 \\
      &  +160.0 $\pm$ 0.6\tablenotemark{c} & $-106.4$ & 14.01 $\pm$ 0.03 &  8.6 $\pm$ 0.9 \\
      &  +183.7 $\pm$ 1.8\tablenotemark{} &  $-82.7$ &  13.57 $\pm$ 0.14 & 8.5 $\pm$ 3.9 \\
      &  +197.3 $\pm$ 1.5\tablenotemark{} &  $-69.1$ & 13.27 $\pm$ 0.21 &   2.8 $\pm$ 3.6 \\
      &  +214.7 $\pm$ 0.9\tablenotemark{c} &  $-51.7$ & 14.17 $\pm$ 0.03 &   11.5 $\pm$ 1.2 \\

S II  & +90.8  & $-175.6$ & $<13.14$ &  \nodata \\
      & +119.3  & $-147.1$ & $<13.51$ &  \nodata \\
      &  +162.0 $\pm$ 1.2\tablenotemark{} & $-104.4$ & 14.33 $\pm$ 0.05 & 10.9 $\pm$ 1.8 \\
      &  +189.9 $\pm$ 2.1\tablenotemark{} &  $-76.5$ & 14.04 $\pm$ 0.12 &  9.3 $\pm$ 3.5 \\
      &  +216.0 $\pm$ 0.6\tablenotemark{} &  $-50.4$ & 14.68 $\pm$ 0.03 & 10.4 $\pm$ 1.0 \\

C IV  &  +140.6 $\pm$ 17.7 & $-125.8$ &  13.91 $\pm$ 0.07 & 86.4 $\pm$ 4.5 \\
      &  +208.9 $\pm$ 3.0 & $-57.5$ &  14.09 $\pm$ 0.08 & 34.8 $\pm$ 4.5  \\
Si IV &  +113.1 $\pm$ 2.1 & $-153.3$ &  12.57 $\pm$ 0.06 & 19.1 $\pm$ 3.2  \\
      &  +154.0 $\pm$ 1.5 & $-112.4$ &  11.92 $\pm$ 0.17 & 4.0 $\pm$ 3.4  \\
      &  +217.2 $\pm$ 1.2 & $-49.2$ &  13.90 $\pm$ 0.02 & 40.5 $\pm$ 1.2  \\
\hline\multicolumn{5}{c}{\textbf{Sk$-$69$^\circ$175}} \\ \hline
O I &  +120.0 $\pm$ 0.9\tablenotemark{c} & $-146.4$ & 15.21 $\pm$ 0.03 &  15.0 $\pm$ 1.8 \\
    &  +164.3 $\pm$ 0.0\tablenotemark{c} & $-102.1$  & 14.60 $\pm$ 0.18 &  13.0 $\pm$ 8.8 \\
    &  +200.9 $\pm$ 2.4\tablenotemark{} &  $-65.5$ & 15.26 $\pm$ 0.06 &  26.1 $\pm$ 5.5 \\

Si II &  +110.2 $\pm$ 6.0\tablenotemark{} & $-156.2$ & 14.63 $\pm$ 0.17 & 19.0 $\pm$ 8.8 \\ 
      &  +163.4 $\pm$ 0.0\tablenotemark{} & $-103.0$ & 13.59 $\pm$ 0.20 &  10.5 $\pm$ 7.0 \\
      &  +202.5 $\pm$ 0.9\tablenotemark{c} &  $-63.9$ & 14.38 $\pm$ 0.03 & 22.4 $\pm$ 1.9 \\ 

Fe II &  +113.7 $\pm$ 0.6\tablenotemark{} & $-152.7$ & 14.26 $\pm$ 0.06 & 8.5 $\pm$ 1.1 \\
      &  +160.5 $\pm$ 0.0\tablenotemark{} & $-106.0$ & 13.93 $\pm$ 0.15 & 31.3 $\pm$ 15.5 \\
      &  +195.3 $\pm$ 4.2\tablenotemark{} &  $-71.1$ &  14.05 $\pm$ 0.09 & 23.1 $\pm$ 5.9 \\

S II &  +115.9 $\pm$ 0.6\tablenotemark{} & $-150.5$ & 14.72 $\pm$ 0.05 &  5.8 $\pm$ 1.0 \\
     &  +183.7 $\pm$ 4.8\tablenotemark{} & $-82.7$ & 14.80 $\pm$ 0.07 &  37.3 $\pm$ 7.6 \\

Al II &  +108.1 $\pm$ 0.3\tablenotemark{c} & $-158.4$ &  12.80 $\pm$ 0.15 & 10.7 $\pm$ 3.3 \\
      &  +192.4 $\pm$ 4.8\tablenotemark{c} &  $-74.0$ &  12.97 $\pm$ 0.07 &  34.4 $\pm$ 7.3 \\

\hline\multicolumn{5}{c}{\textbf{Sk$-$68$^\circ$112}} \\ \hline
O I &   +90.0 $\pm$ 0.3\tablenotemark{} & $-177.9$ & 14.04 $\pm$ 0.06 &  6.0 $\pm$ 0.9 \\
    &  +115.8 $\pm$ 0.3\tablenotemark{c} & $-152.1$ & 14.40 $\pm$ 0.02 &  10.0 $\pm$ 0.4 \\
    &  +166.7 $\pm$ 0.6\tablenotemark{} & $-101.2$ & 13.85 $\pm$ 0.03 & 10.6 $\pm$ 0.9 \\
    &  +205.4 $\pm$ 0.0\tablenotemark{} & $-62.5$ & 14.01 $\pm$ 0.04 & 13.6 $\pm$ 1.6 \\

Si II &   +88.5 $\pm$ 0.6\tablenotemark{} & $-179.4$ & 13.63 $\pm$ 0.03 &   8.5 $\pm$ 0.8 \\
      &  +116.0 $\pm$ 0.6\tablenotemark{c} & $-151.9$ &  13.83 $\pm$ 0.02 & 11.8 $\pm$ 0.7 \\
      &  +166.9 $\pm$ 0.6\tablenotemark{} & $-101.0$ & 13.57 $\pm$ 0.03 & 11.4 $\pm$ 0.9 \\
      &  +205.4 $\pm$ 0.0\tablenotemark{} &  $-62.5$ & 13.59 $\pm$ 0.05 & 15.5 $\pm$ 2.0 \\

Fe II &   +88.6 $\pm$ 1.2\tablenotemark{} & $-179.3$ &  13.20 $\pm$ 0.10 &  5.4 $\pm$ 2.2 \\
      &  +113.6 $\pm$ 1.8\tablenotemark{} & $-154.3$ &  13.34 $\pm$ 0.09 &  11.4 $\pm$ 3.2 \\
      &  +162.6 $\pm$ 0.0\tablenotemark{} & $-105.3$ &  12.83 $\pm$ 0.17 &  3.1 $\pm$ 4.1 \\
      & +205.4  & $-62.5$ & $<12.61$ &  \\
      &  +232.1 $\pm$ 0.9\tablenotemark{c} &  $-35.8$ &  14.16 $\pm$ 0.03 &  14.2 $\pm$ 1.2 \\

S II & +88.5  & $-179.4$ & $<13.49$ &  \nodata \\
     & +116.0 & $-151.9$ & $<13.45$ & \nodata  \\
     &  +169.4 $\pm$ 2.7\tablenotemark{} & $-98.5$ & 13.99 $\pm$ 0.12 & 13.4 $\pm$ 4.7 \\
     &  +205.4 $\pm$ 0.0\tablenotemark{} & $-62.5$ & 13.80 $\pm$ 0.23 &  11.5 $\pm$ 7.3 \\

C IV  &  +105.0 $\pm$ 3.0 & $-162.9$ &  12.89 $\pm$ 0.21 & 15.2 $\pm$ 5.3  \\
      &  +172.2 $\pm$ 8.1 & $-95.7$ &  13.55 $\pm$ 0.14 & 54.1 $\pm$ 19.5  \\
      
Si IV &  +90.2 $\pm$ 8.1  & $-177.7$  & 12.92 $\pm$ 0.10  & 48.0 $\pm$ 11.2  \\ 
      &  +210.8 $\pm$ 1.8 & $-57.1$ &  13.38 $\pm$ 0.03 & 36.1 $\pm$ 1.8  \\
      &  +219.2 $\pm$ 0.3 & $-48.7$  &  13.18 $\pm$ 0.03 & 9.0 $\pm$ 0.6  \\
\hline\multicolumn{5}{c}{\textbf{BI\,173}} \\ \hline
O I &   +97.6 $\pm$ 2.5\tablenotemark{*} & $-167.2$ &  14.62 $\pm$ 0.36 &     7.5 $\pm$ 13.7 \\
    &  +115.6 $\pm$ 3.6\tablenotemark{} & $-149.2$ &  15.01 $\pm$ 0.08 &     8.5 $\pm$ 8.7 \\
    &  +167.3 $\pm$ 0.0\tablenotemark{c} &  $-97.5$ &  15.46 $\pm$ 0.06 &  18.9 $\pm$ 5.1 \\
    
Si II &  +105.8 $\pm$ 2.4\tablenotemark{c} & $-159.0$ &  14.10 $\pm$ 0.11 & 11.8 $\pm$ 1.8 \\
      &  +124.0 $\pm$ 2.7\tablenotemark{c} & $-140.8$ &  13.79 $\pm$ 0.20 &  9.2 $\pm$ 1.9 \\
      &  +167.3 $\pm$ 0.0\tablenotemark{c} &  $-97.5$ &  13.60 $\pm$ 0.11 &  6.5 $\pm$ 1.9 \\
  
Fe II &  +100.1 $\pm$ 1.2\tablenotemark{} & $-164.7$ & 13.68 $\pm$ 0.15 &  6.0 $\pm$ 1.9 \\
      &  +118.2 $\pm$ 2.7\tablenotemark{} & $-146.6$ &  13.90 $\pm$ 0.05 &  13.2 $\pm$ 3.3 \\
      &  +168.3 $\pm$ 1.2\tablenotemark{} &  $-96.5$ &  13.37 $\pm$ 0.10 &  4.3 $\pm$ 2.1 \\
      &  +205.4 $\pm$ 0.0\tablenotemark{s} &  $-59.4$ &  14.96 $\pm$ 0.13 &  15.9 $\pm$ 1.8 \\

S II & +105.8  & $-159.0$ & $<13.37$ & \nodata  \\
     & +124.0  & $-140.8$ & $<13.29$ &  \nodata \\
     &  +167.3 $\pm$ 0.0  \tablenotemark{} &  $-97.5$ & 13.97 $\pm$ 0.14 &  9.9 $\pm$ 4.9 \\
     &  +205.5 $\pm$ 1.5  \tablenotemark{} &  $-59.3$ & 15.33 $\pm$ 0.05 &  13.6 $\pm$ 1.3 \\

Al II &  +100.1 $\pm$ 0.0\tablenotemark{s} & $-164.7$ & 12.84 $\pm$ 0.09 & 12.9 $\pm$ 3.1 \\
      &  +118.2 $\pm$ 0.0\tablenotemark{s} & $-146.6$ & 12.78 $\pm$ 0.11 & 9.6 $\pm$ 3.1 \\
      &  +162.3 $\pm$ 6.6\tablenotemark{} &  $-102.4$ & 12.49 $\pm$ 0.28 & 17.0 $\pm$ 8.6 \\
      
Al III &  +196.3 $\pm$ 2.0\tablenotemark{} & $-68.5$ & 12.43 $\pm$ 0.06 & 9.0 $\pm$ 2.1 \\     
C IV   &  +103.1 $\pm$ 3.9 & $-161.7$  & 13.01 $\pm$ 0.11 & 20.7 $\pm$ 6.4  \\
       &  +173.7 $\pm$ 2.1 & $-91.1$ & 13.68 $\pm$ 0.04 & 29.2 $\pm$ 3.3  \\
       &  +229.0 $\pm$ 1.5 & $-35.8$  & 13.55 $\pm$ 0.04 & 17.8 $\pm$ 1.9  \\
       
Si IV  &  +191.8 $\pm$ 3.6 & $-72.9$  & 13.15 $\pm$ 0.05 & 39.1 $\pm$ 4.4  \\
       &  +229.6 $\pm$ 0.6 & $-35.2$  & 12.90 $\pm$ 0.06 & 8.5 $\pm$ 1.2  \\
\hline\multicolumn{5}{c}{\textbf{Sk$-$69$^\circ$104}} \\ \hline      
O I &   +94.1 $\pm$ 0.6 \tablenotemark{} & $-167.3$ & 14.02 $\pm$ 0.02 & 10.9 $\pm$ 0.7 \\
    &  +123.4 $\pm$ 0.9\tablenotemark{c} & $-137.9$ & 14.27 $\pm$ 0.04 &  7.7 $\pm$ 0.9 \\
    &  +141.2 $\pm$ 0.6\tablenotemark{c} & $-120.2$ & 14.24 $\pm$ 0.05 &  7.8 $\pm$ 1.4 \\
    &  +160.8 $\pm$ 0.6\tablenotemark{} & $-100.6$ & 13.93 $\pm$ 0.05 &  7.3 $\pm$ 1.2 \\
    &  +183.2 $\pm$ 0.6\tablenotemark{} &  $-78.2$ & 13.83 $\pm$ 0.03 &  8.5 $\pm$ 1.0 \\
    &  +231.5 $\pm$ 1.8\tablenotemark{} &  $-29.9$ & 15.04 $\pm$ 0.08 &  15.1 $\pm$ 1.1 \\

Si II &  +106.2 $\pm$ 1.8\tablenotemark{} & $-155.2$ &  13.54 $\pm$ 0.19 &  11.2 $\pm$ 5.9 \\
      &  +124.6 $\pm$ 0.9\tablenotemark{c} & $-136.8$ & 13.71 $\pm$ 0.10 &  5.7 $\pm$ 1.9 \\
      &  +140.3 $\pm$ 0.9\tablenotemark{c} & $-121.1$ & 13.85 $\pm$ 0.06 &   6.5 $\pm$ 1.6 \\
      &  +159.5 $\pm$ 0.9 \tablenotemark{} & $-101.9$ & 13.58 $\pm$ 0.07 &  8.8 $\pm$ 2.1 \\
      &  +184.5 $\pm$ 0.9\tablenotemark{} &  $-76.9$ & 13.56 $\pm$ 0.05 & 11.2 $\pm$ 1.7 \\
      &  +231.7 $\pm$ 0.9\tablenotemark{} &  $-29.7$ & 14.38 $\pm$ 0.04 & 18.2 $\pm$ 0.7 \\

Fe II &  +105.7 $\pm$ 0.0\tablenotemark{} & $-155.7$ & 12.90 $\pm$ 0.30 & 2.2 $\pm$ 7.9 \\
      &  +126.1 $\pm$ 3.6\tablenotemark{} & $-135.3$ & 13.56 $\pm$ 0.18 &  10.2 $\pm$ 5.8 \\
      &  +143.5 $\pm$ 2.4\tablenotemark{} & $-117.9$ & 13.40 $\pm$ 0.23 &  5.3 $\pm$ 3.8 \\
      &  +160.8 $\pm$ 0.0\tablenotemark{} & $-100.6$ & 12.89 $\pm$ 0.39 &  4.5 $\pm$ 7.5 \\
      &  +183.1 $\pm$ 0.0\tablenotemark{} &  $-78.3$ &  13.38 $\pm$ 0.17 & 15.1 $\pm$ 8.0 \\
      &  +229.8 $\pm$ 1.8\tablenotemark{} &  $-31.6$ &  13.91 $\pm$ 0.08 & 10.3 $\pm$ 2.5 \\

S II  & +106.2  & $-155.2$ & $<13.53$ &  \nodata \\    
      & +124.6  & $-136.8$ & $<13.20$ &  \nodata  \\
      & +140.3  & $-121.1$ & $<13.49$ & \nodata  \\
      & +159.5  & $-101.9$ & $<13.60$ & \nodata  \\
      & +184.5  & $-76.9$ & $<13.75$ & \nodata  \\
      &   +230.3 $\pm$ 0.9\tablenotemark{} & $-31.1$ & 14.59 $\pm$ 0.06 &  7.9 $\pm$ 1.4 \\      

Al II &   +92.3 $\pm$ 4.8\tablenotemark{} & $-169.1$ & 12.38 $\pm$ 0.28 & 13.4 $\pm$ 7.3 \\ 
      &  +128.0 $\pm$ 3.0\tablenotemark{s} & $-133.4$ & 12.99 $\pm$ 0.07 & 21.6 $\pm$ 4.7 \\ 
      &  +160.8 $\pm$ 0.0\tablenotemark{} & $-100.6$ & 11.96 $\pm$ 0.29 & 3.5 $\pm$ 2.7 \\
      &  +185.1 $\pm$ 0.0\tablenotemark{} &  $-76.3$ &  12.33 $\pm$ 0.20 & 18.9 $\pm$ 11.1 \\
      &  +219.7 $\pm$ 4.2\tablenotemark{} &  $-41.7$ &  12.47 $\pm$ 0.25 & 10.1 $\pm$ 4.4 \\
      
C IV &  +124.6 $\pm$ 0.0 & $-136.8$ &  13.80 $\pm$ 0.04 & 59.1 $\pm$ 8.0  \\
     &  +131.6 $\pm$ 1.8 & $-129.8$ &  13.04 $\pm$ 0.13 & 10.0 $\pm$ 3.0  \\
     &  +157.1 $\pm$ 0.9 & $-104.3$ &  13.33 $\pm$ 0.05 & 9.4 $\pm$ 1.3  \\
     &  +197.4 $\pm$ 0.0 & $-64.0$  &  12.82 $\pm$ 0.12 & 7.1 $\pm$ 2.6  \\
     
Si IV & +146.5 $\pm$ 3.0 & $-114.9$ &  13.01 $\pm$ 0.05 & 33.6 $\pm$ 3.9  \\ 
      & +156.9 $\pm$ 1.2 & $-104.5$ &  12.46 $\pm$ 0.12 & 6.9 $\pm$ 2.1  \\
      & +203.8 $\pm$ 2.7 & $-57.6$  &  12.36 $\pm$ 0.12 & 13.3 $\pm$ 3.9  \\ \hline
\end{longtable}
\end{ThreePartTable}
\clearpage
  
\begin{figure}
  \centering
  \includegraphics[width=0.50\textwidth, trim={14.0 25 0 0}, clip]{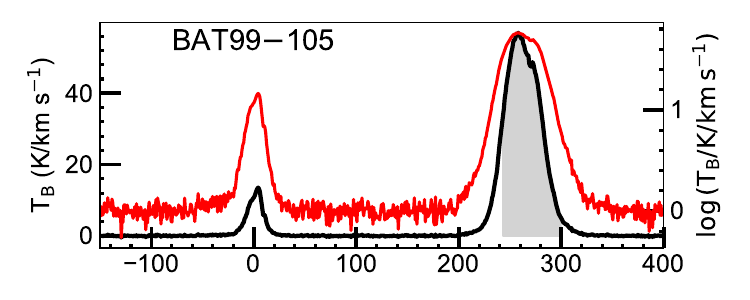} \\
  \includegraphics[width=0.49\textwidth, trim={0 0 0 70}, clip]{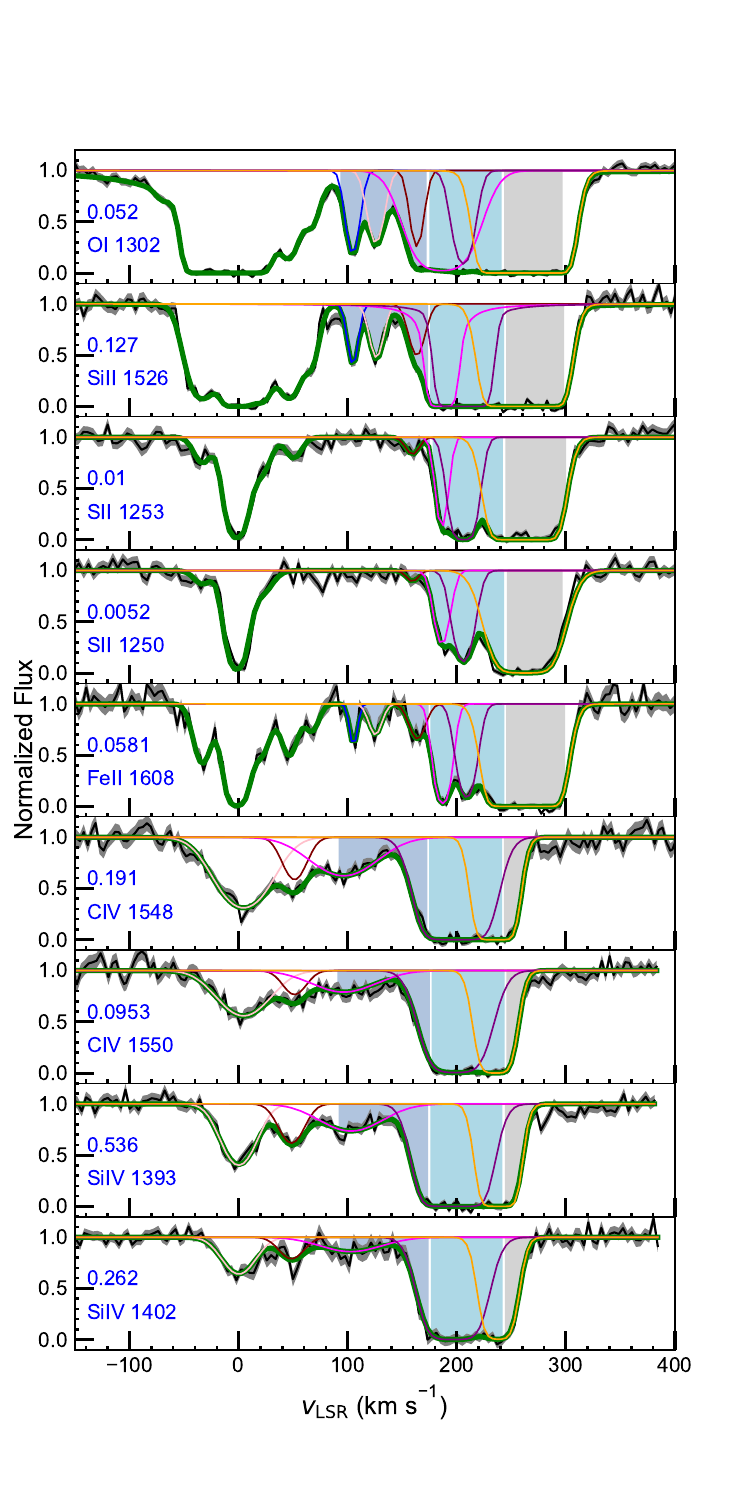}
  \caption{These plotstacks are the same as Figure~\ref{fig:SK-69d246_lines} for sightline ${\rm SK}-69\arcdeg246$, but are for the remaining 7~of the~8 sightlines explored in this study.}
  \label{fig:plotstacks}
\end{figure}

\addtocounter{figure}{-1}
\begin{figure}
  \centering
  \includegraphics[width=0.50\textwidth, trim={14.0 25 0 0}, clip]{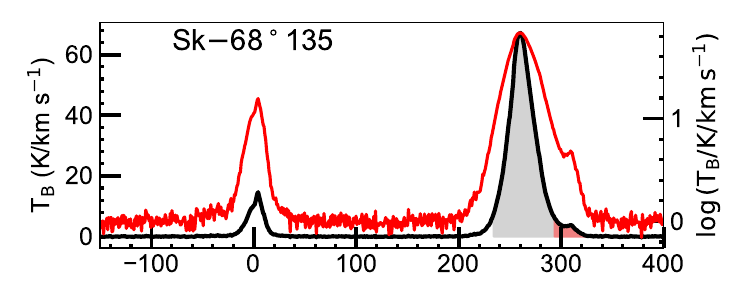}\\
  \includegraphics[width=0.49\textwidth, trim={0 0 0 70}, clip]{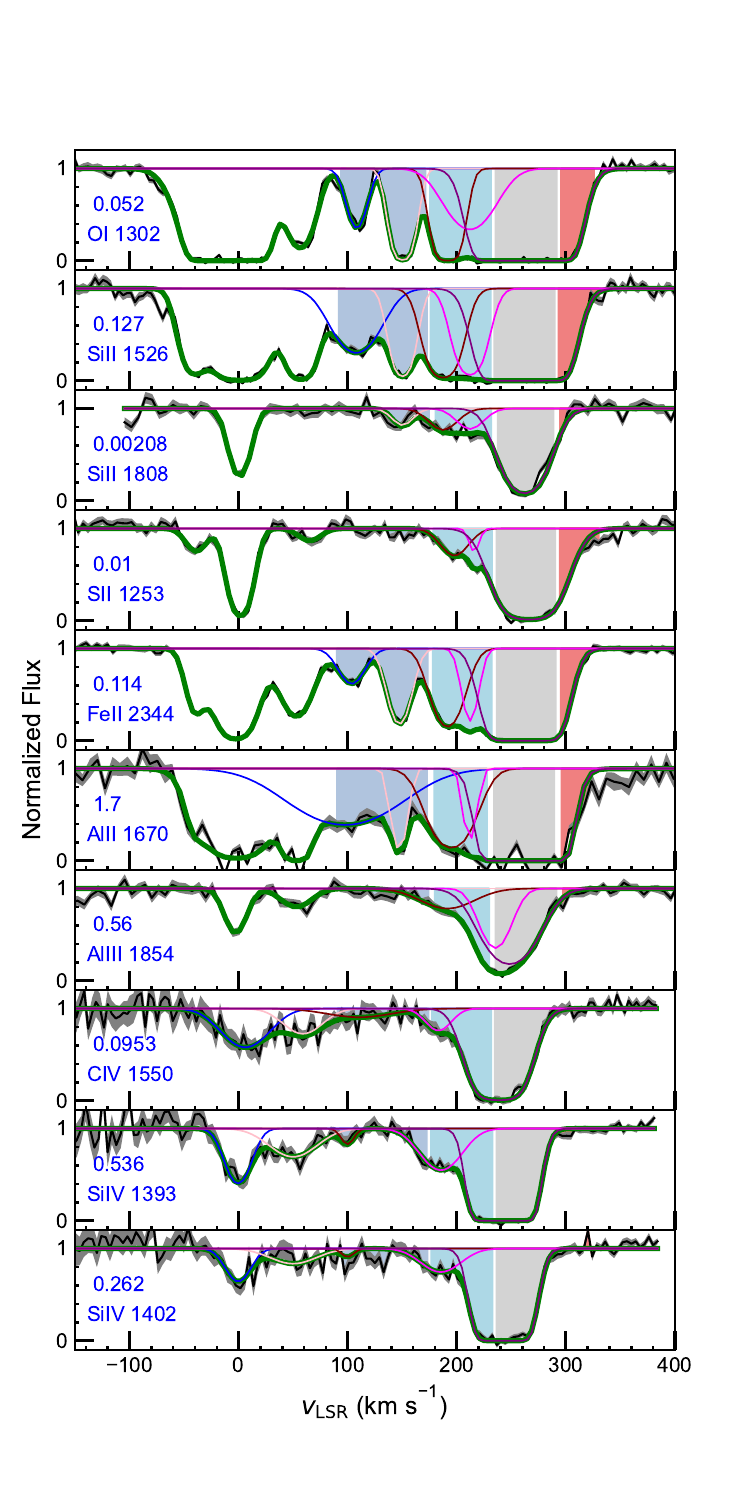}
  \caption{(Continued.)}
\end{figure}

\addtocounter{figure}{-1}
\begin{figure}
  \centering
  \includegraphics[width=0.498\textwidth, trim={8.5 25 0 0}, clip]{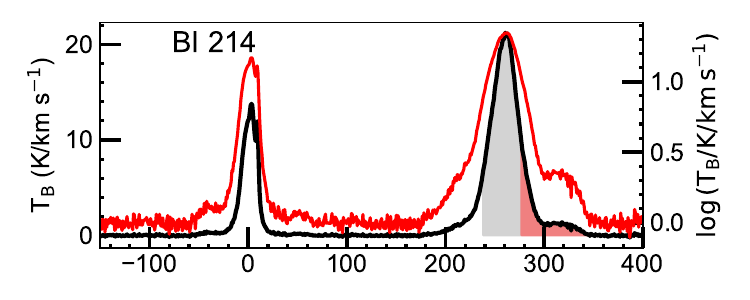}\vspace{0pt}
  \includegraphics[width=0.435\textwidth, trim={7 0 15 70}, clip]{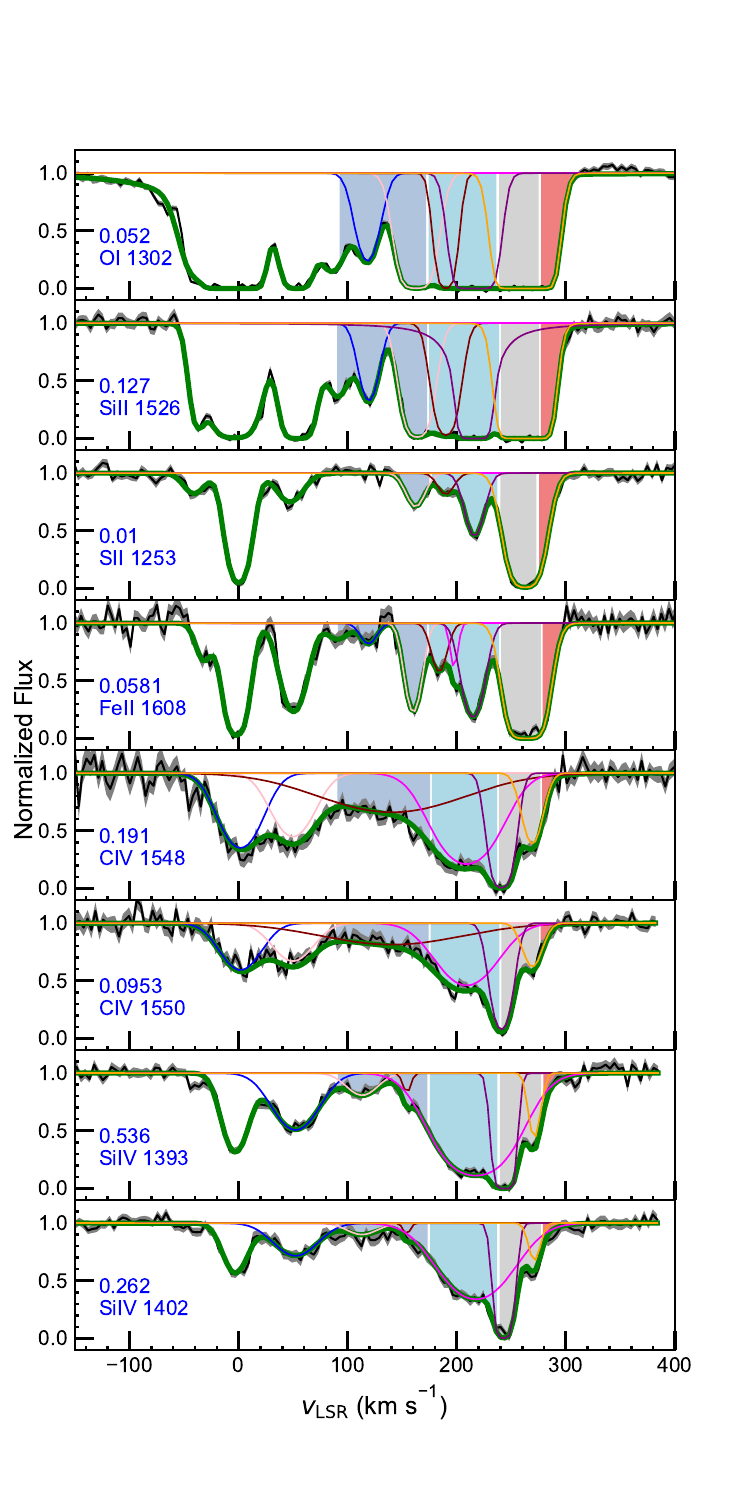}
  \caption{(Continued.)}
\end{figure}

\addtocounter{figure}{-1}
\begin{figure}
  \centering
  \includegraphics[width=0.49\textwidth, trim={10.5 25 0 0}, clip]{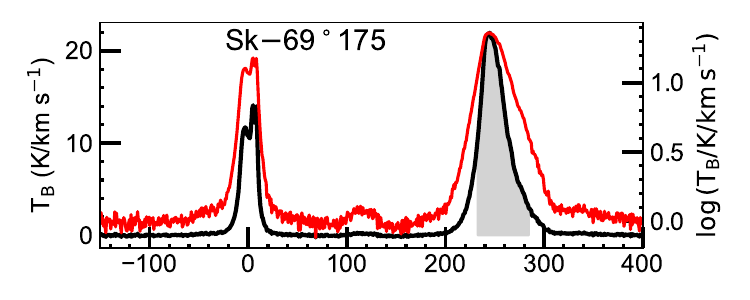}\vspace{0pt}
  \includegraphics[width=0.435\textwidth, trim={10 0 10 70}, clip]{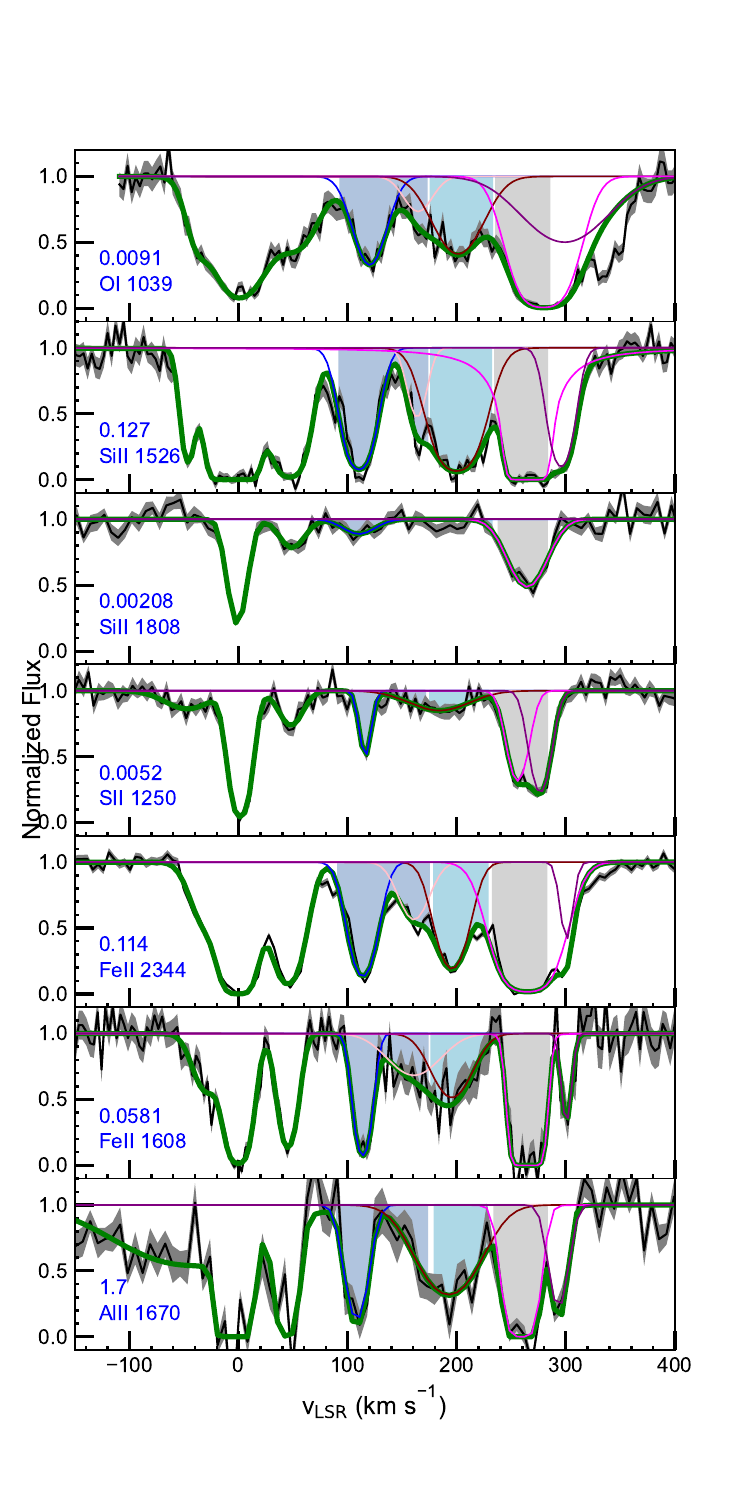}
  \caption{(Continued.)}
\end{figure}

\addtocounter{figure}{-1}
\begin{figure}
  \centering
  \includegraphics[width=0.49\textwidth, trim={11.0 25 0 0}, clip]{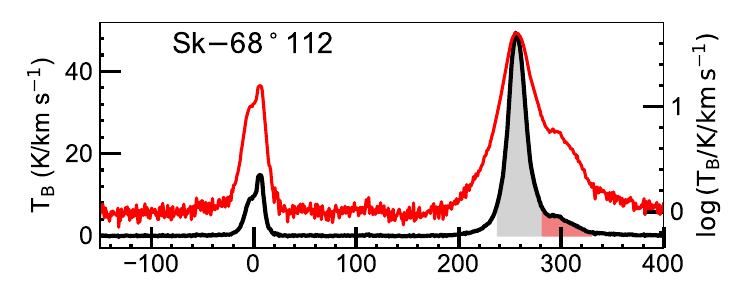}\vspace{0pt}
  \includegraphics[width=0.464\textwidth, trim={0 0 10 70}, clip]{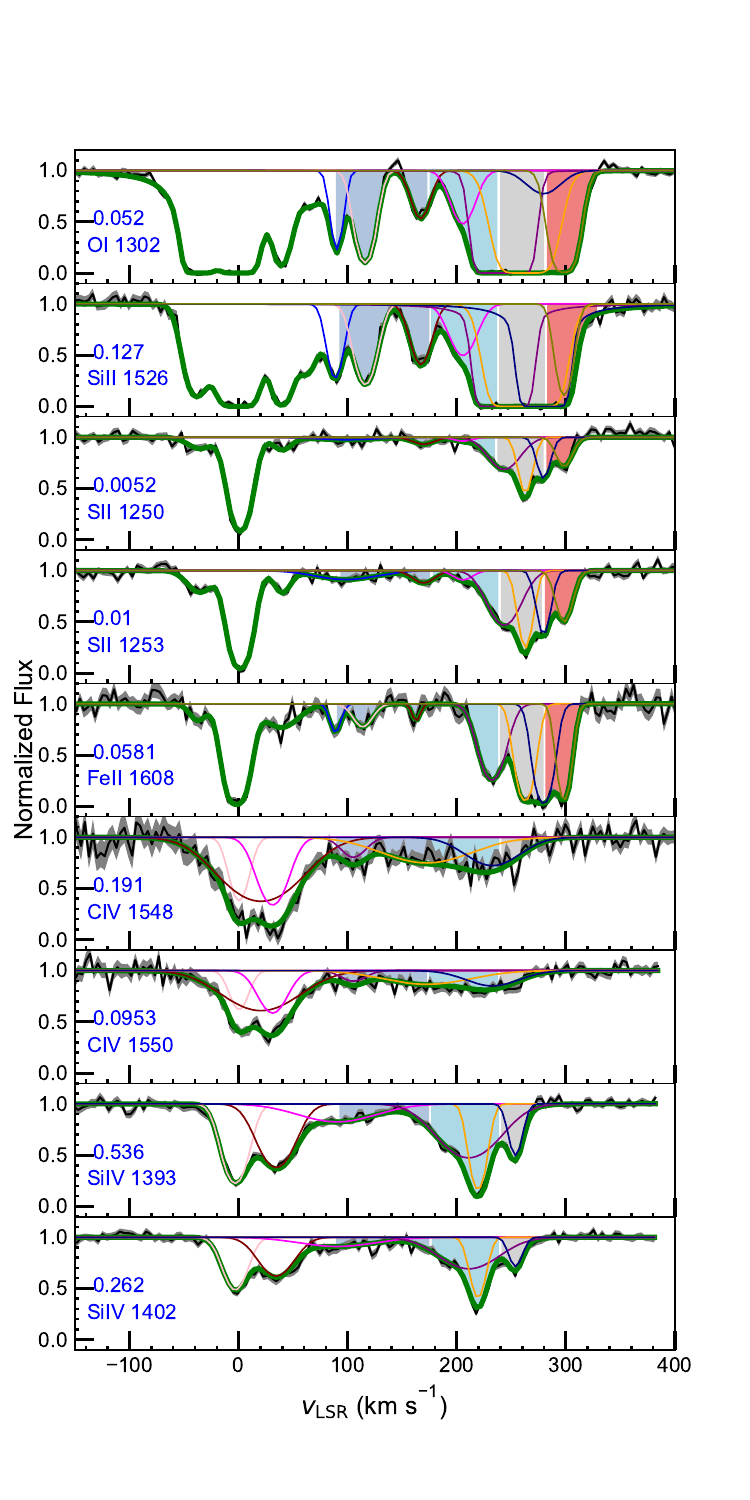}
  \caption{(Continued.)}
\end{figure}

\addtocounter{figure}{-1}
 \begin{figure}
  \centering
  \includegraphics[width=0.485\textwidth, trim={12 25 0 0}, clip]{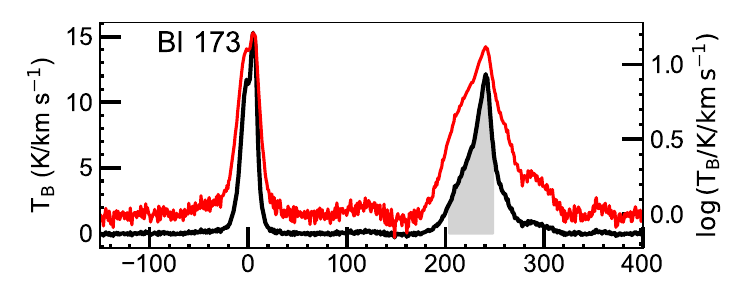}\vspace{0pt}
  \includegraphics[width=0.44\textwidth, trim={10 0 3 70}, clip]{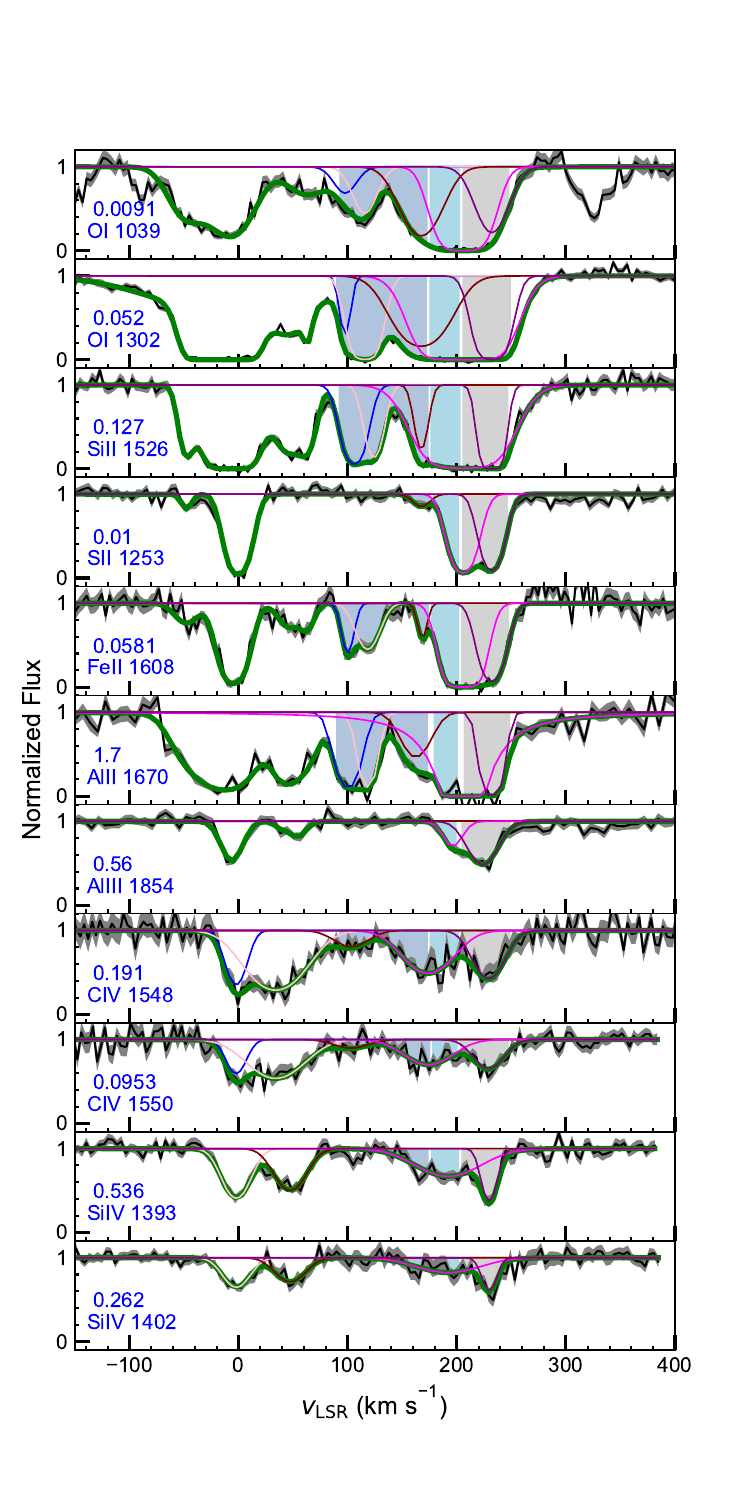}
  \caption{(Continued.)}
\end{figure}

\addtocounter{figure}{-1}
\begin{figure}
  \centering
  \includegraphics[width=0.475\textwidth, trim={8.7 25 13 0}, clip]{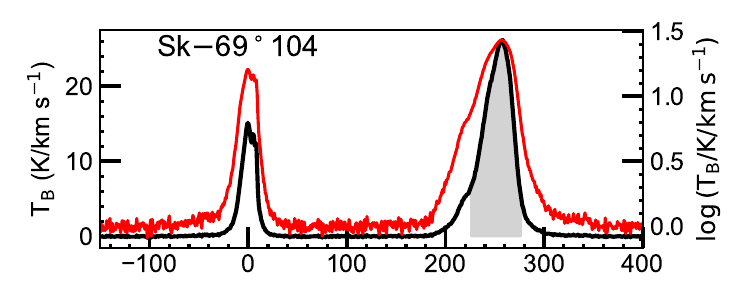}\vspace{0pt}
  \includegraphics[width=0.44\textwidth, trim={5 0 10 70}, clip]{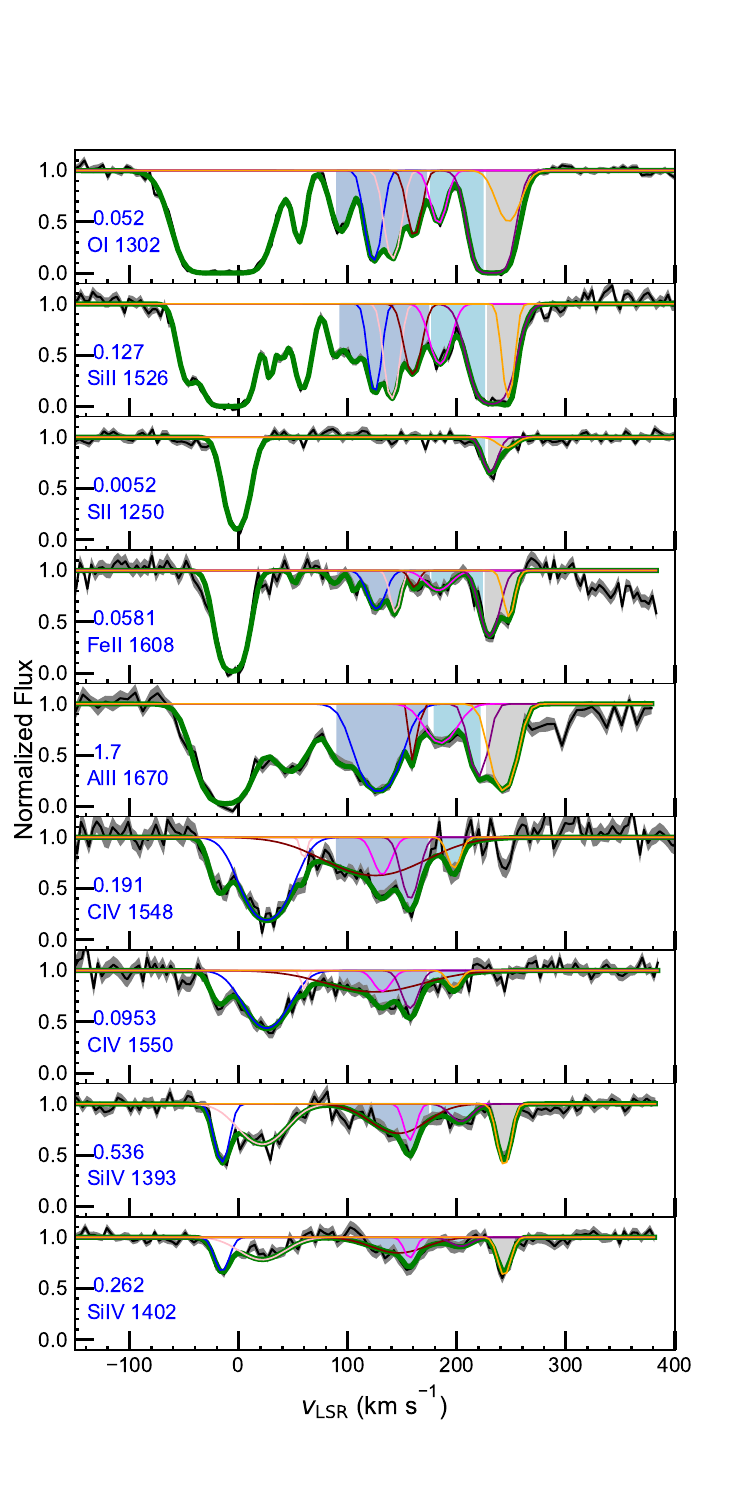}
  \caption{(Continued.)}
\end{figure}

\clearpage

\section{Column density versus velocity}
\label{section:N_v}
\begin{figure}
  \centering
  \includegraphics[width=0.45\textwidth]{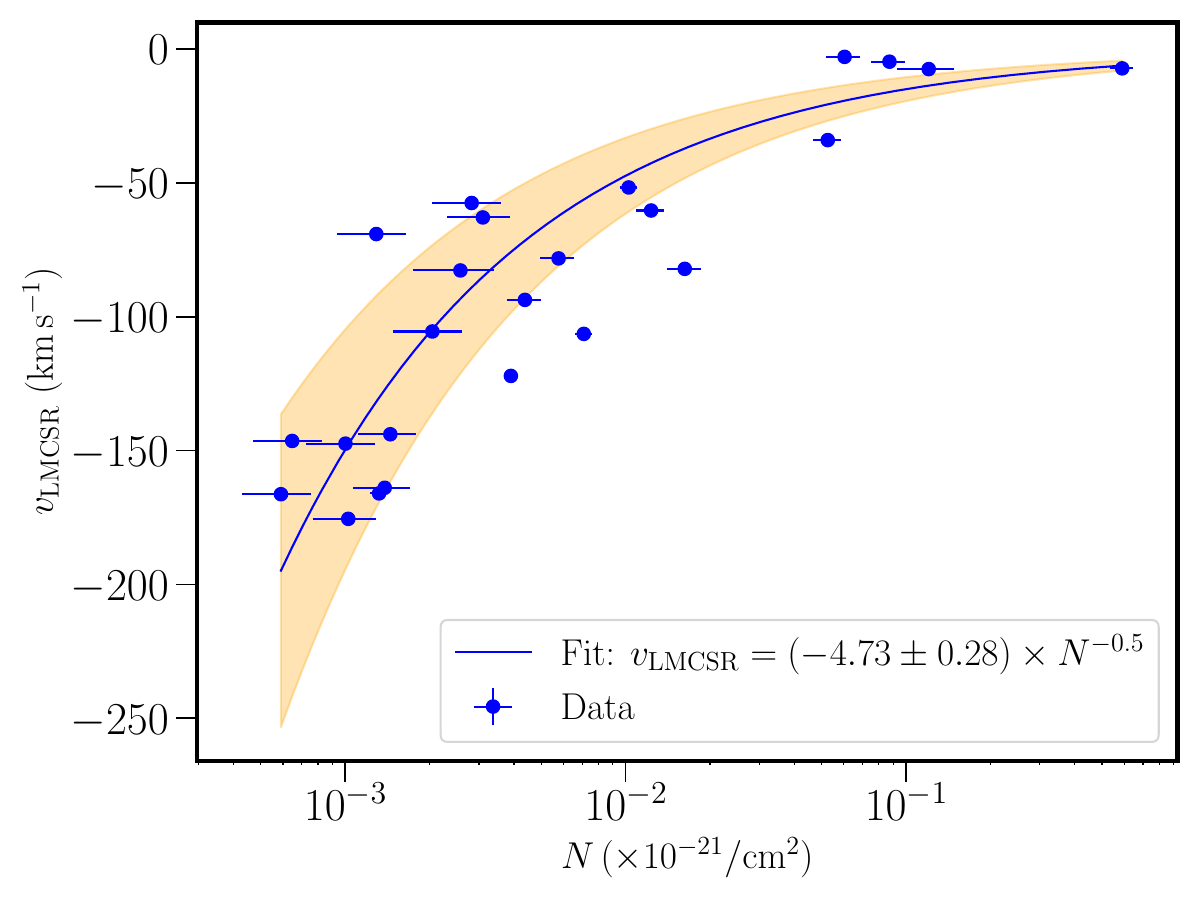}
  \caption{Neutral hydrogen column densities derived from the $N_{\rm Fe\textsc{~ii}}$ column densities assuming an $Z=0.5\,Z_{\odot}$ abundance and no ionization correction. We only include Voigt fitted column densities for the four nearest sightlines to 30~Doradus, with velocities extending to $v_{\mathrm{LMCSR}} \approx -175\, \kms$. The blue trend line and the associated yellow  5$\sigma$ uncertainty envelope represent a fit to the \citet{2002ASPC..254..292H} model that describes the velocity trend of gas clouds that are accelerated by a starburst driven galactic wind.}
  \label{fig:N_v}
\end{figure}

The galactic wind model from \citet{2002ASPC..254..292H} describes starburst-driven, multiphase outflows powered by mechanical energy from supernovae and stellar winds. These winds expel interstellar material into galactic halos and beyond, with cloud velocities (v$_{\mathrm{cloud}}$) determined by ram-pressure acceleration:
    \begin{equation}
        v_{\mathrm{cloud}} \sim 600 \dot{p}_{34}^{1/2} \Omega_w^{-1/2} r_{0}^{-1/2} N_{\mathrm{cloud},21}^{-1/2} \, \text{km/s}
    \end{equation}
    where \( \dot{p}_{34} \) represents the wind’s momentum flux in units of \( 10^{34} \) dynes, \( \Omega_w \) is the wind’s solid angle, \( r_{0}\) is considered to be the radius of the star-forming region in \kpc, and \( N_{\mathrm{cloud},\,21} \) is the cloud’s column density in units of \(10^{21}\)~cm\(^{-2}\). The inverse dependence of \( v_{\mathrm{cloud}} \) on \( N_{\mathrm{cloud},\,21} \)
 implies that denser clouds accelerate more slowly. We explore the relationship between speed of the cloud and component column density (see Figure~\ref{fig:N_v}) following this model and under our assumption that lower column density clouds move faster (see Section~\ref{subsection:MWvsLMC}).We demonstrate that our data can be well-fitted under this assumption, yielding an outflow mass rate comparable to our more conservative estimate (see Section~\ref{section:Discussion}). 

\section{Doppler parameters distribution}
\label{section:b-val}
\begin{figure}
  \centering
  \includegraphics[width=0.45\textwidth]{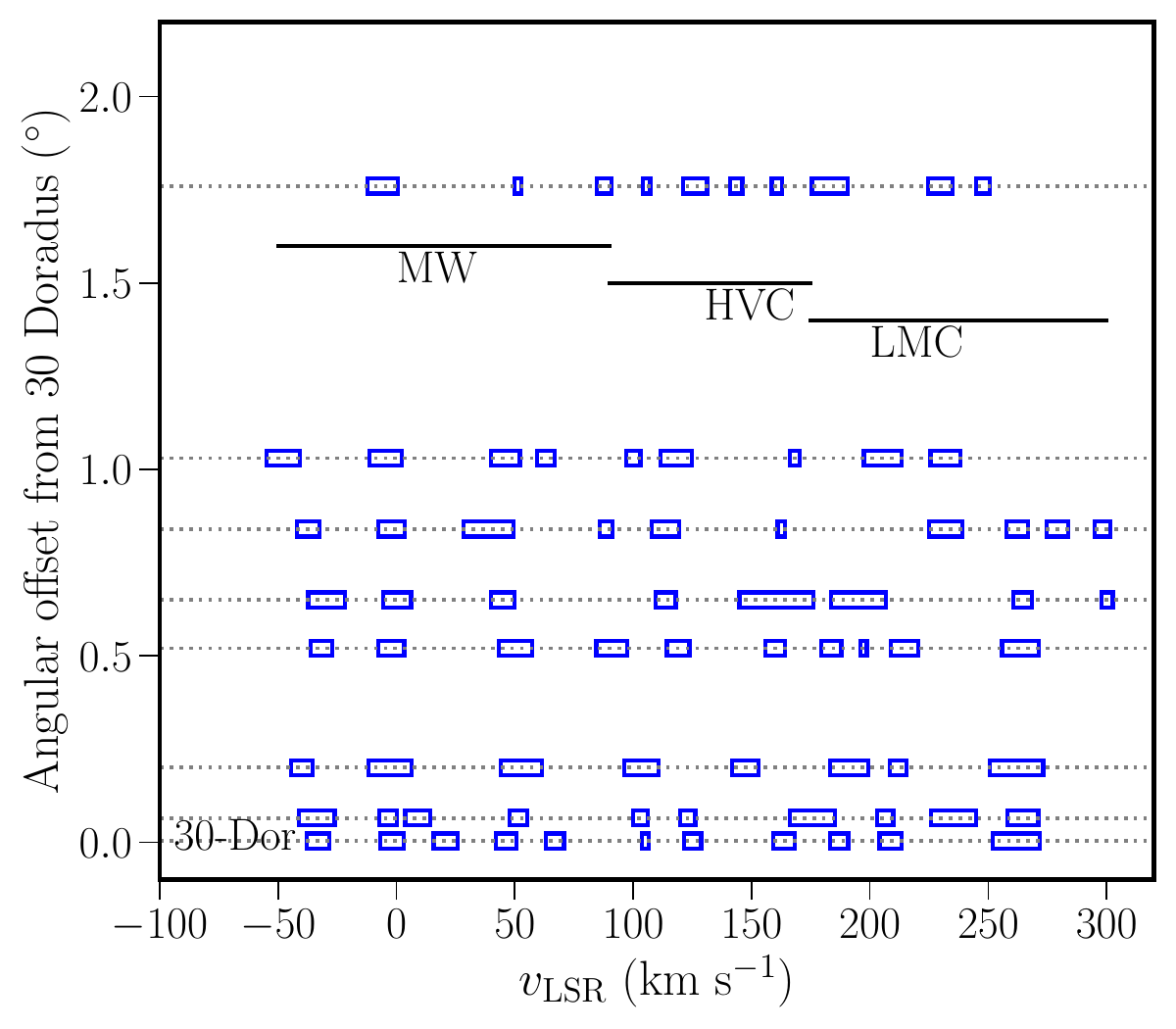}
  \caption{Comparison of Doppler $b$-values distribution across all the 8~sightlines at varying angular separations from 30~Doradus for Fe\textsc{~ii} lines covering the entire velocity range from the MW to the LMC. The width of the rectangular boxes is equal to the magnitude of the Doppler $b$-value. The center of these boxes represents the velocity centroid of the individual components. The horizontal dotted lines are drawn at the corresponding angular separations of these sightlines from the center of 30~Doradus. The three horizontal lines in black show the approximate velocity ranges for the MW, HVC, and the LMC region.}
  \label{fig:doppler_all}
\end{figure}

We compare the magnitude of $b$-values across the entire velocity range from the MW to the LMC using Fe\textsc{~ii} (see  Figure~\ref{fig:doppler_all}). We selected Fe\textsc{~ii} because it offers multiple transitions (Fe\textsc{~ii}\,$\lambda1608$, 2249, 2260, 2344) which aids in obtaining robust measurements for the line widths. Also, the combination of these lines has the advantage of providing a substantial number of detectable components without being too weak or too strong to saturate when compared to other ions. Furthermore, to ensure the reliability of $b$-values, we retained those components that were significantly detected, specifically those with a FWHM greater than the instrumental profile.

To test whether or not the the linewidths in these kinematicaly different regions are drawn from the same underlying population, we used the Anderson-Darling statistical test. We found that there is no statistically significant differences in $b$-values among these regions at the 95\% confidence level. When we subdivide the galactic wind region further into two velocity regimes: (1) a fast-moving component ($+150\lesssim$ $v_{\rm LSR} \lesssim +175\,\kms$) exhibits a mean $b$-value ($10.1 \pm 9.7\,\kms$) and (2) a slower-moving counterpart ($+175 \lesssim v_{\rm LSR} \lesssim +210\, \kms$) exhibits a larger mean $b$-value of $12.5 \pm 6.1\,\kms$. This small contrast in $b$-values between these groups may be attributed to the blending of weaker components within the slower-moving ones. Moreover, we also do not observe any specific trend in the magnitude of $b$-values with respect to the angular distance from the 30~Doradus. We discuss the linewidths of both the low and high ions further in Section~\ref{subsec:b_dist}. Overall, we did not find a statistically significant difference in the mean $b$-values between any of these velocity sub-regions.

\end{document}